\documentclass[twocolumn]{aastex701}
\usepackage{amsmath}	

\usepackage{CJKutf8}
\usepackage{makecell}

\newcommand{\todo}{\ifmmode \text{\color{red}\Huge{\(\bullet\)}} \else {\color{red}{\Huge$\bullet$}}\fi}

\newcommand{\sbsc}[1]{_\mathrm{#1}}

\newcommand{\Ihi}{\ensuremath{I_{_{\mathrm{HI}}}}}
\newcommand{\co}[2]{\mbox{$\mathrm{CO}\,(#1\text{--}#2)$}}
\newcommand{\hi}{\ifmmode {\rm H}\,\textsc{i} \else H\,\textsc{i}\fi}
\newcommand{\htwo}{{H$_2$}}

\newcommand{\alphaCOline}[2]{\alpha\sbsc{CO(#1\text{--}#2)}}

\newcommand{\rgal}{\ensuremath{r_{\mathrm{gal}}}}

\newcommand{\Riso}{r_{25}}

\newcommand{\sigatom}{\ensuremath{\mathrm{\Sigma_{\rm atom}}}}
\newcommand{\sigatomunit}{\ensuremath{\mathrm{M_{\odot}\,pc^{-2}}}}
\newcommand{\sigmol}{\ensuremath{\mathrm{\Sigma_{\rm mol}}}}

\newcommand{\sigsfr}{\ensuremath{{\Sigma_\mathrm{SFR}}}}

\newcommand{\halpha}{\ensuremath{{\mathrm{H\alpha}}}}

\newcommand{\sigsfrunit}{\ensuremath{\mathrm{M_{\odot}\,yr^{-1}\,kpc^{-2}}}}
\newcommand{\siggas}{\ensuremath{\mathrm{\Sigma_{gas}}}}

\newcommand{\Ifuv}{\ensuremath{I_{\mathrm{FUV}}}}

\newcommand{\sigstar}{\ensuremath{\mathrm{\Sigma_{\star}}}}

\newcommand{\tdephi}{\ensuremath{\tau{_{\mathrm{dep}}^{\mathrm{atom}}}}}
\newcommand{\tdepmol}{\ensuremath{\tau{_{\mathrm{dep}}^{\mathrm{mol}}}}}
\newcommand{\tdepgas}{\ensuremath{\tau{_{\mathrm{dep}}^{\mathrm{gas}}}}}

\newcommand{\uI}{\mbox{$\rm MJy~sr^{-1}$}}
\newcommand{\uIco}{\mbox{$\rm K~km~s^{-1}$}}

\newcommand{\uSig}{\mbox{$\rm M_\odot~pc^{-2}$}}

\newcommand{\uSigSFR}{\mbox{$\rm M_\odot~yr^{-1}~kpc^{-2}$}}

\newcommand{\uIha}{\mbox{$\rm erg~s^{-1}\ cm^{-2}\ arcsec^{-2}$}}

\newcommand{\Iwiseone}{I\sbsc{3.4\,\mu\mathrm{m}}}

\newcommand{\MtoLwiseone}{\Upsilon_{3.4\,\mu \mathrm{m}}}

\begin{document}
\title{A Multiwavelength Inventory for the Local Group L-band Survey I: Atlas and Radial Profiles of Local Group Galaxies}
\newcommand{\NRAO}{\affiliation{National Radio Astronomy Observatory, 520 Edgemont Road, Charlottesville, VA 22903, USA}}

\newcommand{\OSU}{\affiliation{Department of Astronomy, The Ohio State University, 140 West 18th Avenue, Columbus, OH 43210, USA}}

\newcommand{\CCAPP}{\affiliation{Center for Cosmology and Astroparticle Physics (CCAPP), 191 West Woodruff Avenue, Columbus, OH 43210, USA}}

\newcommand{\Princeton}{\affiliation{Department of Astrophysical Sciences, Princeton University, 4 Ivy Lane, Princeton, NJ 08544, USA}}

\newcommand{\Kentucky}{\affiliation{Department of Physics and Astronomy, University of Kentucky, 506 Library Drive, Lexington, KY 40506, USA}}

\newcommand{\CfA}{\affiliation{Center for Astrophysics $\mid$ Harvard \& Smithsonian, 60 Garden Street, Cambridge, MA 02138, USA}}

\newcommand{\astron}{\affiliation{Netherlands Institute for Radio Astronomy (ASTRON),  Oude Hoogeveensedijk 4, 7991 PD Dwingeloo, Netherlands}}

\newcommand{\kapeyn}{\affiliation{Kapteyn Astronomical Institute, University of Groningen, PO Box 800, 9700 AV Groningen, The Netherlands}}

\newcommand{\UCT}{\affiliation{Department of Astronomy, University of Cape Town, Private Bag X3, 7701 Rondebosch, South Africa}}

\newcommand{\STScI}{\affiliation{Space Telescope Science Institute, 3700 San Martin Drive, Baltimore, MD 21218, USA}}

\author[0000-0002-1185-2810]{Cosima Eibensteiner}
\email{ceibenst@nrao.edu, cosimaeibensteiner.astro@gmail.com}
\altaffiliation{Jansky Fellow of the National Radio Astronomy Observatory}
\NRAO

\author[0000-0002-2545-1700]{Adam~K.~Leroy}
\email{leroy.42@osu.edu}
\OSU
\CCAPP
\author[0000-0003-0378-4667]{Jiayi~Sun \begin{CJK*}{UTF8}{gbsn}(孙嘉懿)\end{CJK*}}
\email{jiayi.sun@uky.edu}
\Kentucky

\author[0000-0002-5204-2259]{Erik~Rosolowsky}
\email{rosolowsky@ualberta.ca}
\affiliation{Department of Physics, 4-183 CCIS, University of Alberta, Edmonton, AB T6G 2E1, Canada}

\author[0000-0001-9605-780X]{Eric~W.~Koch}
\email{ekoch@nrao.edu}
\NRAO

\author[0000-0001-9504-7386]{Nickolas Pingel}
\email{}
\affiliation{University of Wisconsin–Madison, Department of Astronomy, 475 N Charter St, Madison, WI 53703, USA}
\affiliation{Department of Astronomy, Indiana University, 727 East Third Street, Bloomington, IN 47405, USA}

\author[0000-0003-2896-3725]{Chang-Goo~Kim}
\email{changgookim@gmail.com}
\Princeton

\author[0000-0002-8400-3705]{Laura~B.~Chomiuk}
\email{chomiuk@pa.msu.edu}
\affiliation{Center for Data Intensive and Time Domain Astronomy, Department of Physics and Astronomy, Michigan State University, East Lansing, MI 48824, USA}

\author[0000-0001-8241-7704]{Ryan Chown}
\email{ryan.chown@algomau.ca}
\affiliation{Faculty of Computer Science \& Technology, Algoma University, Sault Ste. Marie, ON P6A 2G4, Canada}

\author[0000-0002-5480-5686]{Alberto D. Bolatto}
\email{bolatto@umd.edu}
\affiliation{Department of Astronomy and Joint Space-Science Institute, University of Maryland, College Park, MD 20742, USA}

\author[0000-0003-4961-6511]{Michael P. Busch}
\email{mpbusch@nrao.edu}
\altaffiliation{Jansky Fellow of the National Radio Astronomy Observatory}
\affiliation{National Radio Astronomy Observatory, 520 Edgemont Road, Charlottesville, VA 22903, USA}

\author[0000-0002-1264-2006]{Julianne J. Dalcanton}
\email{}
\affiliation{Center for Computational Astrophysics, Flatiron Institute, 162 Fifth Avenue, New York, NY 10010, USA}
\affiliation{Department of Astronomy, University of Washington, Box 351580, Seattle, WA 98195, USA}

\author[0000-0002-3227-4917]{Amanda A. Kepley}
\email{akepley@nrao.edu}
\NRAO

\author[0000-0003-0588-7360]{Christina W. Lindberg}
\email{christina.lindberg@cfa.harvard.edu}
\affiliation{Center for Astrophysics $|$ Harvard \& Smithsonian, 60 Garden Street,
Cambridge, MA 02138, USA}

\author[0000-0002-0509-9113]{Eve~C.~Ostriker}
\email{eco@astro.princeton.edu}
\Princeton

\author[0000-0001-8224-1956]{Jürgen Ott}
\email{jott@nrao.edu}
\affiliation{National Radio Astronomy Observatory, 1011 Lopezville Road, Socorro, NM 87801, USA}

\author[0000-0002-4781-7291]{Sumit K. Sarbadhicary}
\email{ssarbad1@jh.edu}
\affiliation{Department of Physics and Astronomy, The Johns Hopkins University, Baltimore, MD 21218 USA}

\author[0000-0003-2599-7524]{Adam Smercina}\thanks{NHFP Hubble Fellow}
\email{asmercina@stsci.edu}
\STScI

\author[0000-0002-3418-7817]{Sne{\v z}ana Stanimirovi{\'c}}
\email{sstanimi@wisc.edu}
\affiliation{University of Wisconsin–Madison, Department of Astronomy, 475 N Charter St, Madison, WI 53703, USA}

\author[0000-0003-1356-1096]{Elizabeth Tarantino} 
\email{etarantino@stsci.edu}
\STScI

\author[0000-0002-5877-379X]{Vicente Villanueva}
\email{vicente.avl365@gmail.com}
\affiliation{Instituto de Estudios Astrof\'isicos, Facultad de Ingenier\'ia y Ciencias, Universidad Diego Portales, Av. Ej\'ercito Libertador 441, 8370191 Santiago, Chile}
\affiliation{Millennium Nucleus for Galaxies, MINGAL}

\author[0000-0001-6320-2230]{Tobin M. Wainer}
\email{tobinw@uw.edu}
\affiliation{Department of Astronomy, University of Washington, Box 351580, Seattle, WA 98195, USA}

\author[0000-0003-4793-7880]{Fabian~Walter}
\email{walter@mpia.de}
\affiliation{Max Planck Institut für Astronomie (MPIA) K\"onigstuhl 17, 69117 Heidelberg, Germany}

\author[0000-0002-0012-2142]{Thomas G. Williams}
\email{thomas.g.williams@manchester.ac.uk}
\affiliation{UK ALMA Regional Centre Node, Jodrell Bank Centre for Astrophysics}
\affiliation{Department of Physics and Astronomy, The University of Manchester, Oxford Road, Manchester M13 9PL, UK}

%
\collaboration{all}{The LGLBS collaboration}

\begin{abstract}
{Resolving atomic gas at $\lesssim100$~pc physical resolution is currently feasible only for the very nearest galaxies.}
We combine 120~pc resolution \hi\ data {using the Karl G. Jansky Very Large Array (VLA) }from the first data release of the Local Group L-Band Survey (LGLBS) with uniformly processed infrared, ultraviolet, H$\alpha$, and CO maps.
We analyze the radial profiles of gas, stars, and recent star formation for six local galaxies: M31, M33, IC~10, IC~1613, NGC~6822, and WLM. 
Across the sample, atomic-gas disks are the most extended component, reaching $r_{\rm HI}=2.8$--$5.8$~kpc in the dwarfs and $13$--$26$~kpc in M33 and M31. Azimuthally averaged \sigatom\ profiles are often flat or only slowly declining compared to the steeper decline of \sigstar\ and \sigsfr, with outer \hi\ scale lengths of $l_{\rm HI}^{\rm outer}\approx1.0$--$8.5$~kpc. The mass-weighted \hi\ surface densities on 120~pc scale generally exceed the azimuthally averaged values by 10--70\%, showing that localized high-column-density atomic structures persist even where azimuthally averaged profiles appear smooth. Atomic-gas depletion times rise from $\tau_{\rm dep}^{\rm atom}\approx0.8$--17~Gyr within the stellar half-mass radius to 10s--100~Gyr in the outer disks. Molecular depletion times are shorter. 
We provide a public release of these multi-wavelength data, which provide a valuable reference for the LGLBS targets and also form the basis for a companion paper that relates star formation, gas phases, and midplane pressure on 120~pc scales.
\end{abstract}



\section{Introduction} 
\label{sec:intro}

Neutral, atomic hydrogen (\hi) represents the dominant component of the interstellar medium (ISM) of galaxies \citep[e.g.,][]{Walter2020,Saintonge2022}. \hi\ serves as the material for molecular cloud formation, and the resulting molecular clouds form stars. The time for star formation to consume the molecular gas (H$_2$) reservoir is short compared to the typical age of $z=0$ galaxies \citep[e.g.,][]{Tacconi2020}. Therefore, galaxy growth requires continuous cycling of \hi\ into H$_2$. But this process does not occur with the same efficiency everywhere. Instead, galaxies exhibit a wide range of molecular gas fractions. For example, the outer parts of disk galaxies and dwarf galaxies can show little molecular gas or star formation activity despite abundant \hi, while the inner parts of disk galaxies often have abundant molecular gas despite sometimes showing \hi\ depressions
\citep[e.g.,][]{Schruba2011,Hunter2024}.

H$_2$ forms and persists in gas that is shielded from dissociating radiation \citep[by themself or by dust; e.g.,][]{Sternberg2014,Sternberg2023}. Attaining the column and volume densities needed for high H$_2$ abundance depends on metallicity as well as dynamical processes that compress diffuse gas, including gravity, converging and/or turbulent flows, and thermal instability. These are opposed by processes that disperse dense clouds, including galactic shear as well as radiative and supernovae feedback from young, massive stars \citep[e.g.,][]{Dobbs2014,Chevance2023}.

The angular resolution of \hi\ maps produced by 21-cm observations is often poorer compared to that of optical or millimeter wave facilities. As a result, star formation self-regulation and the H$_2$-to-\hi\ transition have been studied in less physical detail beyond the Milky Way, e.g., compared to the explosion in ALMA observations of the molecular gas in galaxies \citep[e.g.,][]{Lin2020,Leroy2021_survey,Brown2021}. One way around this limitation is to study 21-cm emission from the nearest galaxies. There, current facilities can reach the $< 100$~pc physical resolution needed to access signatures of cold cloud formation and stellar feedback. This has motivated surveys of the Magellanic Clouds \citep[e.g., GASKAP;][]{Dickey2013,Dempsey2022} and local volume dwarf galaxies \citep[e.g., LITTLETHINGS, VLA-ANGST;][]{Hunter2012,Ott2012}. \citet{Hunter2024} summarizes key results from these surveys.

With this goal of resolving stellar feedback and cloud formation signatures, the Local Group L Band Survey \citep[LGLBS;][]{Koch2025} used all four configurations of the Karl G. Jansky Very Large Array (VLA) to survey the six actively star-forming Local Group ($D < 1.2$~Mpc) galaxies visible from the observatory: M31, M33, and the dwarf galaxies IC~10, IC~1613, NGC~6822, and WLM. The first data release in \citet{Koch2025} already provided sensitive \hi\ maps of all targets at $< 120$~pc resolution, and the full joint imaging products reaches $\lesssim 30$~pc \citep[e.g.,][]{Pingel2024}.

A key step to utilize these data to study star formation self-regulation and the H$_2$-to-\hi\ balance is to combine the \hi\ data with estimates of the molecular gas, gas, stars, and recent star formation. This paper addresses this exact goal. We homogenize \hi, CO, infrared, H$\alpha$, and ultraviolet measurements and construct radial profiles of atomic gas, molecular gas, stellar mass, and recent star formation. Then we use these profiles to measure disk scale lengths, galaxy sizes, gas clumping, integrated masses, star formation rates, and gas depletion times. A companion paper uses these data products to measure the connection between gas depletion time, molecular fraction, and ISM conditions.

\section{Data} \label{sec:data}

\begin{deluxetable*}{lccccccccccc}[t!]
\tabletypesize{\footnotesize}
\tablecaption{\label{tab:physical} Physical Properties of  Local Galaxies L-Band Survey (LGLBS) Targets}
\tablewidth{0pt}
\tablehead{
\colhead{Galaxy} & 
\colhead{Distance} & 
\colhead{R.A.} &
\colhead{Dec.} &
\colhead{Inclination} & 
\colhead{P.A.} &
\colhead{$r_{25}$} &
\colhead{$\log_{10} M_{\rm HI} $} & 
\colhead{$\log_{10} M_\star$} & 
\colhead{$\log_{10} {\rm SFR}$} &
\colhead{$12+\log_{10} {\rm O/H}$} \\
&  [kpc] & (J2000) & (J2000) & [$^{\circ}$] & [$^{\circ}$] & [$''$] & [$ {\rm M_\odot}$]& [$ {\rm M_\odot}$] & [$ {\rm M_\odot~yr^{-1}}$]  \\
&  (1) & (2) & (3) & (4)&  (5) & (6) & (7) & (8) & (9)& (10)
}
\startdata
IC~10     &$770 \pm 100$ &  00:20:23.1 & +59:17:35   & 47    & 65    & 203   &7.88 & 8.7 &  -1.7 & 8.37  \\
IC~1613   &$760 \pm 36$  &  01:04:47.8 & +02:07:04   & 48    & 73    & 546   &7.80 & 7.8 &  -2.2 & 7.73  \\
NGC~6822  &$526 \pm 25$  &  19:44:57.0 & $-$14:48:01 & 66    & 118   & 465   & 8.25 & 8.1& -2.1 &8.23  \\
WLM       &$984 \pm 19$  &  00:01:58.2 & $-$15:27:39 & 74    & 174   & 314    & 7.91& 7.4 & -2.4 & 7.83  \\
M33       &$859 \pm 24$  &  01:33:50.9 & +30:39:36  &  55    & 201   & 1850  & 9.30 & 9.6 & -0.5  &8.59  \\
M31       &$776 \pm 22$  &  00:42:44.3 & +41:16:08  &  77    & 35    & 5335  & 9.80  &10.6 & -0.4 &8.94  \\
\enddata
\tablecomments{(1) LGLBS adopted distances from \cite{Koch2025} that are based on \cite{Dellagli2018,Sanna2008,Gerbrandt2015},\cite{Lee2024dist}, \citet{Savino2022} and \cite{Lee2021dist}. (2)-(5) are from LEDA, (5)-(10) are adopted from \cite{Koch2025}. }
\end{deluxetable*}


\begin{deluxetable*}{llcccc}[t!]
\tabletypesize{\footnotesize}
\tablecaption{\label{tab:literature_obs} Archival Data and Products used in this paper. }
\tablewidth{0pt}
\tablehead{
\colhead{Observable} &
\colhead{Galaxy} & 
\colhead{Resolution} &
\colhead{Provenance} & 
\colhead{Used for} &
\colhead{Reference}  \\
& & &  &    &
}
\startdata
FUV 154~nm & All except IC10 & 7.5 \& 15$^{\prime\prime}$ & GALEX, $z0$MGS$^{(*)}$ & \sigsfr$_{\rm FUV}$, Eq.~\ref{eq:SFR_FUV} & (A)\\
\hline
mid-IR 22$\,\mu$m & All & 7.5 \& 15$^{\prime\prime}$ & WISE~4, $z0$MGS$^{(*)}$ & \sigsfr $_{, \rm WISE4}$ &  (A) \\
 &  & & &  Eq.~\ref{eq:SigSFR_W4ONLY} & \\
 &  & & &  \sigsfr $_{, \rm FUV+WISE4}$ &\\
  &  & & &  Eq.~\ref{eq:sigsfr_fuvw4} & \\
\hline
H$\alpha$ 656~nm & IC10 & 0.5$^{\prime\prime}$ & PT, FP, NCCD & \sigsfr $_{, \rm H\alpha+WISE4}$ & (C) \\
H$\alpha$ 656~nm & IC1613 &  0.5$^{\prime\prime}$ & PT, FP, NCCD & Eq.~\ref{eq:SigSFR_HaW4} &  (C)\\
H$\alpha$ 656~nm & NGC6822 & 0.5$^{\prime\prime}$& HT, 2048 & \sigsfr $_{, \rm H\alpha+WISE4}$ & (C)\\
H$\alpha$ 656~nm & WLM &  0.5$^{\prime\prime}$& PT, FP, NCCD & Eq.~\ref{eq:SigSFR_HaW4} &  (C)\\
H$\alpha$ 656~nm & M33 &  2$^{\prime\prime}$& KPNO$^{(a)}$ & Eq.~\ref{eq:SigSFR_HaW4} & (B)\\
\hline
near-IR 3.4~$\mu$m & All & 7.5 \& 15$^{\prime\prime}$ & WISE~1, $z0$MGS$^{(*)}$ & \sigstar, Eq.~\ref{eq:Sigstar_3p4um} & (A) \\
 \hline
\hi\ rot. curve & IC10 &  6.0$^{\prime\prime}$ &VLA, LITTLE THINGS$^{(c)}$ & $V_c$,  Paper II & (F)\\
\hi\ rot. curve & IC1613 & 6.0$^{\prime\prime}$ &  VLA, LITTLE THINGS & $V_c$,  Paper II & (F)\\
\hi\ rot. curve & NGC6822 &  3.5$^{\prime}$& KAT-7   & $V_c$,  Paper II &  (G)\\
\hi\ rot. curve & WLM &  6.0$^{\prime\prime}$ &  VLA, LITTLE THINGS & $V_c$, Paper II &  (F) \\
\hi\ rot. curve & M33 & 20.0$^{\prime\prime}$ & VLA   & $V_c$,  Paper II & (E)\\
\hi\ rot. curve & M31 &  2.1$^{\prime}$ & WSRT, HI survey$^{(b)}$ & $V_c$,  Paper II &  (D)\\
\hline
\co10 & IC10 &  8.5$^{\prime\prime}$ &CARMA+IRAM(EMIR) & \sigmol, Eq.~\ref{eq:Sigma_mol} & (M)\\
\co21 & IC10 &  12.8$^{\prime\prime}$ &IRAM(EMIR) & & (M)  \\
\co21, f1 & NGC6822 & 1.4$^{\prime\prime}$ &  ALMA & & (N)  \\
\co21, f2 & NGC6822 & 1.2$^{\prime\prime}$ & ALMA &  & (N)  \\
\co21, f3 & NGC6822 & 1.1$^{\prime\prime}$ &  ALMA & & (O)  \\
\co21, f4 & NGC6822 & 1.1$^{\prime\prime}$ &  ALMA & & (N)  \\
\co21 & WLM & 2$^{\prime\prime}$ &  ALMA  &  &  (P) \\ 
\co10 & M33 & &ALMA & & (J)   \\
\co21 & M33 & 11.7$^{\prime\prime}$ &ACA &  & (K) \\
\co21 & M33 & 10.7$^{\prime\prime}$ & IRAM(EMIR) & \sigmol, Eq.~\ref{eq:Sigma_mol} & (L) \\
\co10 & M31 & 5.5$^{\prime\prime}$ &CARMA &  & (H)   \\
\co10 & M31 & 23$^{\prime\prime}$  &IRAM(EMIR) & \sigmol, Eq.~\ref{eq:Sigma_mol} &  (I)\\
\hline
\enddata
\tablecomments{(*): IC 10, IC 1613, NGC 6822, and WLM were not part of the $z0$MGS data release, but we processed these observations and reprocessed M33 and M31 in the same manner for this paper (see \autoref{app:physical}). (a) Telescope, instrument, and detector used for the H$\alpha$ observations. Telescopes: KPNO=Kitt Peak National Observatory; PT=1.8 m Perkins Telescope at Lowell Observatory; HT=1.1 m Hall Telescope at Lowell Observatory;  Instruments: FP=Ohio State University Fabry-P\'{e}rot used as a simple 3:1 focal reducer. Detectors: NCCD=a TI 800; 2048=SITe 2048 ; 2048 CCD. (b) The HI survey that this rotation curve is based on is from \cite{Braun2009}. (c) LITTLE THINGS stands for Local Irregulars That Trace Luminosity Extremes The HI Nearby Galaxy Survey \citep[][]{Hunter2012}. References: (A) \cite{Leroy2019}, (B) \cite{Greenawalt1998} and are described in more detail in \cite{Hoopes2000}, (C) \cite{Hunter2004}, (D) \cite{Corbelli2010}, (E) \cite{Corbelli2014}, (F)  \cite{Oh2015}, (G) For NGC~6822 we list the large-scale \hi\ rotation curve from \cite{Namumba2017} adopted for Paper~II; for a higher-resolution inner-disk rotation curve, see \cite{Weldrake2003}, (H)  \cite{Caldu-Primo2016}, (I) \cite{Nieten2006},  (J) PI: E. Rosolowsky (2022.1.00276.S), 2017.1.00901.S, 2019.1.01182.S, (K) 2018.A.00058.S, 2021.1.0999.S, 2022.1.00403.S, (L) \cite{Druard2014} Gratier+10, 12, (M) PI: A. Leroy, (N) PI: Chown (2024.1.01179.S), (O) \cite{Schruba2017}, (P) \cite{Rubio2015,Archer2024}.}
\end{deluxetable*}

We study the six Local Group galaxies targeted by LGLBS. \autoref{tab:physical} gives their physical properties, distance, and orientation. We analyze a combination of new LGLBS 21-cm data and literature multiwavelength observations to construct maps and radial profiles of atomic gas, molecular gas, stellar mass, and star formation rate surface density.
Here and in Table~\ref{tab:literature_obs} we summarize the data used. Appendix \ref{app:physical} details how we convert from intensity to physical quantities. 

\paragraph{Atomic gas} We trace \hi\ using VLA+GBT 21-cm observations from LGLBS \citep[][]{Koch2025}. We use data products that include data from the VLA's C+D configurations and short spacing corrections from the Green Bank Telescope (GBT). We convolved all cubes to a common physical beamsize of 120~pc (FWHM). The cubes have 2.1 km s$^{-1}$-wide velocity channels. We convert from 21-cm intensity to \hi\ mass surface density, $\Sigma_\mathrm{HI}$, assuming optically thin emission (\S \ref{app:sigatom}). 

\paragraph{Molecular gas} We use \co10 and \co21 observations to trace molecular gas. These data come from IRAM, ALMA, and CARMA. All galaxies except IC~1613 have some CO coverage, and all available CO data reach our $120$~pc working resolution. The noise and spatial coverage of the CO data are heterogeneous, and only IC~10, M31, and M33 yield useful radial profiles. The full-galaxy data for NGC 6822 are shallow, while deep observations targeting individual fields do not cover enough area to reconstruct the full profile. For WLM, the data are sensitive and cover a wide area but the CO emission is so faint and the CO-to-H$_2$ conversion factor so uncertain that we do not consider a radial profile useful. We estimate \sigmol\ either from the CO using a variable $\alpha_{\rm CO}$ (\S \ref{app:alphaco}) or from an assumed star formation depletion time (\S \ref{app:expectedSigmol}). 

\paragraph{Star formation} Our fiducial \sigsfr\ estimates use ultraviolet (UV) data from GALEX \citep{Martin2005} and mid-infrared (mid-IR) data from WISE \citep{Wright2010}. We processed these data following \citet[][]{Leroy2019} as summarized in the Appendix (\S \ref{app:sigsfr}). Then we separately convert the individual bands and their combination to \sigsfr\ estimates, $\sigsfr_{,\rm FUV}$, $\sigsfr_{,\rm FUV+W4}$, $\sigsfr_{,\rm W4}$ (\S \ref{app:sigsfr}).WISE4 is less sensitive than GALEX's FUV imaging. Therefore in outer galaxies or our dwarf targets, we will often show only $\sigsfr_{,\rm FUV}$. This assumes no extinction, and is formally a lower limit, but this should be a reasonable assumption in the low metallicity, dust-poor outer parts of galaxies. 

Where available, we also include H$\alpha$ observations to trace recent star formation. We use H$\alpha$ narrowband maps from LITTLE THINGS\footnote{\url{http://www2.lowell.edu/users/dah/littlethings/}} \citep[][]{Hunter2004,Hunter2012} for IC~10, IC~1613, NGC~6822, and WLM. These have calibration accuracy ${\sim} 4\%$ and have been corrected for the contribution of \ion{N}{2}. For M33 we compare to the map of \citet{Greenawalt1998}. 
As described in Appendix \ref{app:sigsfr}, we estimate \sigsfr\ without ($\sigsfr_{,\rm H\alpha}$) and with ($\sigsfr_{,\rm H\alpha+W4}$) a mid-IR based extinction correction.

We estimate statistical uncertainties in GALEX and WISE maps from the scatter in the map outside $\rgal\ ~ > 1.5\times\Riso$. For H$\alpha$, which has more limited coverage, we estimate uncertainties based on emission-free regions after all other processing. 

\begin{figure*}[ht!]
    \centering
    \includegraphics[width=1.0\linewidth]{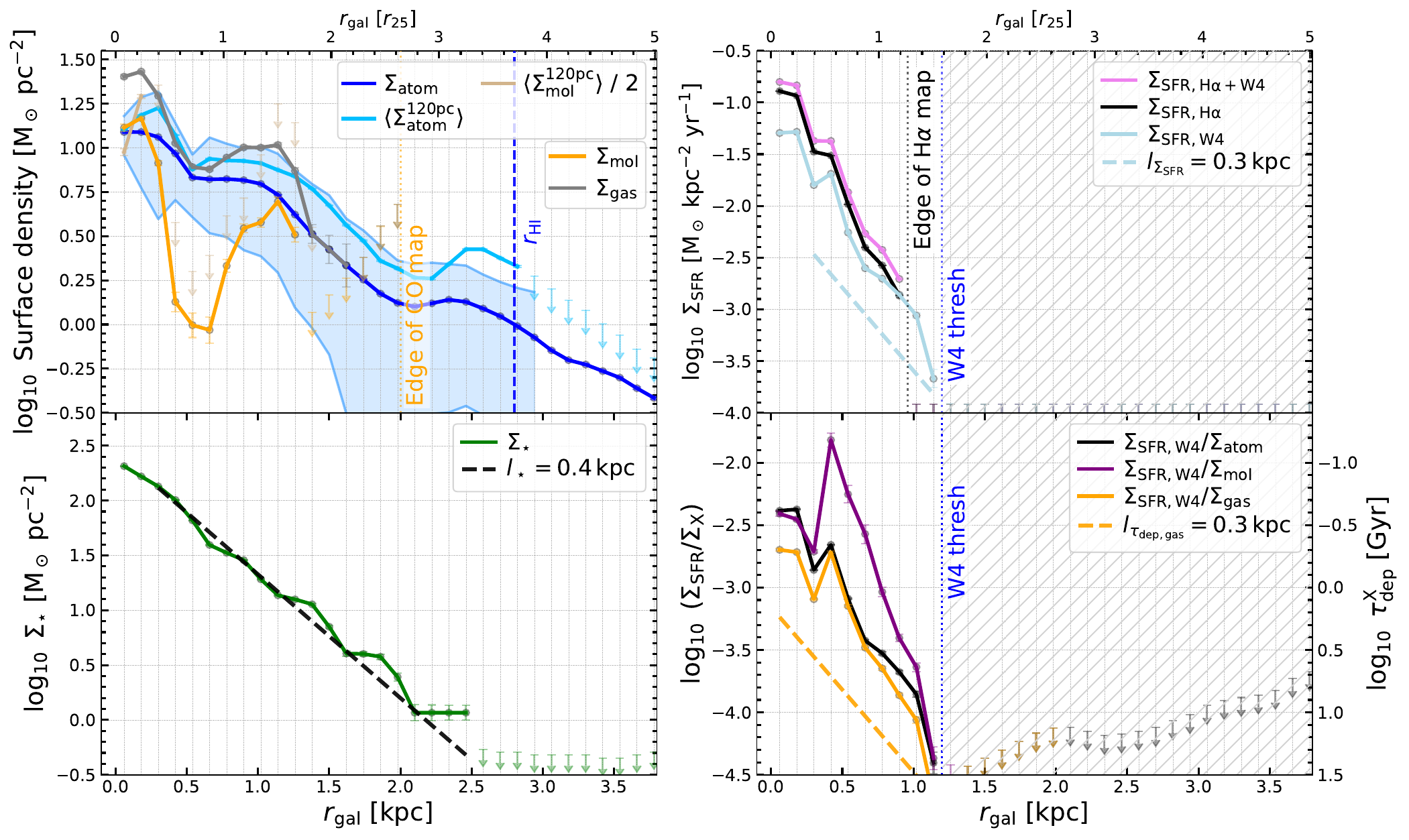}
    \caption{
    Azimuthally-averaged surface density profiles for IC~10. The $x$-axes show galactocentric radius in physical units and normalized to the optical isophotal radius ($r_{25}$). 
    \textit{Upper left}: Gas surface densities. Blue colors show the atomic gas surface density \sigatom\ (\autoref{eq:21cmtosigatom}) including the 16-84\% range as a shaded blue region. The vertical blue dashed line denotes $r_{\rm HI}$, the radius where $\sigatom=1~\uSig$. In light blue colors we show the profile of the mass-weighted atomic gas surface density $\langle \Sigma_{\rm atom}^{\rm 120pc} \rangle$ (\autoref{eq:mass-weighted-sigatom}), where upper limits show radial bins that did not satisfy our 50$\%$ completion criterion. The difference between the two profiles shows how clumpy the atomic gas is. Upper limits in all the other profiles denote non-detection or beyond where their cumulative profiles converge (see \autoref{tab:fluxes} and \autoref{tab:radii}), which we highlight in the SFR profiles as vertical threshold lines. Where available, we show profiles of molecular gas surface density \sigmol\ (\autoref{eq:Sigma_mol}) the mass-weighted molecular gas surface density $\langle \Sigma_{\rm mol}^{\rm 120pc} \rangle$, and total gas surface density \siggas. \textit{Lower left}: Stellar mass surface density, \sigstar\, 
    with a mass-to-light ratios taken from \autoref{tab:fluxes} (\autoref{eq:Sigstar_3p4um}). \textit{Upper right}: \sigsfr\ , from H$\alpha$+W4, FUV+W4(22$\mu$m),  H$\alpha$, FUV, and W4(22$\mu$m) where available. \textit{Lower right}: Ratios of \sigsfr\ to \sigatom, \sigmol, and \siggas , i.e., the star formation efficiency of atomic, molecular, and total neutral gas. The second $y$-axis shows the corresponding gas depletion times, $\tau_{\rm dep}^{\rm atom}$, $\tau_{\rm dep}^{\rm mol}$, $\tau_{\rm dep}^{\rm gas}$ (\autoref{eq:sfe}). We show fits for the scale lengths of stellar mass, star formation activity, and the gas depletion time (Table \ref{tab:tdep}).}
    \label{fig:radial_ic10}
\end{figure*}

\begin{figure*}
    \centering
    \includegraphics[width=1.0\linewidth]{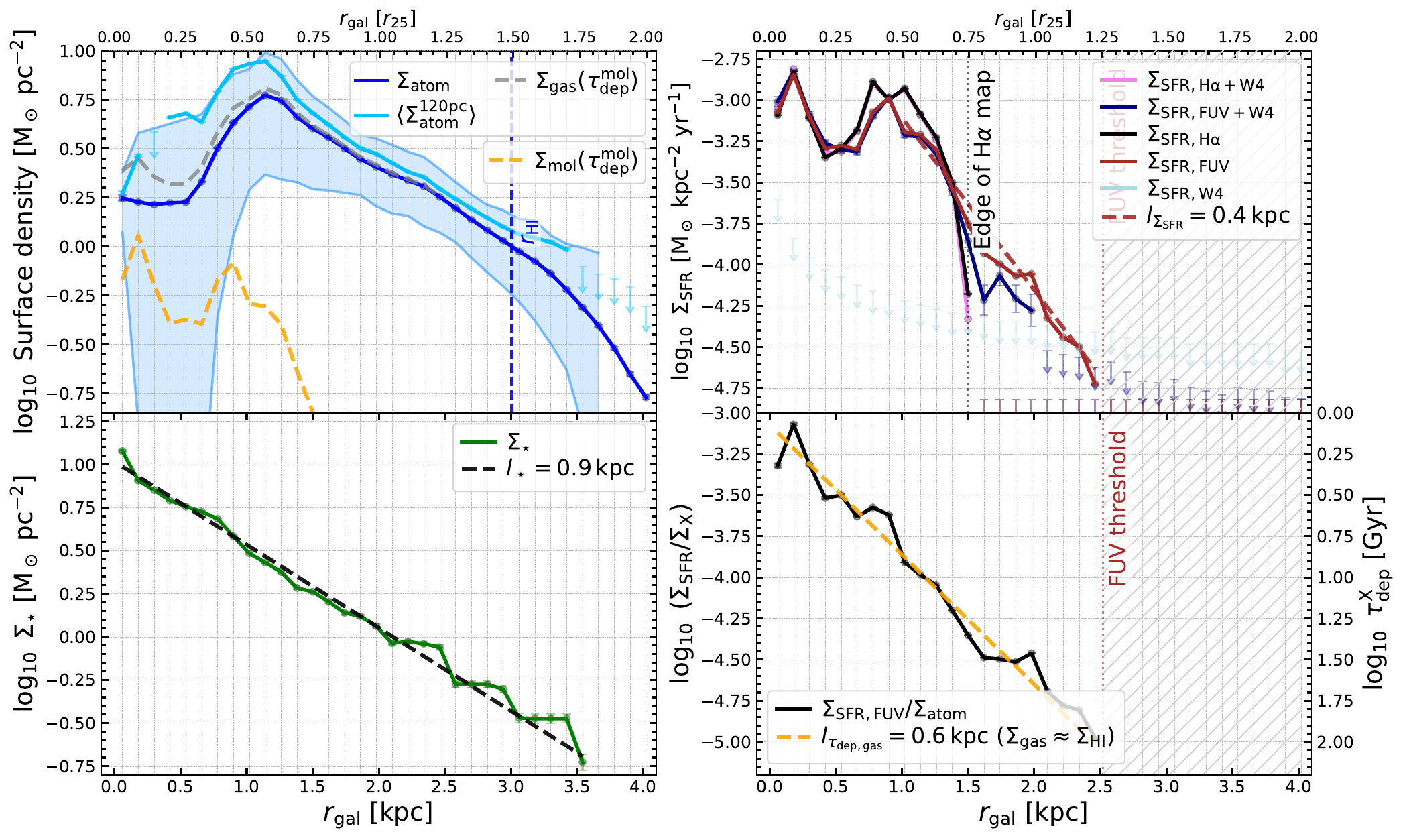}
    \caption{Radial profiles same as Fig. \ref{fig:radial_ic10} but for IC~1613. }
    \label{fig:radial_ic1613}
\end{figure*}

\begin{figure*}
    \centering
    \includegraphics[width=1.0\linewidth]{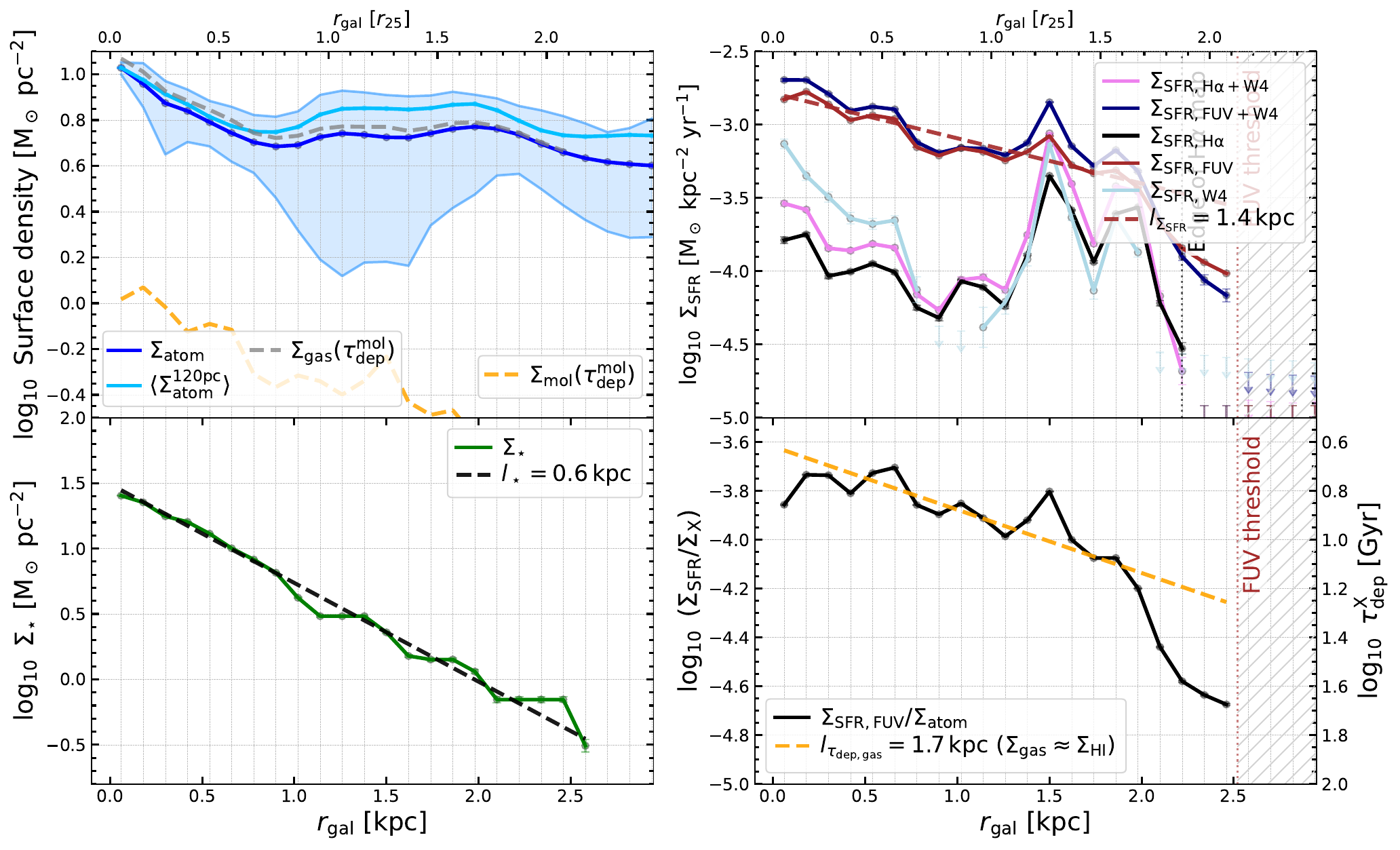}
    \caption{Radial profiles same as Fig. \ref{fig:radial_ic10} but for NGC~6822.}
    \label{fig:radial_ngc6822}
\end{figure*}

\begin{figure*}
    \centering
    \includegraphics[width=1.0\linewidth]{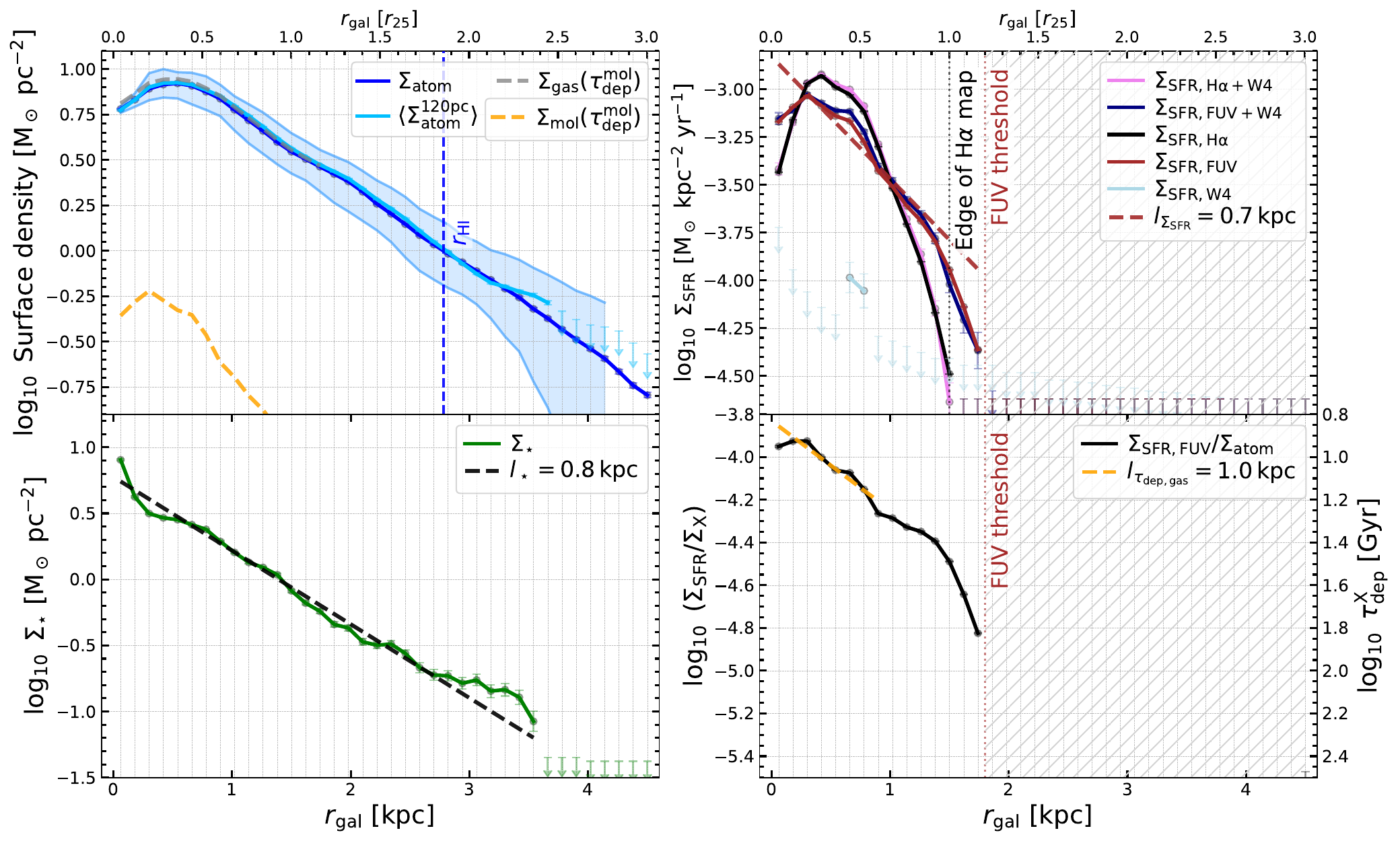}
    \caption{Radial profiles same as Fig. \ref{fig:radial_ic10} but for WLM.}
    \label{fig:radial_wlm}
\end{figure*}

\begin{figure*}
    \centering
    \includegraphics[width=1.0\linewidth]{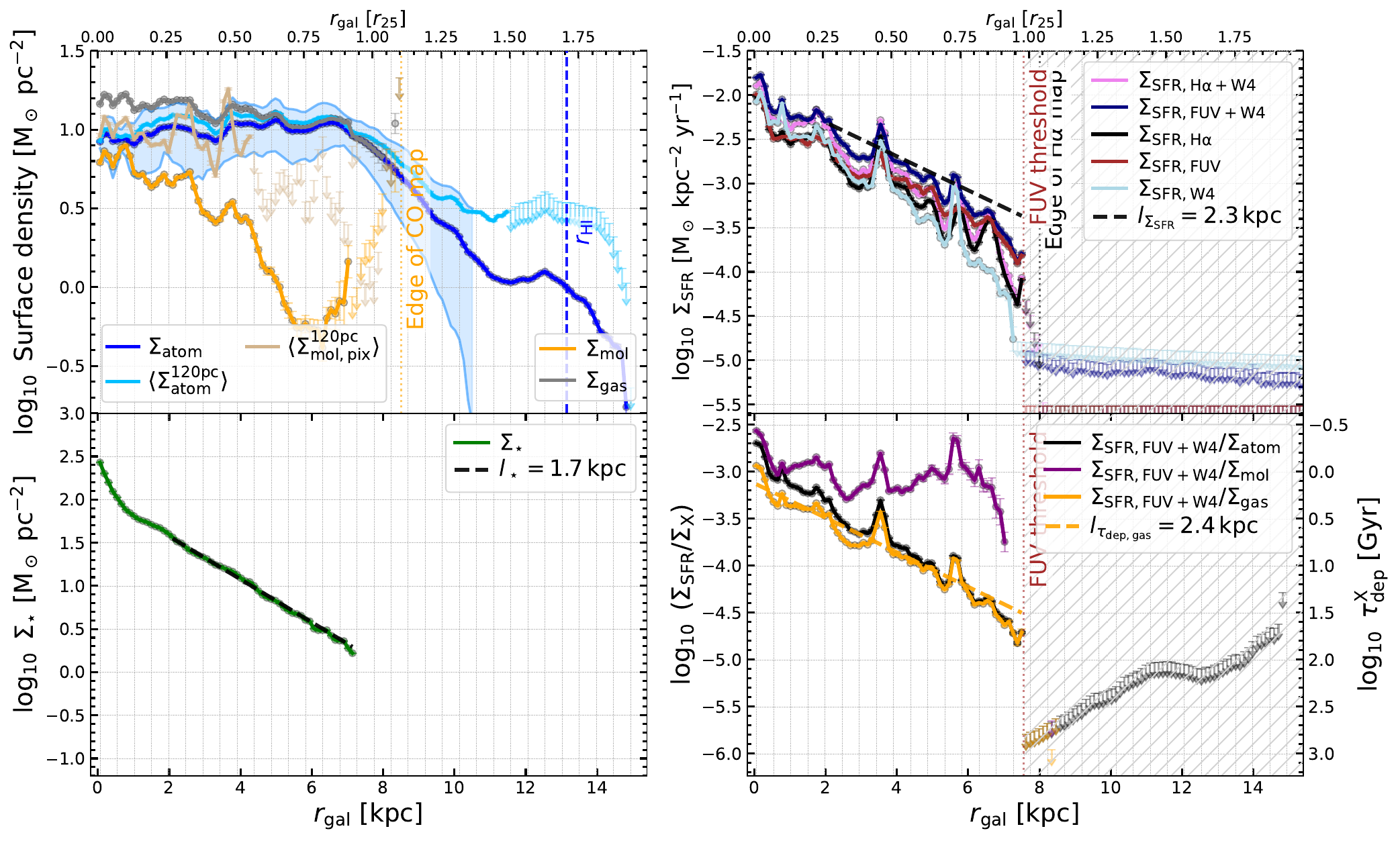}
    \caption{Radial profiles same as Fig. \ref{fig:radial_ic10} but for M33.}
    \label{fig:radial_m33}
\end{figure*}

\begin{figure*}[ht!]
    \centering
    \includegraphics[width=1.0\linewidth]{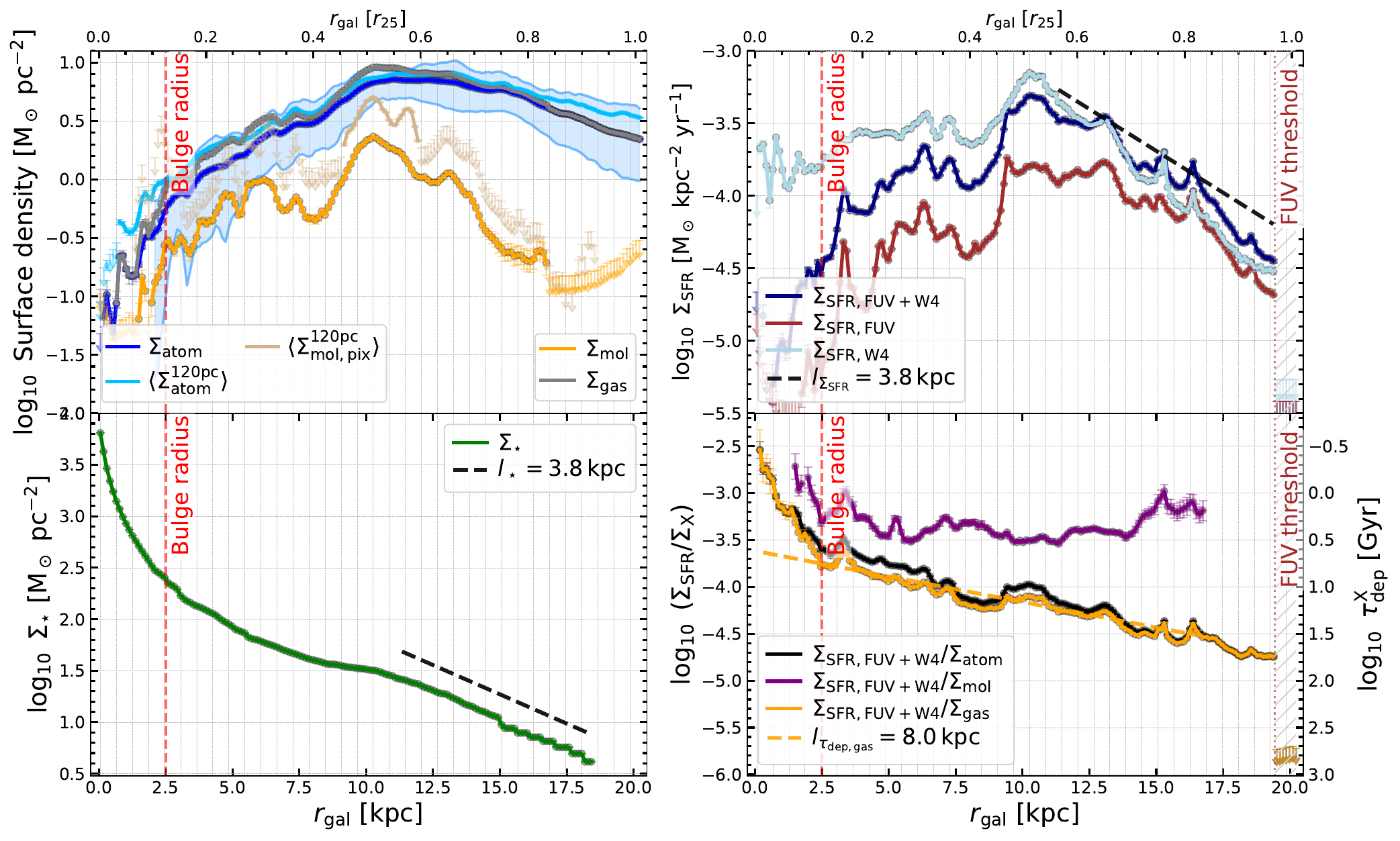}
    \caption{Radial profiles same as Fig. \ref{fig:radial_ic10} but for M31.} 
    \label{fig:radial_m31}
\end{figure*}

{
For each SFR tracer, we define a convergence radius, $r_{\rm conv}$, as the first radius at which the cumulative flux curve satisfies our adopted convergence criteria (Table~\ref{tab:radii}). In the radial profile figures (Figures \ref{fig:radial_ic10}, \ref{fig:radial_ic1613}, \ref{fig:radial_ngc6822}, \ref{fig:radial_wlm}, \ref{fig:radial_m33}, \ref{fig:radial_m31}), these tracer specific radii are indicated by the vertical dashed lines and labeled as ``FUV threshold" (W4 for IC~10). We use $r_{\rm conv}$ when measuring the integrated fluxes, luminosities, and derived quantities reported in Table~\ref{tab:fluxes}.    }


The \sigsfr\ estimates have several associated challenges. IC 10 lacks useful GALEX mapping, and we default to using W4 alone or H$\alpha$ for that galaxy. Both UV and mid-IR emission in the inner portion of M31 have significant contributions from the old stellar population. We correct for this by subtracting a term proportional to near-IR emission, which we take as a tracer of this old stellar population (\S \ref{app:oldstar_correction}). In all targets except M31, the sensitivity of the SFR tracers becomes a limiting factor in the outer disks of our targets so that many rings yield only upper limits.

\paragraph{Stellar mass} We use near-infrared mapping, WISE Band 1 at $3.4\mu$m, to trace the stellar mass of our targets. 
We convert W1 light to stellar mass using a radially varying mass-to-light ratio, $\MtoLwiseone(r)$, derived from exponential fits to the radial W1-light and SFR surface density profiles. The ratio of these fitted profiles provides a smooth radial proxy for the local specific star formation rate, which we convert to $\MtoLwiseone(r)$ following the calibration in \citet{Leroy2019} (see \S~\ref{app:sigstar} for details).
We revisited the background subtraction and masking of foreground stars and artifacts for these specific galaxies. Where needed, we filled missing pixels in the IR (and FUV) maps by assigning them the median of all valid pixels in the same deprojected galactocentric radial bin {\footnote{{We assessed the effect of filling masked foreground star and artifact pixels. 
Within $\Riso$, the masked fraction is 13.4\% in W1 and 11.0\% in W2 across the sample, while the resulting change in the W1-based \sigstar radial profile is only 3.6\% (3.1\% for the dwarf galaxies).}}}, and applied the same procedure to the corresponding uncertainty maps. The final \sigstar\ profiles use the azimuthal mean of the W1 emission within $0.75\,\Riso$ and the median value in each radial bin at larger radii. 

\paragraph{Derived quantities}

Based on the estimated physical quantities, we calculate the star formation efficiency of atomic gas, molecular gas, and total neutral atomic plus molecular gas. Here, star formation efficiency refers to the star formation rate per unit gas mass, which is the inverse of the gas depletion time. That is, 

\begin{eqnarray}
\label{eq:sfe}
\textrm{SFE}_{\rm atom} &=& \frac{\sigsfr}{\sigatom} = \frac{1}{\tau_{\rm dep}^{\rm atom}}, \\
\nonumber \textrm{SFE}_{\rm mol} &=& \frac{\sigsfr}{\sigmol} = \frac{1}{\tau_{\rm dep}^{\rm mol}}, \\
\nonumber \textrm{SFE}_{\rm gas} &=& \frac{\sigsfr}{\sigatom + \sigmol} = \frac{\sigsfr}{\siggas} = \frac{1}{\tau_{\rm dep}^{\rm gas}},
\end{eqnarray} 

\noindent where all of these gas mass surface densities account for the mass of helium and heavier elements (see \autoref{app:sigatom}). In Paper II \citep[][]{Eibensteiner2026b}, we put these profiles of SFE into context and compare them to estimates of dynamical equilibrium pressure $P_{\rm DE}$ and Toomre $Q_{\rm gas}$ parameter. 

\begin{figure*}[ht!]
    \centering
    \includegraphics[width=0.95\linewidth]{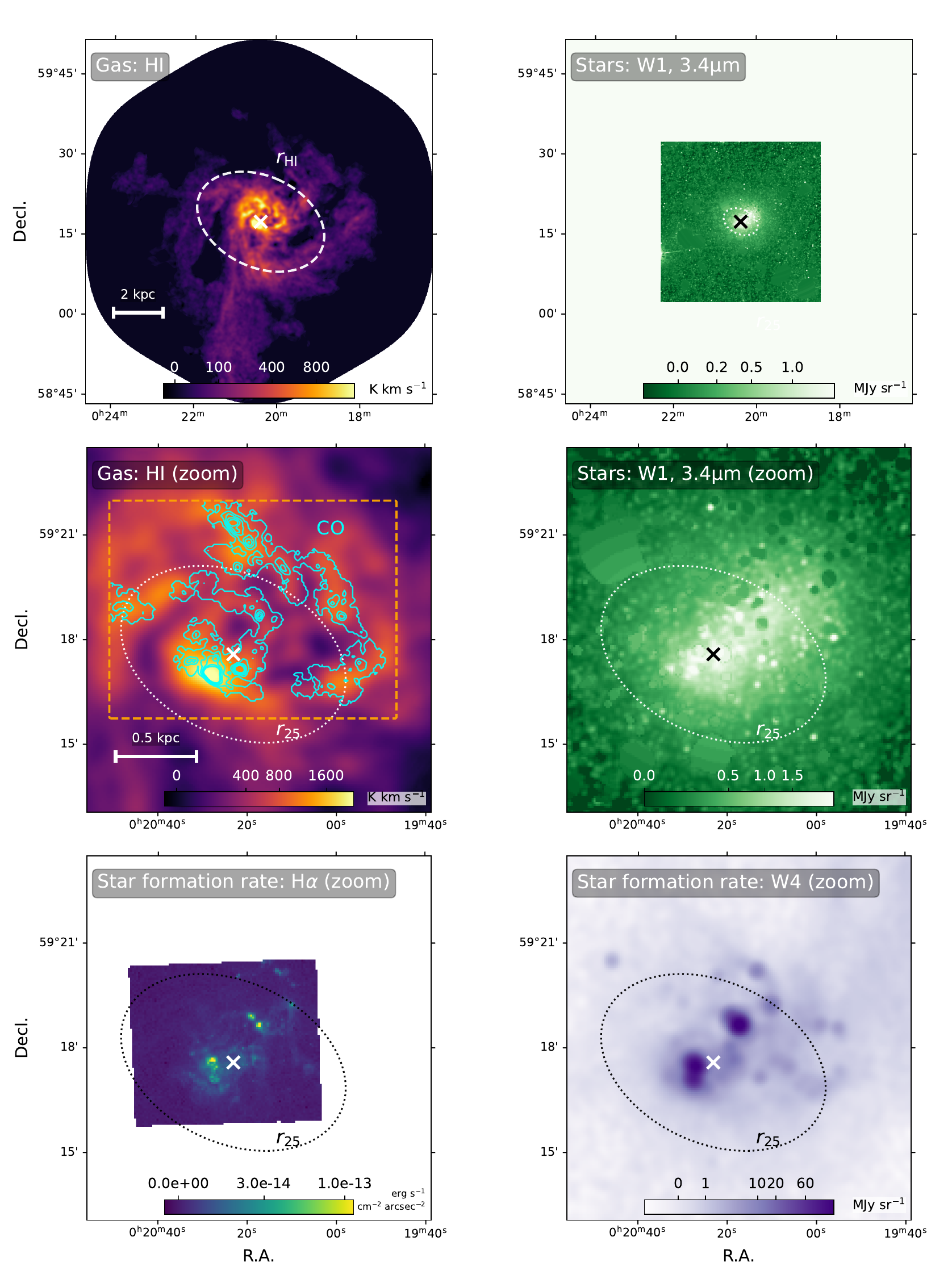}
    \caption{Atlas of gas, stars and star formation for IC~10. The upper two maps show field of views matched to the \hi\ map. The lower four panels show zoomed in versions to better illustrate the connection to the
other tracers. The white dashed circle shows the $r_{\mathrm{HI}}$ taken from \autoref{tab:measured_radial_profile_properties}, and the dotted circle indicates the optical radius $r_{25}$ taken from \autoref{tab:physical}. The orange dashed lines show the FoV of CO observations. Colorbars show intensities in units of K km s$^{-1}$ for \hi, MJy sr$^{-1}$ for W1 and FUV, and erg~s$^{-1}$~cm$^{-2}$~arcsec$^{-2}$ for H$\alpha$.   } \vspace{0.15cm}
    \label{fig:atlas_ic10}
\end{figure*}

\begin{figure*}[ht!]
    \centering
    \includegraphics[width=1.0\linewidth]{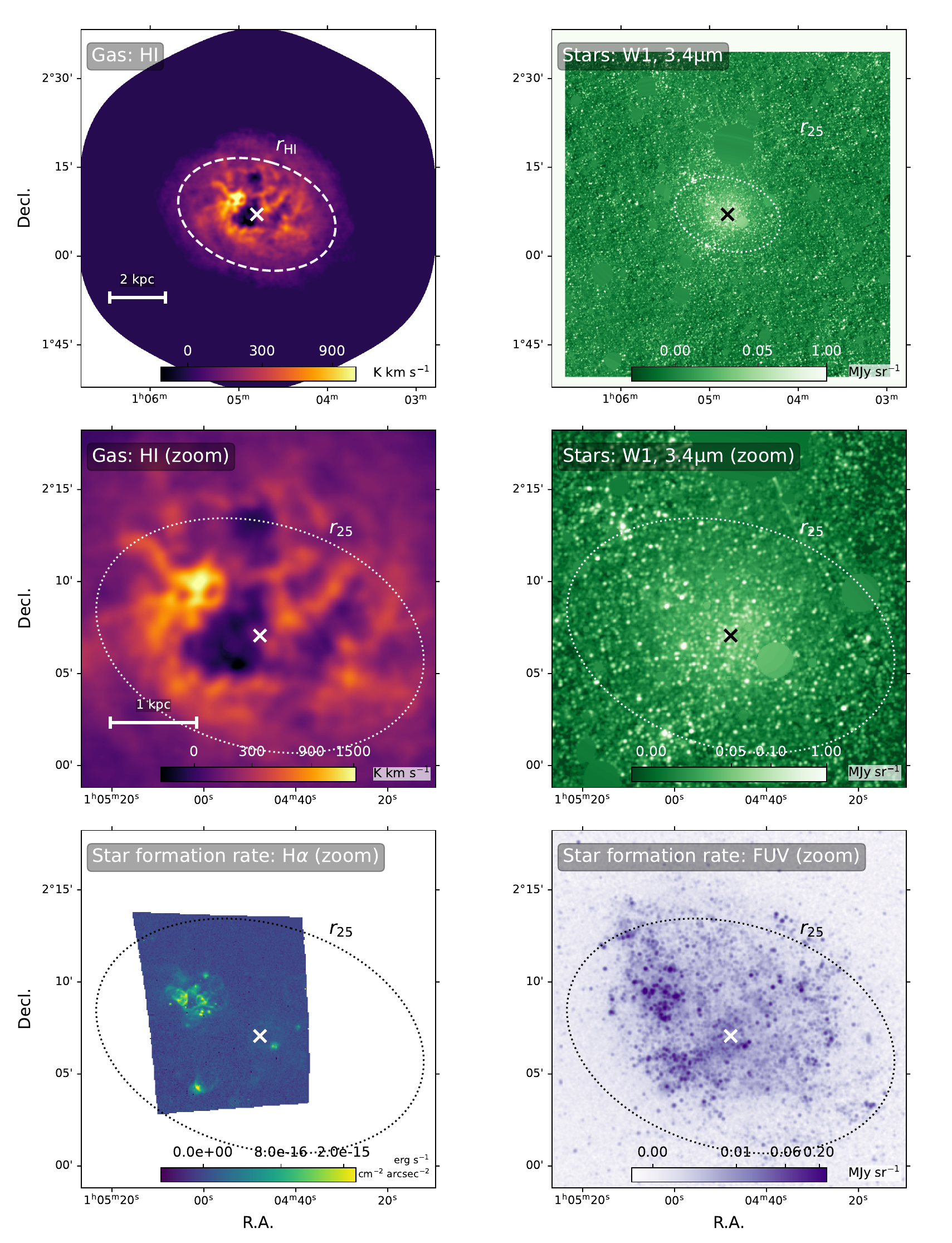}
    \caption{Atlas of gas, stars and star formation, same as \autoref{fig:atlas_ic10} but for IC~1613.} \vspace{0.15cm}
    \label{fig:atlas_ic1613}
\end{figure*}

\begin{figure*}[ht!]
    \centering
    \includegraphics[width=1.0\linewidth]{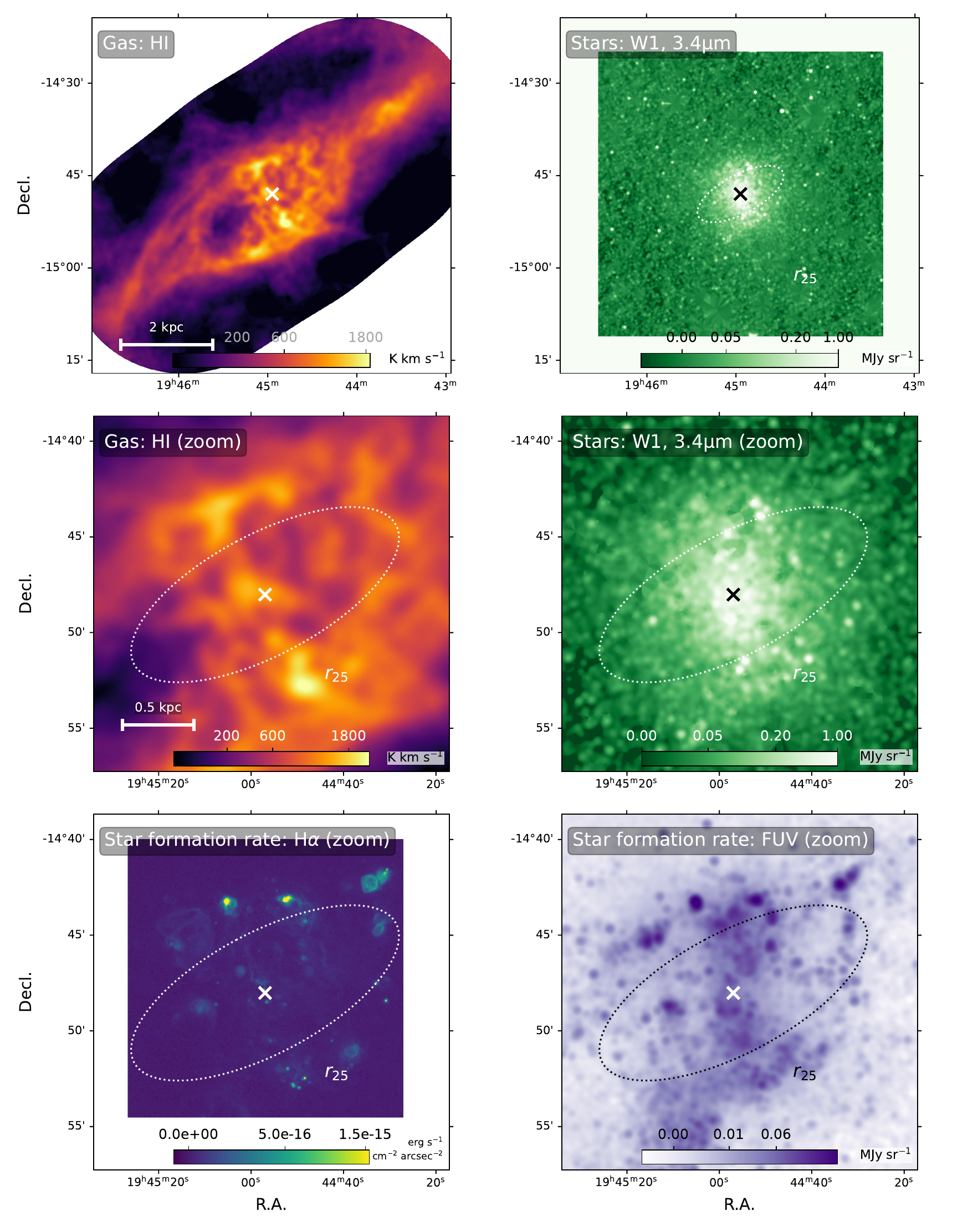}
    \caption{Atlas of gas, stars and star formation, same as \autoref{fig:atlas_ic10} but for NGC~6822. $r_{\rm HI}$ is not reached in NGC~6822.} \vspace{0.15cm}
    \label{fig:atlas_ngc6822}
\end{figure*}

\begin{figure*}[ht!]
    \centering
    \includegraphics[width=1.0\linewidth]{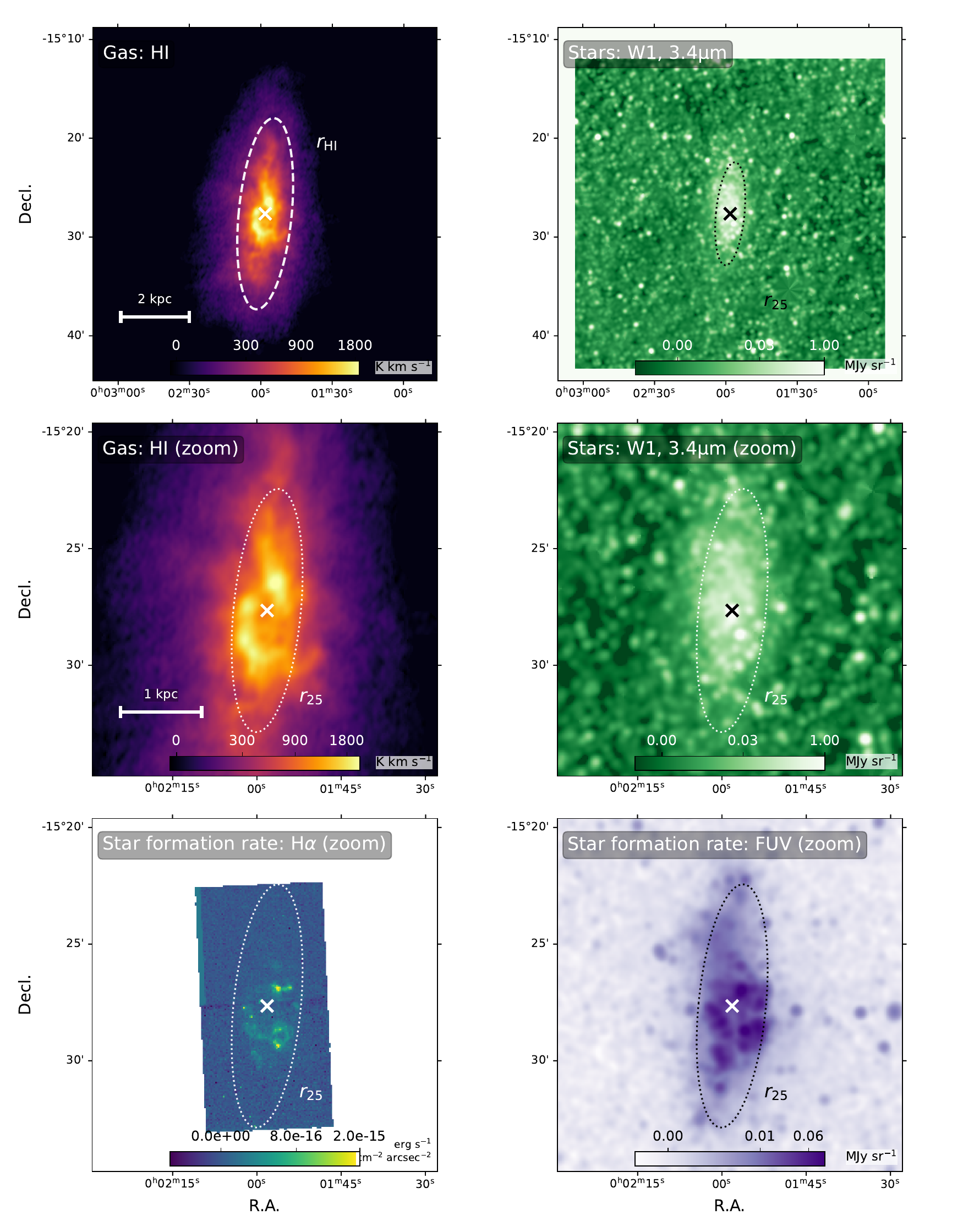}
    \caption{Atlas of gas, stars and star formation, same as \autoref{fig:atlas_ic10} but  for WLM.} \vspace{0.15cm}
    \label{fig:atlas_wlm}
\end{figure*}

\begin{figure*}[ht!]
    \centering
    \includegraphics[width=0.95\linewidth]{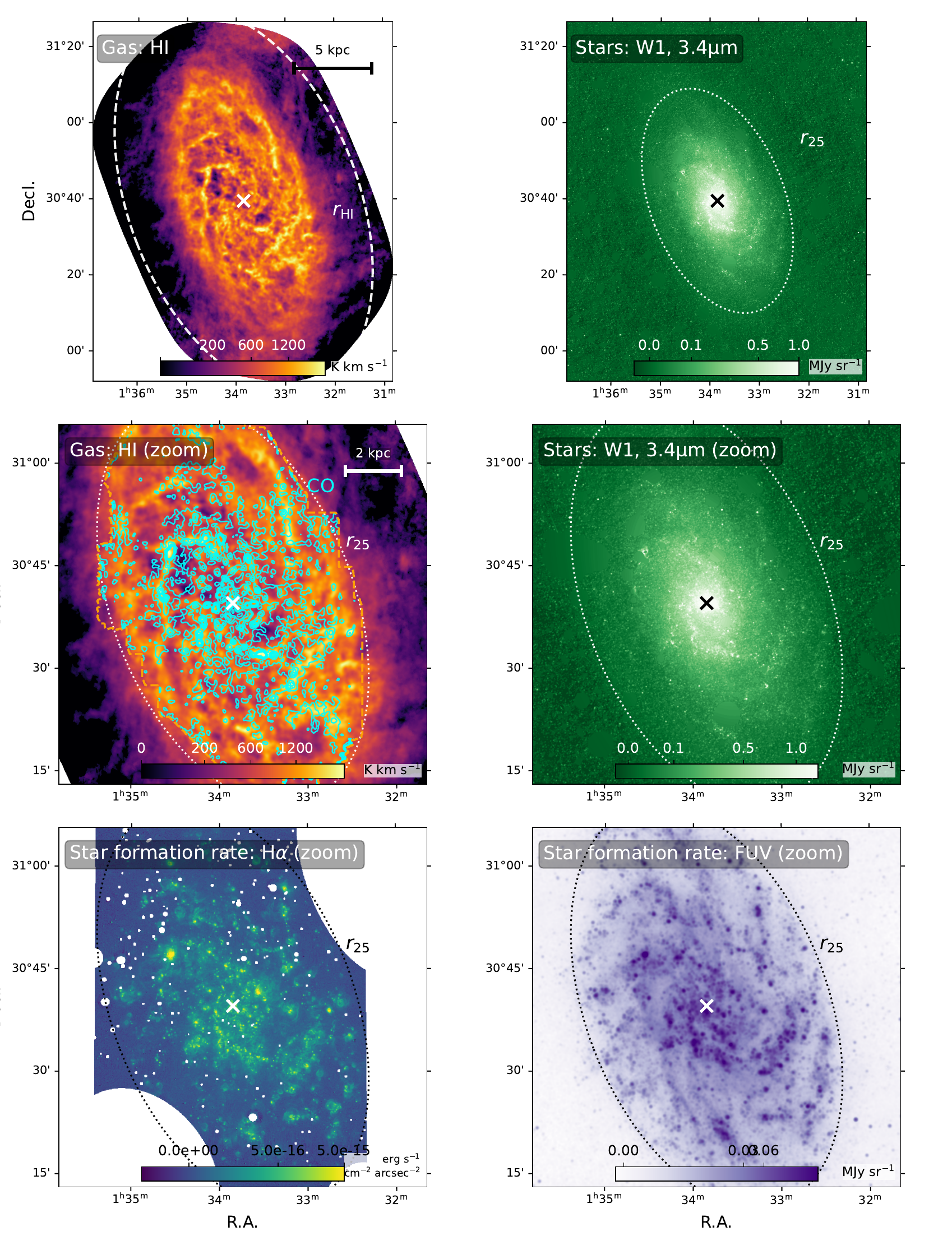}
    \caption{Atlas of gas, stars and star formation, same as \autoref{fig:atlas_ic10} but for M33. For H$\alpha$ we show the unfilled version (i.e., NaNs were not assigned the average of all valid pixels in the same deprojected galactocentric radial bin) for visualization purposes to show how many artefacts we have masked.}  \vspace{0.15cm}
    \label{fig:atlas_m33}
\end{figure*}

\begin{figure*}[ht!]
    \centering
    \includegraphics[width=1.0\linewidth]{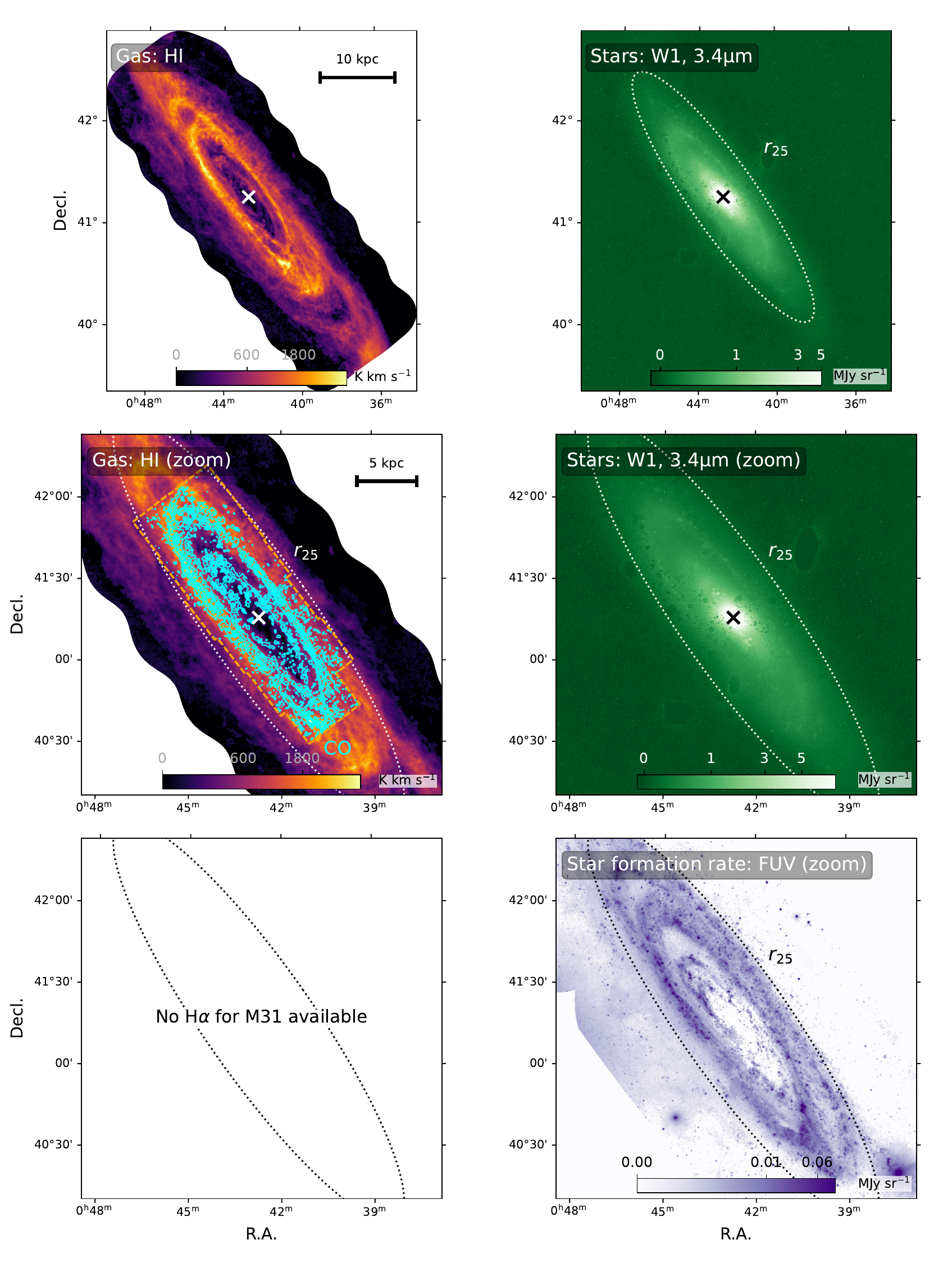}
    \caption{Atlas of gas, stars and star formation, same as \autoref{fig:atlas_ic10} but for M31. No H$\alpha$ map was available for M31. The FUV map is corrected for old stellar contamination (see \autoref{app:oldstar_correction}). } \vspace{0.15cm}
    \label{fig:atlas_m31}
\end{figure*}

\section{Atlas and radial profiles of atomic gas, molecular gas, stellar mass, star formation rate, and depletion time}
\label{sec:radial_profiles}

In Figures \ref{fig:atlas_ic10} to \ref{fig:radial_m31}, we present images and radial profiles of gas (\sigatom, \sigmol, \siggas), stellar mass (\sigstar), star formation (\sigsfr) and depletion time (\tdephi). The images show a zoomed-out view of the 21-cm integrated maps and the near-IR starlight, tracing stellar mass, over the area where LGLBS detects \hi. The zoomed-in lower panels focus on the optical disk of the galaxy. These include contours of CO emission and tracers of recent star formation.

The radial profiles report azimuthal averages\footnote{Or medians in the case of the outer stellar profiles (\autoref{app:sigstar}.} using bins with width and spacing of $120$~pc in galactocentric radius. Within each ring, we also calculate the mass-weighted \hi\, and molecular gas surface densities: \mbox{$\langle \Sigma_{\rm atom}^{120{\rm pc}}\rangle$} and \mbox{$\langle \Sigma_{\rm mol}^{120{\rm pc}}\rangle$} respectively. These indicate the surface density of material from which emission in the ring originates in the high resolution data (\S \ref{app:mass_weighted_sigatom}). The profiles extend out to where \hi\ is detected in the averaged radial profile with a signal-to-noise of $\ge$3. For NGC~6822 and M31 this criteria is met out to the edge of the LGLBS \hi\ field of view. We make a similar set of measurements within $500$~pc diameter hexagonal apertures that cover each galaxy. 

The mechanical calculation of the profiles follows \citet{Sun2022}. The processed survey images, profiles, and the region-average measurements are publicly available accompanying this paper\footnote{\textcolor{red}{\url{https://dx.doi.org/10.11570/26.0020}}}. The radial extent of gas, stars and star formation activity are summarized in \autoref{tab:measured_radial_profile_properties} and gas depletion times in \autoref{tab:tdep}. Integrated fluxes, luminosities, and physical quantities measured from the radial profiles are summarized in Table~\ref{tab:fluxes}. Measurements of atomic gas surface densities are summarized in \autoref{tab:hi_props}.

\subsection{IC~10}

IC~10 (Figure \ref{fig:atlas_ic10}) shows a complex \hi\ morphology, including multiple holes in the inner disk and an extended outer envelope with multiple streamers \citep[see also earlier studies by][]{Shostak1989,Brinks1990,Wilcots1998}. CO emission follows the brightest \hi\ peaks and is also detected in the eastern stream beyond $\Riso$. These CO features trace the main star-forming complexes visible in H$\alpha$ and W4.

The radial profile of \sigatom\ (Figure \ref{fig:radial_ic10}, top left) declines from its central \sigatom\ $\approx 12$ \uSig\ with an exponential scale length of $l_{\rm HI}^{\rm outer}=1.0$~kpc, $\approx 3$ times the stellar scale length and $\approx 4$ times the $\Sigma_{\rm SFR}$ scale length. The mass-weighted $\langle \Sigma_{\rm atom}^{\rm 120pc} \rangle$ also declines but remains well above the mean. This matches the visual impression that the \hi\ is clumpy ($\langle \Sigma_{\rm atom}^{\rm 120pc} \rangle > \Sigma_{\rm atom}$) and the brightness of individual features declines with radius ($\langle \Sigma_{\rm atom}^{\rm 120pc} \rangle$ declines). The broad $16{-}84\%$ range of $\Sigma_{\rm atom}$ reflects the giant shells in the inner disk and the low covering fraction of the outer streamers. 

In the outer disk, $\langle \Sigma_{\rm atom}^{\rm 120pc} \rangle$ remains high as the azimuthally averaged $\Sigma_{\rm atom}$ falls. This highlights that even where at the canonical $r_{\rm HI}$ where $\Sigma_{\rm atom} = 1~M_\odot~{\rm pc}^{-2}$, the outer disk still contains localized high-column-density \hi\ structures and the divergence between the mass- and area-weighted profiles reflects the declining filling factor of atomic gas. 

IC 10 has abundant CO and molecular gas \citep[see][]{Leroy2006,Kepley2018}, and $\Sigma_{\rm mol} > \Sigma_{\rm atom}$ only in the inner ring where our adopted galaxy center lies near the brightest star-forming region in the galaxy, IC10 SE. This leads to a central bump in the $\Sigma_{\rm mol}$ radial profile so that $\Sigma_{\rm mol} \approx \Sigma_{\rm atom}$ out $r_{\rm gal} \approx 0.3$ kpc.

Integrated over the whole galaxy, IC~10 has $\tau_{\rm dep}^{\rm gas} \approx 2.7$~Gyr, but within the radius containing 50\% of the stellar mass ($r_{50,\star}$), $\tau_{\rm dep}^{\rm gas}$ drops to $\approx 0.7$~Gyr and $\tau_{\rm dep}^{\rm atom} \approx 0.5$~Gyr. In the inner rings dominated by IC10 SE and the other molecular clouds, $\tau_{\rm dep}^{\rm mol}$ is $\sim 0.8$~Gyr. 
Away from the bright complex, $\tau_{\rm dep}^{\rm gas}$ declines with increasing $r_{\rm gal}$, though the lack of a UV map prevents us from probing far into IC10's outer disk. The short $\tau_{\rm dep}$ near $r_{\rm gal} \sim 0.5$~kpc may reflect that the star-forming regions there (especially the one to the southeast of the galaxy center) are in a more evolved state, showing more SF tracer emission relative to their current molecular gas content.

\subsection{IC~1613}

In IC~1613 the \hi, \halpha , and FUV in \autoref{fig:atlas_ic1613} all show a prominent off-center star-forming complex. The inner \hi\ disk also contains numerous \hi\ holes that appear essentially empty in \halpha, consistent with earlier \hi\ and \halpha\ imaging that reveals a network of shells and cavities \citep[][]{Lake1989,Meaburn1988,Brinks1990,Pokhrel2020}. These \hi\ cavities surrounded by higher-column density shells resembles that seen in IC 10 and NGC 6822 and other star-forming dwarfs, including IC 2574 \citep{walter1999}, and Holmberg II \citep[][]{Weisz2009}. This is typically interpreted as the imprint of stellar feedback on a low-pressure atomic gas disk \citep[e.g.][]{Lake1989,Silich2006,Weisz2009}. Consistent with this idea, lower intensity FUV emission pervades the disk, even in the areas covered by the \hi\ holes and where H$\alpha$ emission is weak or absent. This likely indicates widespread star formation activity on a somewhat longer timescale (${\sim}~100$Myr) than traced by H$\alpha$.

In the radial profiles (\autoref{fig:radial_ic1613}), the inner $\rgal\lesssim0.5$~kpc has lower \sigatom\ and a large 16-84\% range due to the holes. The \sigatom\ profile rises to $\approx 6$ $\uSig$ at $\rgal{{\sim}}1$~kpc and then declines with a $l_{\rm HI}^{\rm outer} \approx 1.0$~kpc exponential scale length. This $l_{\rm HI}^{\rm outer}$ is similar to the scale length of FUV emission ($1.1$~kpc) and slightly larger than the stellar scale length ($0.9$~kpc). The mass-weighted \hi\ profile, $\langle \Sigma_{\rm atom}^{120pc} \rangle$ lies above the annular mean in the inner disk, showing that the emission arises from shells and clumps of high column gas even where the azimuthal average is depressed by the holes. 
Most of the atomic mass is therefore concentrated in the inner shell system, with $r_{50,\rm atom}\approx0.9$~kpc and $r_{90,\rm atom}\approx1.1$~kpc, even though the \hi\ disk extends to $r_{\rm HI}\approx3.0$~kpc, or $\sim1.5\,r_{25}$. 

We do not have CO-based \sigmol\ profiles. The expected \sigmol\ profile based on $\Sigma_{\rm SFR}$ (dashed orange lines; see \autoref{app:expectedSigmol}) suggests that molecular gas will be subdominant at all radii. $\Sigma_{\rm mol}$ does seem likely to be high locally in the main star-forming complex  (this will be explored in E. Tarantino et al. in preparation). 

The optical disk ($r_{\rm gal} < r_{50,*}$) shows depletion times  $\tau_{\rm dep}^{\rm gas} \approx 6$~Gyr, though this reaches as short as $\approx 1$~Gyr in individual rings of the inner galaxy. In the outer disk, \hi\ remains detected well past where FUV emission is seen, implying long \hi\ depletion times of tens to $\gtrsim100$~Gyr. These estimates agree with the global gas consumption timescales of $\sim10$-$20$~Gyr  inferred from by \citet[][]{Hashemi2019} and the inefficient star formation inferred for IC 1613 and other low-mass dwarfs by \citet{Grossi2007,Hunter2012}.

\subsection{NGC~6822}\label{sec:NGC6822}
The \hi\ disk of NGC~6822 (\autoref{fig:atlas_ngc6822}) harbors a giant \hi\ hole and extended features along the major axis that reach beyond the LGLBS field of view. Previous studies have documented these and suggested that NGC 6822 was influenced by a recent tidal interaction \citep[][]{deBlok2000,deBlok2003,Weldrake2003,deBlok2006}.

The $\Sigma_{\rm atom}$ profile (\autoref{fig:radial_ngc6822}) peaks in the center with $\sigatom \approx 11$~\uSig. The prominent \hi\ hole leads to a wide 16-84\% range in the profile between $r_{\rm gal} \approx 0.9{-}1.9$~kpc. The 120~pc mass-weighted \hi\ profile stays well above the annular mean through this region, again emphasizing that the emission is structured within the ring.

The \hi\ is extended and the azimuthally averaged \sigatom\ never reaches $1~M_\odot~{\rm pc}^{-2}$ within the VLA map. We therefore quote an extrapolated $r_{\rm HI}\approx5.8$~kpc in Table~\ref{tab:measured_radial_profile_properties}. {An alternative extrapolation based on sectors near the minor axis yields a smaller value ($r_{\rm HI,ext}\approx4.0$~kpc) indicating that the exact extrapolated radius is sensitive to the asymmetric outer \hi\ structure in this galaxy.} 
By contrast, stellar mass profile is compact, with $l_\star \approx 0.6$~kpc. As a result, the atomic component is more extended than the stellar one, with effective radius $r_{50,\rm atom} \approx 2.3$~kpc for the \hi\ compared to $r_{50,\star} \approx 0.8$~kpc for the stellar mass. The recent star formation is intermediate, with $r_{50,\rm SFR} \approx 1.3$~kpc and many prominent star forming regions visible in \halpha\ and FUV emission beyond $\Riso$ \citep[see][]{Schruba2017,Chown2025}.

NGC~6822 shows the clearest offset between SFR tracers in our sample. Inside roughly the optical disk, the FUV-based \sigsfr\ exceeds the H$\alpha$-based values by $\approx 0.5$-$0.7$~dex. At $r_{\rm gal} \approx 1{-}1.5$~kpc, coincident with the bright outer disk star forming regions, the different SFR tracers converge to within $\lesssim0.2$~dex. This FUV H$\alpha$ mismatch is qualitatively consistent with previous results for low-mass galaxies and outer galaxy disks \citep[][]{Lee2009,Grossi2015,Lee2016,FloresVelazquez2021,Padave2025}. The mismatch emerges because H$\alpha$ is driven by to the most massive, short-lived stars and so fluctuates in the presence of bursty star formation histories (SFHs). Stochastic sampling of the IMF or leakage of ionizing photons from the galaxy can also affect H$\alpha$. Meanwhile FUV reflects photospheric emission from lower mass stars, and still emerges from stellar populations up to $\sim100$~Myr old. In this picture, the inner disk of NGC6822 underwent a significant starburst in the recent past but harbors limited present-day star formation \citep[][]{Efremova2011}.

NGC~6822 has no deep, wide-field CO coverage, although ALMA has targeted the bright star-forming regions \citep[][]{Schruba2017,Chown2025}. The expected \sigmol\ profile (dashed orange line; see \autoref{app:expectedSigmol}) is based on the FUV profile. Given the implied recent star formation history, this might be best interpreted as an estimate of how much molecular gas was present in the inner galaxy in the recent past. The ALMA observations show locally bright molecular clouds coincident with the prominent outer disk \ion{H}{2} regions, which are also visible in the FUV.

Inside $\rgal\sim1.5$~kpc $\tau_{\rm dep}^{\rm atom}$ calculated using the FUV-based tracer remains relatively flat at$\approx 6{-}10$~Gyr. Beyond this radius, $\tau_{\rm dep}$ rises with an average of $\approx 17$~Gyr outside $r_{50,\star}$ and reaching $> 30$~Gyr by the edge of our SFR profiles. By contrast, \citet{Schruba2017} found $\tdepmol\sim2$~Gyr for individual outer-disk star-forming complexes, but the covering fraction of these regions is small. Thus NGC~6822 contains localized efficient star formation, but the formation of cold, dense material appears to be inefficient and so represents the bottleneck to star formation in the extended atomic gas disk. Supporting this, LGLBS-based absorption work indicates that only a small fraction of the atomic gas in the extended disk resides in the cold phase \citep{Pingel2024}. 

\subsection{WLM}

WLM is inclined, with the \hi\ (\autoref{fig:atlas_wlm}) and star formation activity traced by H$\alpha$ and bright FUV concentrated near the center and southern side of the disk \citep[e.g.,][]{Leaman2012,Khademi2021}. Fainter diffuse FUV emission pervades the disk. The \sigatom\ profile is high in the inner disk, with maximum $\sigatom \approx 8.4$~\uSig\ near $\rgal\sim0.5$~kpc. At larger radius, \sigatom\ shows an exponential decline in good agreement with LITTLE THINGS \citep[][]{Hunter2012}. 

CO is detected in WLM, but confined to a handful of compact clouds \citep[][]{Rubio2015,Hunt2015,Archer2024}. 
The depletion time-based estimated \sigmol\ profile shows that the galaxy is dominated by atomic gas at all radii.

WLM shows long depletion times, $\tau_{\rm dep}^{\rm atom} \approx 16$~Gyr overall. $\tau_{\rm dep}^{\rm atom}$ increases with an exponential scale length of $\approx 1$~kpc from $\sim 10$~Gyr near the galaxy center to $\gtrsim 30$~Gyr at the edge of our SFR profiles. 

\subsection{M33}

The \hi\ map in M33 shows multiple, flocculent spiral-arm features, which also host the molecular gas (\autoref{fig:atlas_m33}) as well as multiple small, feedback-driven \hi\ holes \citep{Deul1987}. Our field of view covers almost the entire $r_\mathrm{HI}$ radius, although additional lower column density \hi\ emission is present to the northwest and southeast of the main disk, consistent with earlier wide-field \hi\ mapping \citep{Deul1987,Brinks1990,Putman2009}. The two spiral arms that dominate the W1 stellar-mass image show the grand design spiral structure seen in the older stellar population \citep{smercina2023}. The \halpha\ and FUV emission extend out to $r_{25}$ and follow the flocculent structure of the gas.

In M33, \sigatom\ stays above ${\sim}$10~\uSig\ out to $\rgal\ {\sim}7$~kpc and declines steeply beyond, showing a saturation at this surface density consistent with previous work \citep[e.g.][]{Corbelli2003,Putman2009,Gratier2010,Kam2017,Koch2018}.
The 120~pc mass-weighted \hi\ profile lies above the annular mean, but the excess is smaller in the inner galaxy than what we see in the dwarfs. This matches the visual impression that the \hi\ in the disk of M33 is smoother than the emission in the dwarf galaxies. This situation changes in the outer spiral arms, which show a clumpy structure not captured in the flat azimuthal average. 

In terms of enclosed mass, the atomic disk is more extended than both the stellar and SFR distributions, with $r_{50,\rm atom}\approx5.5$~kpc compared to  $r_{50,\star}\approx2.6$~kpc and $r_{50,\rm SFR}\approx3.2$~kpc.
The azimuthally averaged molecular profile declines steeply until $\rgal\sim6$~kpc, but the 120-pc mass weighted average remains comparable to the atomic gas density where it can be reliably measured.
The stellar profile is well described by an exponential disk with $l_\star\approx1.8$~kpc, while the recent star formation is slightly more extended with $l_{\Sigma_{\rm SFR}}\approx2.3$~kpc, which is comparable to the molecular gas scale length $l_\mathrm{mol}=2.1~\mathrm{kpc}$ \citep{Druard2014}. 

All SFR tracers agree well to about $0.75\,r_{25}$ in M33. Beyond this radius, W4 drops relative to the other tracers. This indicates less extinction, likely due to falling dust-to-gas ratios and lower gas column densities.

Inside $r_{\rm gal} \sim 6$-$7$~kpc, $\tdepmol$ is flat at $\approx 0.5$-$1.5$~Gyr. This mirrors previous studies that found a constant or mildly varying $\tdepmol$ \citep[e.g.][]{Rosolowsky2007,Gardan2007,Gratier2010}. Meanwhile, $\tdephi$ and $\tdepgas$ are both longer than $\tdepmol$ and rise with radius, which implies that the main bottleneck for star formation in M33 is the conversion of \hi\ to \htwo.
As the disk becomes increasingly \hi-dominated at larger radii, $\tdepgas$ rises with an exponential scale lengths of $\approx 2.4$~kpc from $\tau_{\rm dep}^{\rm atom} \approx 1.7$~Gyr within $r_{50,\star}$ to $\tau_{\rm dep}^{\rm atom} \gtrsim 13$~Gyr in the outer galaxy.

\subsection{M31}

The large scale morphology of M31's \hi\ (\autoref{fig:atlas_m31}) agrees well with previous mapping \citep[][]{Brinks1984,Brinks1990,Chemin2009,Braun2009}. The most prominent feature is the ${{\sim}} 10$~kpc ring. There is an ongoing debate about whether the prominent arcs and loops outside this ring represent segments of spiral arms or resonant rings associated with the bar \citep[][]{AthanassoulaBeaton2006,Tenjes2017}. Whatever their origin, star formation activity and CO emission concentrate in these ring-like structures \citep[e.g.][]{Nieten2006,Gordon2006,Tabatabaei2010,Groves2012}. This results in the wavy appearance of the corresponding profiles.

The central bulge shows a lack of \hi\ and CO seen in the \sigatom\ profile (\autoref{fig:radial_m31}), which shows a central depression and then rises to a maximum of $\approx 7.2$~\uSig\ near $\rgal\sim10$~kpc before smoothly declining. Previous wide-field observations have shown a warped, asymmetric \hi\ disk that extends far beyond the optical radius \citep[e.g.][]{Braun1991,Braun2009,Chemin2009,Corbelli2010}. 

The 120~pc mass-weighted \hi\ profile is higher than the annular mean. As with M33, the contrast between mass-weighted and mean profile and 16-84\% range shows that M31 has a smoother \hi\ distribution than the shell-dominated dwarfs. The high inclination of the galaxy likely contributes to this apparent smoothness. 

Molecular gas is detected out to near the edge of the \citet{Nieten2006} map, though atomic gas dominates the ISM mass budget at all radii. The large extent of CO emission to $r_{\rm gal} \gtrsim 15$~kpc resembles that seen for NGC 4565 \citep{Krahm2026} or the Milky Way \citep{Heyer2015} and may be a general feature of high mass disk galaxies.

The atomic gas is more extended than the stellar mass distribution, with effective radii of $r_{50,\rm atom} \approx 13.8$~kpc and $r_{50,\star}=5.2$~kpc. Beyond $r_{\rm gal} \approx 14$~kpc the atomic gas mass in a ring exceeds the stellar mass. The distribution of SFR activity is intermediate between the gas and stars, with effective radius $r_{\rm 50, SFR} \approx 11.3$~kpc, reflecting the influence of the 10~kpc ring. 

The FUV+W4-based $\Sigma_{\rm SFR}$ and our CO-based $\Sigma_{\rm mol}$ imply molecular depletion times $\tdepmol\sim2$-$4$~Gyr with a weak radial increase. This is longer than the other galaxies with CO coverage in our sample, longer than typically found in disk galaxies \citep{Leroy2013,Sun2023}, and about two times longer than the ${\sim}1$~Gyr average between $6{-}17$~kpc found by \citet{Tabatabaei2010}. They find an integrated SFR $\sim 0.3$~M$_\odot$~yr$^{-1}$ (from H$\alpha$ and far-IR) and we use the same CO map, so the difference reflects our adopted CO-to-H$_2$ conversion factor. We estimate $M_{\rm mol} \approx 7 \times 10^8$~M$_\odot$, about twice the estimate in \citet{Tabatabaei2010} or \citet{Nieten2006}. This longer $\tau_{\rm dep}^{\rm mol}$ may reflect a genuine suppression of star formation efficiency. This would be consistent with M31's status as a ``green-valley,'' relatively low-SFR massive spiral that may be in a transition toward lower star formation \citep[e.g.][]{Mutch2011,Lehner2015,Boardman2020}. However, note that \citet{Lewis2017} measure the SFR directly from stellar populations and suggest that SFR may be underestimated by the standard UV and IR tracers.

The atomic and total gas depletion times are $\gtrsim 0.5$~dex ($3$ times) longer than $\tau_{\rm dep}^{\rm mol}$ near the 10~kpc ring. Thus, even in the most massive galaxy in the sample, in all rings, the regulating factor for star formation is the abundance of molecular gas relative to atomic gas. The $\tau_{\rm dep}^{\rm atom}$ and $\tau_{\rm dep}^{\rm gas}$ become even longer at larger radii, with an exponential decline beyond $\rgal \approx 9$~kpc with a scale length of $\approx 8$~kpc.

\begin{deluxetable*}{lcccccc}[t!]
\tabletypesize{\footnotesize}
\tablecaption{\label{tab:measured_radial_profile_properties} Radial Extent of Gas, Stars, and Star Formation Activity}
\tablewidth{0pt}
\tablehead{
\colhead{Quantity} &
\colhead{IC10} & \colhead{IC1613} & \colhead{WLM} & \colhead{NGC6822} & \colhead{M33} & \colhead{M31}
}
\startdata
\cutinhead{Isophotal radii [kpc]} 
$r_{25}$ (1) & 0.76 & 2.01 & 1.50 & 1.18 & 7.70 & 20.07 \\
$r_{\mathrm{HI}}$ (2) & $2.80$ & $3.01$ & $2.78$ & 5.78\tablenotemark{a} & $13.14$ & 25.45\tablenotemark{a} \\
$r (\sigatom>\sigstar)$ (3) & $2.10$ & $0.90$ & $0.18$ & $1.02$ & $4.38$ & n.r. \\
\cutinhead{Effective radii [kpc]} 
$r_{50,\star}$ (4) & 0.61 & 1.36 & 1.21 & 0.83 & 2.55 & 5.19 \\
$r_{90,\star}$ (5) & 1.52 & 2.83 & 2.78 & 1.94 & 5.38 & 13.47 \\
$r_{50,\mathrm{atom}}$ (6) & 1.33 & 0.94 & 1.30 & 2.27 & 5.54 & 13.75 \\
$r_{90,\mathrm{atom}}$ (7) & 2.85 & 1.13 & 2.29 & n.r. & 7.49 & 18.66 \\
$r_{50,\mathrm{SFR}}$ (8) & 0.30 & 0.96 & 0.74 & 1.32 & 3.15 & 11.28 \\
$r_{90,\mathrm{SFR}}$ (9) & 0.70 & 1.72 & 1.37 & 1.98 & 5.70 & 16.17 \\
\cutinhead{Disk scale lengths [kpc]} 
$l_{\star}$ (10) & $0.39$ & $0.90$ & $0.78$ & $0.58$ & $1.75$ & $3.82$ \\
$l_{\mathrm{\Sigma_{SFR}}}$ (11) & $0.27$ & $0.42$ & $0.68$ & $1.41$ & $2.26$ & $3.77$ \\
\cutinhead{Outer \hi\ scale length [kpc]} 
$l_{\mathrm{HI}}^{\rm outer}$ (12) & $1.00$ & $1.02$ & $1.03$ & $2.09$ & $2.05$ & $8.50$ \\
Start of decline (13) & 0.06 & 1.14 & 0.54 & 2.00 & 7.00 & 10.98 \\
\enddata
\tablenotetext{a}{We estimated $r_\mathrm{HI}$ by extrapolating the outer \hi\ radial profile constructed from sectors near the major axis. To stabilize the profile in the outskirts, we measured the mean \sigatom\  in the major-axis sectors of each annulus and then averaged that over the full ellipse, effectively assuming the minor-axis sectors have \sigatom =0. We then fitted an exponential decline to this corrected outer profile and determined the radius where \sigatom=1\uSig. {For NGC~6822 where the major axis is dominated by large non-axisymmetric features (probably related to an interaction, see Section \ref{sec:NGC6822}), we constructed a comparison profile from sectors near the minor axis yielding a smaller extrapolated value, $r_{\rm HI,ext}\approx4.0$~kpc, indicating that the exact extrapolated radius is sensitive to the asymmetric outer \hi\ structure of NGC~6822.}} 
\tablecomments{Radial extent of atomic gas, stellar mass, and star formation activity in our targets. Rows: (1) 25$^{\rm th}$ magnitude $B$-band isophote from HyperLEDA \citep{Makarov2014}, (2) isophotal radius where $\Sigma_{\rm HI} > 1$~M$_\odot$~pc$^{-2}$, (3) radius where $\Sigma_{\rm atom} = \Sigma_\star$, (4)-(9) radius including $50\%$ and $90\%$ of the total stellar mass ($\star$), atomic gas mass, and star formation activity. Total masses come from Table~\ref{tab:fluxes}. (10) exponential scale length of stellar mass measured in the exponential part of the disk, (11) scale length of star formation rate surface density over the same range. (12) Exponential scale length of atomic gas mass surface density in the outer part of the disk. (13) Radius beyond which the atomic gas surface density declines and $l_{\rm HI}^{\rm outer}$ is measured.}
\end{deluxetable*}

\begin{deluxetable*}{lcccccc}[t!]
\tabletypesize{\footnotesize}
\tablecaption{\label{tab:tdep} Depletion Times in Local Group Galaxies}
\tablewidth{0pt}
\tablehead{
\colhead{Quantity} &
\colhead{IC10} & \colhead{IC1613} & \colhead{WLM} & \colhead{NGC6822} & \colhead{M33} & \colhead{M31}
}
\startdata
$\tau_{\rm dep}^{\rm gas}$ [Gyr] (1) & 2.74 & 17.00\tablenotemark{a} & 16.44 & 15.02\tablenotemark{a} & 8.85 & 20.11\tablenotemark{b} \\
$\tau_{\rm dep}^{\rm gas} (r < r_{\rm 50, \star})$ [Gyr] (2) & 0.73 & 5.93\tablenotemark{a} & 13.83 & 5.83\tablenotemark{a} & 2.66 & 6.83\tablenotemark{b}\\
$\tau_{\rm dep}^{\rm atom} (r < r_{\rm 50, \star})$ [Gyr] (3) & 0.49 & 5.93 & 13.80 & 5.83 & 1.73 & 4.79\tablenotemark{b}\\
$\tau_{\rm dep}^{\rm gas} (r > r_{\rm 50, \star})$ [Gyr] (4) & 13.08 & 49.85\tablenotemark{a} & 25.71 & 18.08\tablenotemark{a} & 13.01 & 20.95\\
$\tau_{\rm dep}^{\rm mol}$ [Gyr] (5) & 0.76 & \nodata & 0.02 & \nodata & 1.23 & 2.35\\
$l_{\rm \tau , gas}$ [kpc] (6) & 0.33 & 0.55 & 1.04 & 1.68 & 2.36 & 7.99\\
\enddata
\tablenotetext{a}{No independent molecular gas estimate available, so we take $\Sigma_{\rm gas} = \Sigma_{\rm atom}$.}
\tablenotetext{b}{Depletion time measurements for M31 ignore the bulge region {(r $< 2.5$). The values in rows (1)–(3) therefore refer to the disk outside the bulge. 
For reference, the total gas depletion times in} the inner region ($2.5$ kpc $< r < 6$ kpc) is  $\tau_{\rm dep}^{\rm gas}=7.7~\mathrm{Gyr}$ , for the star forming ring ($6 $  kpc $< r < 13$ kpc) {we find} $\tau_{\rm dep}^{\rm gas}=15.5~\mathrm{Gyr}$ and {in the }outer disk ($r>13$~kpc)   $\tau_{\rm dep}^{\rm gas} = 31.8~\mathrm{Gyr}$.}
\tablecomments{Gas depletion times (Eq. \ref{eq:sfe}). We quote gas depletion times of M31 for specific radial bins, because the bulge impacts depletion time estimates in the inner 2.5~kpc.
(1) Galaxy-integrated total gas depletion time, 
(2) gas depletion time within the stellar effective radius, 
(3) atomic gas depletion time within the stellar effective radius, 
(4) total gas depletion time outside the stellar effective radius, allowing comparison with (2) and (3) to assess the impact of molecular gas on $\tau_{\rm dep}^{\rm gas}$, 
(5) molecular gas depletion time for galaxies with a complete CO map, 
(6) exponential scale length describing \textit{increase} of $\tau_{\rm dep}^{\rm gas}$ with radius. $\Sigma_{\rm gas} \approx \Sigma_{\rm atom}$ for all targets without an independent molecular gas estimate. 
}
\end{deluxetable*}

\begin{deluxetable*}{lcccccc}[t!]
\tabletypesize{\footnotesize}
\tablecaption{\label{tab:fluxes}Integrated fluxes, luminosities, and physical quantities}
\tablewidth{0pt}
\tablehead{
\colhead{Quantity} & \colhead{IC10} & \colhead{IC1613} & \colhead{WLM} & \colhead{NGC6822} & \colhead{M33} & \colhead{M31}
}
\startdata
\cutinhead{Broad band fluxes}
W1 [Jy] (1) & $2.66 \pm 0.08$ & $0.32 \pm 0.01$ & $0.11$ & $1.31 \pm 0.03$ & $15.50 \pm 0.37$ & $223.74 \pm 10.17$\\
W2 [Jy] (1) & $1.58 \pm 0.06$ & $0.19 \pm 0.01$ & $0.07$ & $0.75 \pm 0.02$ & $9.04 \pm 0.25$ & $1.90 \pm 0.09$\\
W3 [Jy] (1) & $3.12 \pm 0.16$ & bkg$^{a}$ & bkg$^{a}$ & $0.49 \pm 0.04$ & $32.11 \pm 1.45$ & $116.95 \pm 10.78$\\
W4 [Jy] (1) & $11.42 \pm 0.66$ & bkg$^{a}$ & bkg$^{a}$ & $2.32 \pm 0.15$ & $49.96 \pm 2.86$ & $110.33 \pm 11.30$\\
FUV [Jy] (1) & \nodata & $0.10$ & $0.03$ & $0.38 \pm 0.02$ & $2.51 \pm 0.13$ & $1.78 \pm 0.51$\\
NUV [Jy] (1) & \nodata & $0.10 \pm 0.01$ & $0.04$ & $0.41 \pm 0.03$ & $2.95 \pm 0.15$ & $2.59 \pm 0.58$\\
\cutinhead{Broad band luminosities}
W1 [$10^8$ L$_{\nu,\odot}$] (2) & $10.05 \pm 0.30$ & $1.19 \pm 0.03$ & $0.65 \pm 0.02$ & $2.35 \pm 0.06$ & $72.99 \pm 1.75$ & $866.77 \pm 39.38$\\
W2 [$10^8$ L$_{\nu,\odot}$] (2) & $10.98 \pm 0.43$ & $1.29 \pm 0.04$ & $0.74 \pm 0.02$ & $2.48 \pm 0.07$ & $78.14 \pm 2.19$ & $13.53 \pm 0.64$\\
W3 [$10^{10}$ L$_{\nu,\odot}$] (2) & $1.26 \pm 0.06$ & bkg$^{a}$ & bkg$^{a}$ & $0.09 \pm 0.01$ & $16.13 \pm 0.73$ & $48.33 \pm 4.45$\\
W4 [$10^{10}$ L$_{\nu,\odot}$] (2) & $16.69 \pm 0.96$ & bkg$^{a}$ & bkg$^{a}$ & $1.60 \pm 0.11$ & $91.11 \pm 5.21$ & $165.52 \pm 16.95$\\
FUV [$10^{12}$ L$_{\nu,\odot}$] (2) & \nodata & $1.29 \pm 0.06$ & $0.75 \pm 0.04$ & $2.46 \pm 0.14$ & $42.46 \pm 2.12$ & $24.83 \pm 7.08$\\
NUV [$10^9$ L$_{\nu,\odot}$] (2) & \nodata & $1.93 \pm 0.10$ & $1.28 \pm 0.06$ & $3.72 \pm 0.23$ & $69.70 \pm 3.49$ & $50.25 \pm 11.34$\\
\cutinhead{Line fluxes}
H$\alpha$ [$10^{-12}$ erg s$^{-1}$ cm$^{-2}$] (3) & $124.80 \pm 6.24$ & $12.33 \pm 0.62$ & $4.53 \pm 0.23$ & $12.93 \pm 0.70$ & $295.40 \pm 14.77$ & \nodata\\
H$\alpha$ aperture correction (4) & $1.00$ & $0.97$ & $0.92$ & $0.96$ & $0.96$ & $1.00$\\
CO(1-0) [$10^3$ Jy km s$^{-1}$] (3) & $1.75 \pm 0.18$ & \nodata & \nodata & \nodata & \nodata & $422.23 \pm 52.33$\\
CO(1-0) aperture correction (4) & $1.36$ & $1.00$ & $1.00$ & $1.00$ & $1.00$ & $0.98$\\
CO(2-1) [$10^3$ Jy km s$^{-1}$] (3) & \nodata & \nodata & $0.01$ & \nodata & $59.77 \pm 25.40$ & \nodata\\
CO(2-1) aperture correction (4) & $1.00$ & $1.00$ & $1.62$ & $1.00$ & $0.98$ & $1.00$\\
HI [$10^3$ Jy km s$^{-1}$] (5) & $0.54 \pm 0.01$ & $0.13 \pm 0.01$ & $0.33$ & $2.87 \pm 0.15$ & $11.48 \pm 0.57$ & $39.77 \pm 0.70$\\
\cutinhead{Line luminosities}
H$\alpha$ [$10^{38}$ erg s$^{-1}$] (6) & $88.53 \pm 4.43$ & $8.52 \pm 0.43$ & $5.20 \pm 0.26$ & $4.35 \pm 0.24$ & $261.40 \pm 13.07$ & \nodata\\
CO(1-0) [$10^{6}$ K km s$^{-1}$ pc$^2$] (6) & $2.54 \pm 0.26$ & \nodata & \nodata & \nodata & \nodata & $628.32 \pm 77.87$\\
CO(2-1) [$10^{6}$ K km s$^{-1}$ pc$^2$] (6) & \nodata & \nodata & \nodata & \nodata & $27.03 \pm 11.49$ & \nodata\\
HI [$10^{9}$ K km s$^{-1}$ pc$^2$] (6) & $5.13 \pm 0.07$ & $1.23 \pm 0.07$ & $5.06 \pm 0.07$ & $12.99 \pm 0.68$ & $136.75 \pm 6.84$ & $389.72 \pm 6.84$\\
\cutinhead{Physical properties}
$M_{\star}$ [$10^6$ M$_\odot$] (7) & $91.99 \pm 0.24$ & $17.26 \pm 0.07$ & $4.20 \pm 0.02$ & $27.00 \pm 0.07$ & $1582.69 \pm 0.24$ & $9723.83 \pm 0.24$\\
$\mathrm{SFR}$ [$10^{-3}$ M$_\odot$ yr$^{-1}$] (7) & $15.88$ & $3.16$ & $0.74$ & $1.56$ & $156.83$ & $48.33$\\
$M_{\rm atom}$ [$10^6$ M$_\odot$] (5) & $101 \pm 7$ & $85 \pm 5$ & $100 \pm 7$ & $260 \pm 30$ & $2700 \pm 200$ & $7700 \pm 200$\\
$M_{\rm mol}$ [$10^6$ M$_\odot$] (7) & $6.14 \pm 0.14$ & $25.88^{b}$ & $0.02 \pm 0.10^{b}$ & $9.09 \pm 0.01^{b}$ & $204.32 \pm 4.32$ & $710.97 \pm 5.73$\\
$M/L$ (8) & 0.31 & 0.34 & 0.31 & 0.28 & 0.28 & 0.46  \\
\enddata
\tablenotetext{a}{Only background is visible i.e. empty map}
\tablenotetext{b}{When CO measurements are not available or not constraining, we estimate $M_{\rm mol}$ from the \sigsfr\ profile and $\tau_{\rm dep}^{\rm mol}$.}
\tablecomments{Fluxes, luminosities, and physical quantities are from integrating annulus profiles measured
within a selected outer radius. We use the convergence radius ($r_{\rm conv}$) when the cumulative profile shows convergence, and background radius ($r_{\rm bkg}$) for background-corrected W1/W2 profiles.  The full list of selected radii and selection criteria is provided in Table \ref{tab:radii}.
1) Broad band fluxes from integrating radial profiles. Uncertainties include statistical noise, an uncertainty on the background level, and a flux calibration uncertainty. (2) Broad band luminosities adopting the distances from Table \ref{tab:physical} and the Solar absolute magnitudes from \citet{Willmer2018}:
GALEX FUV 17.30, GALEX NUV 10.16, WISE1 5.91, WISE2 6.57, WISE3 8.48, and WISE4 9.88. (3) Line fluxes from annulus-profile integration. (4) Aperture correction factors based on FUV for H$\alpha$ and W3 for CO, comparing the reference profile at the line-map edge to the converged reference integral. (5) \hi\ flux, luminosity, and mass from \citet{Koch2025} but including an additional factor of $1.35$ to account for the mass of helium. (6) Line luminosity adopting the distances from Table \ref{tab:physical} and applying the aperture correction. (7) Integrated stellar mass, star formation rate, and molecular gas mass from the profiles. These include variable conversion factors, mass-to-light ratios, and aperture corrections as described in the text.  We omit the per-galaxy SFR uncertainties from the table body because the propagated profile-based errors are much smaller than the display precision. The mean fractional SFR uncertainty across the sample is 0.077\%. (8) Mass-to-light ratio at 3.4 $\mu$m, defined here as the ratio of the integrated stellar mass to the integrated W1 luminosity.}
\end{deluxetable*}

\begin{deluxetable*}{lcccccc}[t!]
\tabletypesize{\footnotesize}
\tablecaption{\label{tab:hi_props} Atomic Gas Surface Densities}
\tablewidth{0pt}
\tablehead{
\colhead{Quantity} &
\colhead{IC10} & \colhead{IC1613} & \colhead{WLM} & \colhead{NGC6822} & \colhead{M33} & \colhead{M31}
}
\startdata
$\Sigma_{\rm atom, max}$ [M$_\odot$~pc$^{-2}$] (1) & 12.30 & 5.93 & 8.35 & 10.67 & 11.53 & 7.18\\
\hline
\multicolumn{7}{c}{Inner disk ($r<r_{50, \star}$)} \\
\hline
$\Sigma_{\rm atom, mean}$ [M$_\odot$~pc$^{-2}$] (2) & 10.44 & 3.16 & 7.08 & 7.27 & 9.34 & 1.04\tablenotemark{a}\\
$\log_{10}\Sigma_{\rm atom}$ range [dex] (3) & 0.45 & 0.96 & 0.15 & 0.21 & 0.32 & 0.78\tablenotemark{a}\\
$\left< \Sigma_{\rm atom}^{120\,\mathrm{pc}} \right>$ [M$_\odot$~pc$^{-2}$] (4) & 12.83 & 5.58 & 7.23 & 7.68 & 10.50 & 1.63\tablenotemark{a}\\
$\left<c_{\rm HI,120pc}\right>$ (5) & 1.27 & 1.94 & 1.04 & 1.07 & 1.13 & 1.57\tablenotemark{a}\\
\hline
\multicolumn{7}{c}{Outer disk ($r_{50, \star}<r<r_{\rm HI}$)} \\
\hline
$\Sigma_{\rm atom, mean}$ [M$_\odot$~pc$^{-2}$] (6) & 3.11 & 2.45 & 2.49 & 5.06 & 6.19 & 4.55\\
$\log_{10}\Sigma_{\rm atom}$ range [dex] (7) & 0.85 & 0.37 & 0.27 & 0.52 & 0.44 & 0.42\\
$\left< \Sigma_{\rm atom}^{120\,\mathrm{pc}} \right>$ [M$_\odot$~pc$^{-2}$] (8) & 4.63 & 2.83 & 2.61 & 6.34 & 7.63 & 5.41\\
$\left<c_{\rm HI,120pc}\right>$ (9) & 1.65 & 1.25 & 1.12 & 1.27 & 1.62 & 1.23\\
\enddata
\tablenotetext{a}{Surface densities for M31 ignore the bulge region (the inner 2.5 kpc).}
\tablecomments{Atomic gas surface densities including a factor of $1.35$ to account for helium based on the radial profile measurements. 
(1) Maximum of $\Sigma_{\rm atom}$ across all annuli in the radial profile. 
(2) Mean and (3) 16-84\% range in dex in $\Sigma_{\rm atom}$ within the stellar mass effective radius. 
(4) The average mass-weighted $\Sigma_{\rm atom}$ and (5) the mean clumping factor within the stellar effective radius. 
(6) Mean and (7) 16-84\% range in dex outside the stellar effective radius and inside $r_{\rm HI}$.
(8) The average mass-weighted mean $\Sigma_{\rm atom}$ and (9) the mean clumping factor outside the stellar effective radius and inside $r_{\rm HI}$
(also see \autoref{tab:measured_radial_profile_properties}).
Rows (2) \& (6), (3) \& (7), (4) \& (8), (5) \& (9) contrast the inner versus outer behavior of these quantities.}
\end{deluxetable*}

\section{Conclusions}

We assemble a database tracing gas, star formation activity, and stellar mass across the six star-forming Local Group galaxies targeted by the Local Group L-Band Survey \citep[LGLBS;][]{Koch2025}. Our methods follow \citet{Sun2022} and combine the LGLBS \ion{H}{1} maps with reprocessed WISE near- and mid-IR data and GALEX UV data with archival H$\alpha$ and CO observations. The distinguishing feature of LGLBS is the high physical resolution, $\sim 120$~pc atomic gas data. We publicly release the processed images of stellar mass, atomic and molecular gas, and star formation along with their radial profiles, and region measurements accompanying this paper.
In Paper II \citep{Eibensteiner2026a} we use these measurements to study the regulation of star formation in \ion{H}{1}-dominated systems.

\begin{enumerate}
\item \textbf{Atomic gas makes up most of the ISM over our whole sample and exceeds the stellar mass in the outer regions of our targets}. 
\end{enumerate}

Except for the innermost part of the dwarf starburst IC 10, $\sigatom > \sigmol$ in every azimuthally averaged ring in our sample. This stands in contrast with the frequently studied population of star-forming massive disk galaxies, where often $\sigmol > \sigatom$ in the inner regions \citep{Wong2002,Leroy2008}. This means that LGLBS represents an ideal sample to study environments where the balance of atomic to molecular gas is the key factor regulating star formation activity.

Meanwhile, except for IC 10 all targets show $\sigatom > \sigstar$ at large $\rgal$. In IC 1613 and NGC 6822 this occurs at $\rgal \gtrsim 1$~kpc, while WLM appears atomic gas dominated over the whole disk. M33 is gas-dominated outside about 4~kpc, and M31 is gas-dominated beyond about 14~kpc. The situation in IC 10 is ambiguous because of the difficulty tracing its stellar disk to large radii in the presence of Galactic foreground stars.

\begin{enumerate}
\setcounter{enumi}{1}
\item \textbf{The atomic and total gas depletion times increase as a function of increasing $\rgal$}.
\end{enumerate}

In all of our targets, $\tdepgas$ increases as a function of $\rgal$ (though in NGC 6822 this depends sensitively on our choice of SFR tracer). This increase often appears exponential and relatively smooth rather than, e.g., associated with a threshold. We provide best-fit scale lengths in Table \ref{tab:tdep}. 

By contrast, in the three targets where we have complete CO coverage (IC 10, M33, and M31) the molecular gas depletion time, \tdepmol , shows diverse behavior. It rises with $\rgal$ in IC 10, remains about constant in M33, and shows a weak decline in M31. \tdepmol\ is always less than \tdephi . Thus, the molecular-to-atomic gas balance regulates star formation across LGLBS and the efficiency of this regulation varies within each galaxy. 

\begin{enumerate}
\setcounter{enumi}{2}
\item \textbf{The gas depletion time becomes extremely long in the outer regions of our targets.}
\end{enumerate}

Outside the stellar effective radius, $r_{\rm 50, \star}$, $\tdephi$ is always $> 10$~Gyr (in WLM the whole galaxy has $\tdephi > 17$~Gyr).  Yet the \hi\ disk extends $1.6-8\times$ beyond $r_{50,\star}$ across our sample (Table~\ref{tab:measured_radial_profile_properties}), meaning the bulk of the atomic gas by area resides in this inefficient regime. The depletion time increases exponentially with scale lengths of $\approx0.3-8.0$~kpc (Table~\ref{tab:tdep}). 
This agrees with the observations of inefficient star formation for the outer disks of galaxies \citep{Bigiel2010_Ineff} and dwarf galaxies \citep{Hunter2024}. LGLBS accesses these inefficient outer disks at high physical resolution.

\begin{enumerate}
\setcounter{enumi}{3}
\item \textbf{Dwarf galaxies show prominent shells and cavities that lead to significant clumping and wide ranges of column densities within in each ring.} 
\end{enumerate}

We calculate mass-weighted $\sigatom$ profiles that reflect the 120~pc resolution \sigatom\ and record the 16-84\% range of column densities in each ring. In IC~10 and IC~1613, the mass-weighted $\sigatom$ is enhanced relative to the azimuthal average by a factor of $\approx1.3-1.7$, and the 16-84\% range of \sigatom\ on 120~pc scale spans up to 0.9~dex. This reflects the visible presence of large shells and cavities in the \hi\ maps of these galaxies. M33 and M31 show less clumping and a modest range of \sigatom\ especially towards the \hi-rich outer disk. This agrees with previous works suggesting increased porosity for the ISM in dwarf galaxies \citep{Bagetakos2011,Pokhrel2020}.

\begin{enumerate}
\setcounter{enumi}{4}
\item \textbf{Star formation rate estimates agree reasonably well with a few notable outliers and shortcomings in available data.} 
\end{enumerate}

In IC10, IC1613, WLM, and M33 FUV and H$\alpha$-based SFR estimates show good agreement, and including WISE mid-IR emission suggests reasonable extinction corrections (or very little extinction in the case of the WISE4 non-detections for IC 1613 and WLM). In NGC 6822 the inner galaxy shows significant FUV emission but little H$\alpha$, suggesting a recent burst of star formation. The tracers agree better for the bright star-forming complexes in the outer galaxy. In IC10 the lack of an FUV map prevents probing the SFR in the outer galaxy. In M31, the lack of a public H$\alpha$ map with good continuum subtraction and sensitivity to diffuse emission prevents such a comparison \citep[the][map does not appear to be accessible]{Devereux1994}. Our measurements agree with historic H$\alpha$-based ones for M31, but we highlight the previously identified tension between color-magnitude diagram-based and UV+IR-based estimates in M31 \citep{Lewis2017}.

\begin{acknowledgments}
    CE acknowledges support by the Jansky Fellowship, which is awarded by the National Radio Astronomy Observatory (NRAO), which is operated by the Associated Universities Inc. (AUI).
    The LGLBS collaboration is grateful for support from the National Science Foundation, AST-2205628 and AST-2205630. 
    JS acknowledges support by the National Aeronautics and Space Administration (NASA) through the NASA Hubble Fellowship grant HST-HF2-51544 awarded by the Space Telescope Science Institute (STScI), which is operated by the Association of Universities for Research in Astronomy, Inc., under contract NAS~5-26555.
    ER acknowledges the support of the Natural Sciences and Engineering Research Council of Canada (NSERC), funding reference number RGPIN-2022-03499.
    TGW gratefully acknowledges support from the UK ALMA Regional Centre (ARC) Node, which is supported by the Science and Technology Facilities Council grant number ST/Y004108/1.
    MPB acknowledges support by the Jansky Fellowship, which is awarded by the National Radio Astronomy Observatory (NRAO), which is operated by the Associated Universities Inc. (AUI).
    The National Radio Astronomy Observatory and Green Bank Observatory are facilities of the U.S. National Science Foundation operated under cooperative agreement by Associated Universities, Inc
    This work is based on observations carried out under project number 198-13 with the IRAM 30-meter telescope. IRAM is supported by INSU/CNRS (France), MPG (Germany) and IGN (Spain).
    This paper makes use of the following ALMA data: 2022.1.00276.S, 2017.1.00901.S, 2019.1.01182.S, 2018.A.00058.S, 2021.1.0999.S, 2022.1.00403.S ALMA is a partnership of ESO (representing its member states), NSF (USA) and NINS (Japan), together with NRC (Canada), MOST and ASIAA (Taiwan), and KASI (Republic of Korea), in cooperation with the Republic of Chile. The Joint ALMA Observatory is operated by ESO, AUI/NRAO and NAOJ.
    This paper makes use of CARMA projects c1995 and c1211.
    This work was supported by a collaborative NSF Astronomy \& Astrophysics Grant, AST-2205631.

\end{acknowledgments}

\begin{contribution}

CE led the project, wrote the original drafts of the manuscript, carried out the statistical analysis, made the figures, processed the multi-wavelength data products, and updated the MegaTable code for use with LGLBS. AKL, JS, EWK, ER, and CK contribution to the revision and editing of the manuscript. AKL contributed to the generation of the new GALEX and WISE data products. JS contributed to the use of the MegaTable code to produce the database. AKL, JS, EWK, and ER contributed quality assurance for the multi-wavelength data and the numerical results. ECO, JO, RC, ADB, MPB, JJD, AAK, CWL, SKS, AS, ET, VV, TMW, and TGW provided comments on the drafts. EWK and NP were responsible for imaging the LGLBS \hi\ data with additional contributions to the survey detailed in Koch et al. (2025). All coauthors provided regular input at weekly telecons.


\end{contribution}

\facilities{VLA, GALEX, WISE, KAT-7, CARMA, IRAM(EMIR), ALMA, PT, HT}

\software{
        \texttt{NumPy} \citep{NumPy2020},
        \texttt{SciPy} \citep{SciPy2020},
        \texttt{Astropy} \citep{Astropy2013,Astropy2018,Astropy2022},
        \texttt{radio-astro-tools} (spectral-cube, radio-beam) \citep{SPECTRALCUBE2020},
        \texttt{pandas} \citep{Pandas_1.3.4},
        \texttt{Matplotlib} \citep{Matplotlib2007},
        \texttt{CASA} \citep{CASATeam2022},
        \texttt{astrometry} \url{https://nova.astrometry.net/}
          }

\bibliography{LGLBS}

\begin{thebibliography}{}
\expandafter\ifx\csname natexlab\endcsname\relax\def\natexlab#1{#1}\fi
\providecommand{\url}[1]{\href{#1}{#1}}
\providecommand{\dodoi}[1]{doi:~\href{http://doi.org/#1}{\nolinkurl{#1}}}
\providecommand{\doeprint}[1]{\href{http://ascl.net/#1}{\nolinkurl{http://ascl.net/#1}}}
\providecommand{\doarXiv}[1]{\href{https://arxiv.org/abs/#1}{\nolinkurl{https://arxiv.org/abs/#1}}}

\bibitem[{H.~N. {Archer} {et~al.}(2024){Archer}, {Hunter}, {Elmegreen}, {Rubio}, {Cigan}, {Windhorst}, {Cort{\'e}s}, \& {Jansen}}]{Archer2024}
{Archer}, H.~N., {Hunter}, D.~A., {Elmegreen}, B.~G., {et~al.} 2024, \bibinfo{title}{{Probing the Relationship Between Early Star Formation and CO in the Dwarf Irregular Galaxy WLM with JWST},} \aj, 167, 274, \dodoi{10.3847/1538-3881/ad3f18}

\bibitem[{ {Astropy Collaboration} {et~al.}(2013){Astropy Collaboration}, {Robitaille}, {Tollerud}, {Greenfield}, {Droettboom}, {Bray}, {Aldcroft}, {Davis}, {Ginsburg}, {Price-Whelan}, \& et~al.}]{Astropy2013}
{Astropy Collaboration}, {Robitaille}, T.~P., {Tollerud}, E.~J., {et~al.} 2013, \bibinfo{title}{{Astropy: A community Python package for astronomy},} \aap, 558, A33, \dodoi{10.1051/0004-6361/201322068}

\bibitem[{ {Astropy Collaboration} {et~al.}(2018){Astropy Collaboration}, {Price-Whelan}, {Sip{\H{o}}cz}, {G{\"u}nther}, {Lim}, {Crawford}, {Conseil}, {Shupe}, {Craig}, {Dencheva}, {Ginsburg}, {VanderPlas}, {Bradley}, {P{\'e}rez-Su{\'a}rez}, {de Val-Borro}, {Aldcroft}, {Cruz}, {Robitaille}, {Tollerud}, {Ardelean}, {Babej}, {Bach}, {Bachetti}, {Bakanov}, {Bamford}, {Barentsen}, {Barmby}, {Baumbach}, {Berry}, {Biscani}, {Boquien}, {Bostroem}, {Bouma}, {Brammer}, {Bray}, {Breytenbach}, {Buddelmeijer}, {Burke}, {Calderone}, {Cano Rodr{\'\i}guez}, {Cara}, {Cardoso}, {Cheedella}, {Copin}, {Corrales}, {Crichton}, {D'Avella}, {Deil}, {Depagne}, {Dietrich}, {Donath}, {Droettboom}, {Earl}, {Erben}, {Fabbro}, {Ferreira}, {Finethy}, {Fox}, {Garrison}, {Gibbons}, {Goldstein}, {Gommers}, {Greco}, {Greenfield}, {Groener}, {Grollier}, {Hagen}, {Hirst}, {Homeier}, {Horton}, {Hosseinzadeh}, {Hu}, {Hunkeler}, {Ivezi{\'c}}, {Jain}, {Jenness}, {Kanarek}, {Kendrew}, {Kern}, {Kerzendorf}, {Khvalko}, {King}, {Kirkby}, {Kulkarni},
  {Kumar}, {Lee}, {Lenz}, {Littlefair}, {Ma}, {Macleod}, {Mastropietro}, {McCully}, {Montagnac}, {Morris}, {Mueller}, {Mumford}, {Muna}, {Murphy}, {Nelson}, {Nguyen}, {Ninan}, {N{\"o}the}, {Ogaz}, {Oh}, {Parejko}, {Parley}, {Pascual}, {Patil}, {Patil}, {Plunkett}, {Prochaska}, {Rastogi}, {Reddy Janga}, {Sabater}, {Sakurikar}, {Seifert}, {Sherbert}, {Sherwood-Taylor}, {Shih}, {Sick}, {Silbiger}, {Singanamalla}, {Singer}, {Sladen}, {Sooley}, {Sornarajah}, {Streicher}, {Teuben}, {Thomas}, {Tremblay}, {Turner}, {Terr{\'o}n}, {van Kerkwijk}, {de la Vega}, {Watkins}, {Weaver}, {Whitmore}, {Woillez}, {Zabalza}, \& {Astropy Contributors}}]{Astropy2018}
{Astropy Collaboration}, {Price-Whelan}, A.~M., {Sip{\H{o}}cz}, B.~M., {et~al.} 2018, \bibinfo{title}{{The Astropy Project: Building an Open-science Project and Status of the v2.0 Core Package},} \aj, 156, 123, \dodoi{10.3847/1538-3881/aabc4f}

\bibitem[{ {Astropy Collaboration} {et~al.}(2022){Astropy Collaboration}, {Price-Whelan}, {Lim}, {Earl}, {Starkman}, {Bradley}, {Shupe}, {Patil}, {Corrales}, {Brasseur}, \& et~al.}]{Astropy2022}
{Astropy Collaboration}, {Price-Whelan}, A.~M., {Lim}, P.~L., {et~al.} 2022, \bibinfo{title}{{The Astropy Project: Sustaining and Growing a Community-oriented Open-source Project and the Latest Major Release (v5.0) of the Core Package},} \apj, 935, 167, \dodoi{10.3847/1538-4357/ac7c74}

\bibitem[{E. Athanassoula \& R.~L. Beaton(2006)Athanassoula \& Beaton}]{AthanassoulaBeaton2006}
Athanassoula, E., \& Beaton, R.~L. 2006, \bibinfo{title}{Unravelling the mystery of the M31 bar,} MNRAS, 370, 1499

\bibitem[{I. {Bagetakos} {et~al.}(2011){Bagetakos}, {Brinks}, {Walter}, {de Blok}, {Usero}, {Leroy}, {Rich}, \& {Kennicutt}}]{Bagetakos2011}
{Bagetakos}, I., {Brinks}, E., {Walter}, F., {et~al.} 2011, \bibinfo{title}{{The Fine-scale Structure of the Neutral Interstellar Medium in Nearby Galaxies},} \aj, 141, 23, \dodoi{10.1088/0004-6256/141/1/23}

\bibitem[{F. {Belfiore} {et~al.}(2023{\natexlab{a}}){Belfiore}, {Leroy}, {Sun}, {Barnes}, {Boquien}, {Cao}, {Congiu}, {Dale}, {Egorov}, {Eibensteiner}, {Glover}, {Grasha}, {Groves}, {Klessen}, {Kreckel}, {Neumann}, {Querejeta}, {Sanchez-Blazquez}, {Schinnerer}, \& {Williams}}]{Belfiore2023}
{Belfiore}, F., {Leroy}, A.~K., {Sun}, J., {et~al.} 2023{\natexlab{a}}, \bibinfo{title}{{Calibration of hybrid resolved star formation rate recipes based on PHANGS-MUSE H{\ensuremath{\alpha}} and H{\ensuremath{\beta}} maps},} \aap, 670, A67, \dodoi{10.1051/0004-6361/202244863}

\bibitem[{F. {Belfiore} {et~al.}(2023{\natexlab{b}}){Belfiore}, {Leroy}, {Williams}, {Barnes}, {Bigiel}, {Boquien}, {Cao}, {Chastenet}, {Congiu}, {Dale}, {Egorov}, {Eibensteiner}, {Emsellem}, {Glover}, {Groves}, {Hassani}, {Klessen}, {Kreckel}, {Neumann}, {Neumann}, {Querejeta}, {Rosolowsky}, {Sanchez-Blazquez}, {Sandstrom}, {Schinnerer}, {Sun}, {Sutter}, \& {Watkins}}]{Belfiore2023b}
{Belfiore}, F., {Leroy}, A.~K., {Williams}, T.~G., {et~al.} 2023{\natexlab{b}}, \bibinfo{title}{{Calibrating mid-infrared emission as a tracer of obscured star formation on H II-region scales in the era of JWST},} \aap, 678, A129, \dodoi{10.1051/0004-6361/202347175}

\bibitem[{F. {Bigiel} {et~al.}(2010){Bigiel}, {Leroy}, {Walter}, {Blitz}, {Brinks}, {de Blok}, \& {Madore}}]{Bigiel2010_Ineff}
{Bigiel}, F., {Leroy}, A., {Walter}, F., {et~al.} 2010, \bibinfo{title}{{Extremely Inefficient Star Formation in the Outer Disks of Nearby Galaxies},} \aj, 140, 1194, \dodoi{10.1088/0004-6256/140/5/1194}

\bibitem[{N. {Boardman} {et~al.}(2020){Boardman}, {Zasowski}, {Newman}, {Andrews}, {Fielder}, {Bershady}, {Brinkmann}, {Drory}, {Krishnarao}, {Lane}, {Mackereth}, {Masters}, \& {Stringfellow}}]{Boardman2020}
{Boardman}, N., {Zasowski}, G., {Newman}, J.~A., {et~al.} 2020, \bibinfo{title}{{Are the Milky Way and Andromeda unusual? A comparison with Milky Way and Andromeda analogues},} \mnras, 498, 4943, \dodoi{10.1093/mnras/staa2731}

\bibitem[{A.~D. {Bolatto} {et~al.}(2013){Bolatto}, {Wolfire}, \& {Leroy}}]{Bolatto2013}
{Bolatto}, A.~D., {Wolfire}, M., \& {Leroy}, A.~K. 2013, \bibinfo{title}{{The CO-to-H$_{2}$ Conversion Factor},} \araa, 51, 207, \dodoi{10.1146/annurev-astro-082812-140944}

\bibitem[{R. {Braun}(1991){Braun}}]{Braun1991}
{Braun}, R. 1991, \bibinfo{title}{{The Distribution and Kinematics of Neutral Gas in M31},} \apj, 372, 54, \dodoi{10.1086/169954}

\bibitem[{R. {Braun} {et~al.}(2009){Braun}, {Thilker}, {Walterbos}, \& {Corbelli}}]{Braun2009}
{Braun}, R., {Thilker}, D.~A., {Walterbos}, R.~A.~M., \& {Corbelli}, E. 2009, \bibinfo{title}{{A Wide-Field High-Resolution H I Mosaic of Messier 31. I. Opaque Atomic Gas and Star Formation Rate Density},} \apj, 695, 937, \dodoi{10.1088/0004-637X/695/2/937}

\bibitem[{E. {Brinks}(1990){Brinks}}]{Brinks1990}
{Brinks}, E. 1990, \bibinfo{title}{{The cool phase of the interstellar medium - Atomic gas},} in Astrophysics and Space Science Library, Vol. 161, The Interstellar Medium in Galaxies, ed. H.~A. {Thronson}, Jr. \& J.~M. {Shull}, 39--65, \dodoi{10.1007/978-94-009-0595-5_3}

\bibitem[{E. {Brinks} \& W.~W. {Shane}(1984){Brinks} \& {Shane}}]{Brinks1984}
{Brinks}, E., \& {Shane}, W.~W. 1984, \bibinfo{title}{{A high resolution hydrogen line survey of Messier 31.I. Observations and data reduction.},} \aaps, 55, 179

\bibitem[{T. {Brown} {et~al.}(2021){Brown}, {Wilson}, {Zabel}, {Davis}, {Boselli}, {Chung}, {Ellison}, {Lagos}, {Stevens}, {Cortese}, {Bah{\'e}}, {Bisaria}, {Bolatto}, {Cashmore}, {Catinella}, {Chown}, {Diemer}, {Elahi}, {Hani}, {Jim{\'e}nez-Donaire}, {Lee}, {Leidig}, {Mok}, {Olsen}, {Parker}, {Roberts}, {Smith}, {Spekkens}, {Thorp}, {Tonnesen}, {Vienneau}, {Villanueva}, {Vogel}, {Wadsley}, {Welker}, \& {Yoon}}]{Brown2021}
{Brown}, T., {Wilson}, C.~D., {Zabel}, N., {et~al.} 2021, \bibinfo{title}{{VERTICO: The Virgo Environment Traced in CO Survey},} \apjs, 257, 21, \dodoi{10.3847/1538-4365/ac28f5}

\bibitem[{A. {Cald{\'u}-Primo} \& A. {Schruba}(2016){Cald{\'u}-Primo} \& {Schruba}}]{Caldu-Primo2016}
{Cald{\'u}-Primo}, A., \& {Schruba}, A. 2016, \bibinfo{title}{{Molecular Gas Velocity Dispersions in the Andromeda Galaxy},} \aj, 151, 34, \dodoi{10.3847/0004-6256/151/2/34}

\bibitem[{D. {Calzetti} {et~al.}(2007){Calzetti}, {Kennicutt}, {Engelbracht}, {Leitherer}, {Draine}, {Kewley}, {Moustakas}, {Sosey}, {Dale}, {Gordon}, {Helou}, {Hollenbach}, {Armus}, {Bendo}, {Bot}, {Buckalew}, {Jarrett}, {Li}, {Meyer}, {Murphy}, {Prescott}, {Regan}, {Rieke}, {Roussel}, {Sheth}, {Smith}, {Thornley}, \& {Walter}}]{Calzetti2007}
{Calzetti}, D., {Kennicutt}, R.~C., {Engelbracht}, C.~W., {et~al.} 2007, \bibinfo{title}{{The Calibration of Mid-Infrared Star Formation Rate Indicators},} \apj, 666, 870, \dodoi{10.1086/520082}

\bibitem[{ {CASA Team} {et~al.}(2022){CASA Team}, {Bean}, {Bhatnagar}, {Castro}, {Donovan Meyer}, {Emonts}, {Garcia}, {Garwood}, {Golap}, {Gonzalez Villalba}, {Harris}, {Hayashi}, {Hoskins}, {Hsieh}, {Jagannathan}, {Kawasaki}, {Keimpema}, {Kettenis}, {Lopez}, {Marvil}, {Masters}, {McNichols}, {Mehringer}, {Miel}, {Moellenbrock}, {Montesino}, {Nakazato}, {Ott}, {Petry}, {Pokorny}, {Raba}, {Rau}, {Schiebel}, {Schweighart}, {Sekhar}, {Shimada}, {Small}, {Steeb}, {Sugimoto}, {Suoranta}, {Tsutsumi}, {van Bemmel}, {Verkouter}, {Wells}, {Xiong}, {Szomoru}, {Griffith}, {Glendenning}, \& {Kern}}]{CASATeam2022}
{CASA Team}, {Bean}, B., {Bhatnagar}, S., {et~al.} 2022, \bibinfo{title}{{CASA, the Common Astronomy Software Applications for Radio Astronomy},} \pasp, 134, 114501, \dodoi{10.1088/1538-3873/ac9642}

\bibitem[{L. {Chemin} {et~al.}(2009){Chemin}, {Carignan}, \& {Foster}}]{Chemin2009}
{Chemin}, L., {Carignan}, C., \& {Foster}, T. 2009, \bibinfo{title}{{H I Kinematics and Dynamics of Messier 31},} \apj, 705, 1395, \dodoi{10.1088/0004-637X/705/2/1395}

\bibitem[{M. {Chevance} {et~al.}(2023){Chevance}, {Krumholz}, {McLeod}, {Ostriker}, {Rosolowsky}, \& {Sternberg}}]{Chevance2023}
{Chevance}, M., {Krumholz}, M.~R., {McLeod}, A.~F., {et~al.} 2023, \bibinfo{title}{{The Life and Times of Giant Molecular Clouds},} in Astronomical Society of the Pacific Conference Series, Vol. 534, Protostars and Planets VII, ed. S.~{Inutsuka}, Y.~{Aikawa}, T.~{Muto}, K.~{Tomida}, \& M.~{Tamura}, 1, \dodoi{10.48550/arXiv.2203.09570}

\bibitem[{R. {Chown} {et~al.}(2025){Chown}, {Leroy}, {Bolatto}, {Chastenet}, {Glover}, {Indebetouw}, {Koch}, {Donovan Meyer}, {Pingel}, {Rosolowsky}, {Sandstrom}, {Sutter}, {Tarantino}, {Bigiel}, {Boquien}, {Chiang}, {Dale}, {Dalcanton}, {Egorov}, {Eibensteiner}, {Grasha}, {Hassani}, {He}, {Kim}, {Meidt}, {Pathak}, {Sarbadhicary}, {Stanimirovic}, {Villanueva}, \& {Williams}}]{Chown2025}
{Chown}, R., {Leroy}, A.~K., {Bolatto}, A.~D., {et~al.} 2025, \bibinfo{title}{{Relationships between Polycyclic Aromatic Hydrocarbons, Small Dust Grains, H$_{2}$, and H I in Local Group Dwarf Galaxies NGC 6822 and WLM Using JWST, ALMA, and the VLA},} \apj, 987, 91, \dodoi{10.3847/1538-4357/add73a}

\bibitem[{E. {Corbelli}(2003){Corbelli}}]{Corbelli2003}
{Corbelli}, E. 2003, \bibinfo{title}{{Dark matter and visible baryons in M33},} \mnras, 342, 199, \dodoi{10.1046/j.1365-8711.2003.06531.x}

\bibitem[{E. {Corbelli} {et~al.}(2010){Corbelli}, {Lorenzoni}, {Walterbos}, {Braun}, \& {Thilker}}]{Corbelli2010}
{Corbelli}, E., {Lorenzoni}, S., {Walterbos}, R., {Braun}, R., \& {Thilker}, D. 2010, \bibinfo{title}{{A wide-field H I mosaic of Messier 31. II. The disk warp, rotation, and the dark matter halo},} \aap, 511, A89, \dodoi{10.1051/0004-6361/200913297}

\bibitem[{E. {Corbelli} {et~al.}(2014){Corbelli}, {Thilker}, {Zibetti}, {Giovanardi}, \& {Salucci}}]{Corbelli2014}
{Corbelli}, E., {Thilker}, D., {Zibetti}, S., {Giovanardi}, C., \& {Salucci}, P. 2014, \bibinfo{title}{{Dynamical signatures of a {\ensuremath{\Lambda}}CDM-halo and the distribution of the baryons in M 33},} \aap, 572, A23, \dodoi{10.1051/0004-6361/201424033}

\bibitem[{T.~A. Davis {et~al.}(2014)Davis, Young, Crocker, {et~al.}}]{Davis2014}
Davis, T.~A., Young, L.~M., Crocker, A.~F., {et~al.} 2014, \bibinfo{title}{The role of old stellar populations in heating dust: separating ‘cirrus’ and star-forming emission in nearby galaxies,} Monthly Notices of the Royal Astronomical Society, 444, 3427, \dodoi{10.1093/mnras/stu1686}

\bibitem[{W.~J.~G. {de Blok} \& F. {Walter}(2000){de Blok} \& {Walter}}]{deBlok2000}
{de Blok}, W.~J.~G., \& {Walter}, F. 2000, \bibinfo{title}{{Evidence for Tidal Interaction and a Supergiant H I Shell in the Local Group Dwarf Galaxy NGC 6822},} \apjl, 537, L95, \dodoi{10.1086/312777}

\bibitem[{W.~J.~G. {de Blok} \& F. {Walter}(2003){de Blok} \& {Walter}}]{deBlok2003}
{de Blok}, W.~J.~G., \& {Walter}, F. 2003, \bibinfo{title}{{Young stars in the outer HI disc of NGC 6822},} \mnras, 341, L39, \dodoi{10.1046/j.1365-8711.2003.06669.x}

\bibitem[{W.~J.~G. {de Blok} \& F. {Walter}(2006){de Blok} \& {Walter}}]{deBlok2006}
{de Blok}, W.~J.~G., \& {Walter}, F. 2006, \bibinfo{title}{{The Stellar Population and Interstellar Medium in NGC 6822},} \aj, 131, 343, \dodoi{10.1086/497829}

\bibitem[{F. {Dell'Agli} {et~al.}(2018){Dell'Agli}, {Di Criscienzo}, {Ventura}, {Limongi}, {Garc{\'\i}a-Hern{\'a}ndez}, {Marini}, \& {Rossi}}]{Dellagli2018}
{Dell'Agli}, F., {Di Criscienzo}, M., {Ventura}, P., {et~al.} 2018, \bibinfo{title}{{Evolved stars in the Local Group galaxies - II. AGB, RSG stars, and dust production in IC10},} \mnras, 479, 5035, \dodoi{10.1093/mnras/sty1614}

\bibitem[{J. {Dempsey} {et~al.}(2022){Dempsey}, {McClure-Griffiths}, {Murray}, {Dickey}, {Pingel}, {Jameson}, {D{\'e}nes}, {van Loon}, {Leahy}, {Lee}, {Stanimirovi{\'c}}, {Breen}, {Buckland-Willis}, {Gibson}, {Imai}, {Lynn}, \& {Tremblay}}]{Dempsey2022}
{Dempsey}, J., {McClure-Griffiths}, N.~M., {Murray}, C., {et~al.} 2022, \bibinfo{title}{{GASKAP-HI Pilot Survey Science III: An unbiased view of cold gas in the Small Magellanic Cloud},} \pasa, 39, e034, \dodoi{10.1017/pasa.2022.18}

\bibitem[{E.~R. {Deul} \& J.~M. {van der Hulst}(1987){Deul} \& {van der Hulst}}]{Deul1987}
{Deul}, E.~R., \& {van der Hulst}, J.~M. 1987, \bibinfo{title}{{A survey of the neutral atomic hydrogen in M 33.},} \aaps, 67, 509

\bibitem[{N.~A. {Devereux} {et~al.}(1994){Devereux}, {Price}, {Wells}, \& {Duric}}]{Devereux1994}
{Devereux}, N.~A., {Price}, R., {Wells}, L.~A., \& {Duric}, N. 1994, \bibinfo{title}{{Two Views of the Andromeda Galaxy H(alpha) and Far Infrared},} \aj, 108, 1667, \dodoi{10.1086/117188}

\bibitem[{J.~M. {Dickey} {et~al.}(2013){Dickey}, {McClure-Griffiths}, {Gibson}, {G{\'o}mez}, {Imai}, {Jones}, {Stanimirovi{\'c}}, {Van Loon}, {Walsh}, {Alberdi}, {Anglada}, {Uscanga}, {Arce}, {Bailey}, {Begum}, {Wakker}, {Bekhti}, {Kalberla}, {Winkel}, {Bekki}, {For}, {Staveley-Smith}, {Westmeier}, {Burton}, {Cunningham}, {Dawson}, {Ellingsen}, {Diamond}, {Green}, {Hill}, {Koribalski}, {McConnell}, {Rathborne}, {Voronkov}, {Douglas}, {English}, {Ford}, {Lockman}, {Foster}, {Gomez}, {Green}, {Bland-Hawthorn}, {Gulyaev}, {Hoare}, {Joncas}, {Kang}, {Kerton}, {Koo}, {Leahy}, {Lo}, {Migenes}, {Nakashima}, {Zhang}, {Nidever}, {Peek}, {Tafoya}, {Tian}, \& {Wu}}]{Dickey2013}
{Dickey}, J.~M., {McClure-Griffiths}, N., {Gibson}, S.~J., {et~al.} 2013, \bibinfo{title}{{GASKAP-The Galactic ASKAP Survey},} \pasa, 30, e003, \dodoi{10.1017/pasa.2012.003}

\bibitem[{C.~L. {Dobbs} {et~al.}(2014){Dobbs}, {Krumholz}, {Ballesteros-Paredes}, {Bolatto}, {Fukui}, {Heyer}, {Low}, {Ostriker}, \& {V{\'a}zquez-Semadeni}}]{Dobbs2014}
{Dobbs}, C.~L., {Krumholz}, M.~R., {Ballesteros-Paredes}, J., {et~al.} 2014, \bibinfo{title}{{Formation of Molecular Clouds and Global Conditions for Star Formation},} in Protostars and Planets VI, ed. H.~{Beuther}, R.~S. {Klessen}, C.~P. {Dullemond}, \& T.~{Henning}, 3--26, \dodoi{10.2458/azu_uapress_9780816531240-ch001}

\bibitem[{C. {Druard} {et~al.}(2014){Druard}, {Braine}, {Schuster}, {Schneider}, {Gratier}, {Bontemps}, {Boquien}, {Combes}, {Corbelli}, {Henkel}, {Herpin}, {Kramer}, {van der Tak}, \& {van der Werf}}]{Druard2014}
{Druard}, C., {Braine}, J., {Schuster}, K.~F., {et~al.} 2014, \bibinfo{title}{{The IRAM M 33 CO(2-1) survey. A complete census of molecular gas out to 7 kpc},} \aap, 567, A118, \dodoi{10.1051/0004-6361/201423682}

\bibitem[{B.~V. {Efremova} {et~al.}(2011){Efremova}, {Bianchi}, {Thilker}, {Neill}, {Burgarella}, {Wyder}, {Madore}, {Rey}, {Barlow}, {Conrow}, {Forster}, {Friedman}, {Martin}, {Morrissey}, {Neff}, {Schiminovich}, {Seibert}, \& {Small}}]{Efremova2011}
{Efremova}, B.~V., {Bianchi}, L., {Thilker}, D.~A., {et~al.} 2011, \bibinfo{title}{{The Recent Star Formation in NGC 6822: An Ultraviolet Study},} \apj, 730, 88, \dodoi{10.1088/0004-637X/730/2/88}

\bibitem[{C. {Eibensteiner} {et~al.}(2026{\natexlab{a}}){Eibensteiner}, {Leroy}, {Sun}, {Koch}, \& LGLBS}]{Eibensteiner2026b}
{Eibensteiner}, C., {Leroy}, A.~K., {Sun}, J., {Koch}, E., \& LGLBS. 2026{\natexlab{a}}, \bibinfo{title}{{Adding mul},} \apj

\bibitem[{C. {Eibensteiner} {et~al.}(2026{\natexlab{b}}){Eibensteiner}, {Leroy}, {Sun}, {Koch}, \& LGLBS}]{Eibensteiner2026a}
{Eibensteiner}, C., {Leroy}, A.~K., {Sun}, J., {Koch}, E., \& LGLBS. 2026{\natexlab{b}}, \bibinfo{title}{{Adding mul},} \apj

\bibitem[{J.~A. {Flores Vel{\'a}zquez} {et~al.}(2021){Flores Vel{\'a}zquez}, {Gurvich}, {Faucher-Gigu{\`e}re}, {Bullock}, {Starkenburg}, {Moreno}, {Lazar}, {Mercado}, {Stern}, {Sparre}, {Hayward}, {Wetzel}, \& {El-Badry}}]{FloresVelazquez2021}
{Flores Vel{\'a}zquez}, J.~A., {Gurvich}, A.~B., {Faucher-Gigu{\`e}re}, C.-A., {et~al.} 2021, \bibinfo{title}{{The time-scales probed by star formation rate indicators for realistic, bursty star formation histories from the FIRE simulations},} \mnras, 501, 4812, \dodoi{10.1093/mnras/staa3893}

\bibitem[{E. {Gardan} {et~al.}(2007){Gardan}, {Braine}, {Schuster}, {Brouillet}, \& {Sievers}}]{Gardan2007}
{Gardan}, E., {Braine}, J., {Schuster}, K.~F., {Brouillet}, N., \& {Sievers}, A. 2007, \bibinfo{title}{{Particularly efficient star formation in M{\,}33},} \aap, 473, 91, \dodoi{10.1051/0004-6361:20077711}

\bibitem[{R. {Genzel} {et~al.}(2012){Genzel}, {Tacconi}, {Combes}, {Bolatto}, {Neri}, {Sternberg}, {Cooper}, {Bouch{\'e}}, {Bournaud}, {Burkert}, {Comerford}, {Cox}, {Davis}, {F{\"o}rster Schreiber}, {Garcia-Burillo}, {Gracia-Carpio}, {Lutz}, {Naab}, {Newman}, {Saintonge}, {Shapiro}, {Shapley}, \& {Weiner}}]{Genzel2012}
{Genzel}, R., {Tacconi}, L.~J., {Combes}, F., {et~al.} 2012, \bibinfo{title}{{The Metallicity Dependence of the CO {\textrightarrow} H$_{2}$ Conversion Factor in z >= 1 Star-forming Galaxies},} \apj, 746, 69, \dodoi{10.1088/0004-637X/746/1/69}

\bibitem[{S.~A.~N. {Gerbrandt} {et~al.}(2015){Gerbrandt}, {McConnachie}, \& {Irwin}}]{Gerbrandt2015}
{Gerbrandt}, S. A.~N., {McConnachie}, A.~W., \& {Irwin}, M. 2015, \bibinfo{title}{{The red extended structure of IC 10, the nearest blue compact galaxy},} \mnras, 454, 1000, \dodoi{10.1093/mnras/stv2029}

\bibitem[{A. {Ginsburg} {et~al.}(2019){Ginsburg}, {Koch}, {Robitaille}, {Beaumont}, {Adamginsburg}, {Sip{\H{o}}cz}, {ZuHone}, {Patra}, {Jones}, {Lim}, {Stern}, {Rosolowsky}, {Earl}, {De Val-Borro}, {Jrobbfed}, {Shuokong}, {Kepley}, {Sokolov}, {Badger}, {Maret}, {Garrido}, {Booker}, \& {Tollerud}}]{SPECTRALCUBE2020}
{Ginsburg}, A., {Koch}, E., {Robitaille}, T., {et~al.} 2019, {radio-astro-tools/spectral-cube: Release v0.4.5}, v0.4.5 Zenodo, \dodoi{10.5281/zenodo.3558614}

\bibitem[{K.~D. {Gordon} {et~al.}(2006){Gordon}, {Bailin}, {Engelbracht}, {Rieke}, {Misselt}, {Latter}, {Young}, {Ashby}, {Barmby}, {Gibson}, {Hines}, {Hinz}, {Krause}, {Levine}, {Marleau}, {Noriega-Crespo}, {Stolovy}, {Thilker}, \& {Werner}}]{Gordon2006}
{Gordon}, K.~D., {Bailin}, J., {Engelbracht}, C.~W., {et~al.} 2006, \bibinfo{title}{{Spitzer MIPS Infrared Imaging of M31: Further Evidence for a Spiral-Ring Composite Structure},} \apjl, 638, L87, \dodoi{10.1086/501046}

\bibitem[{P. {Gratier} {et~al.}(2010){Gratier}, {Braine}, {Rodriguez-Fernandez}, {Schuster}, {Kramer}, {Xilouris}, {Tabatabaei}, {Henkel}, {Corbelli}, {Israel}, {van der Werf}, {Calzetti}, {Garcia-Burillo}, {Sievers}, {Combes}, {Wiklind}, {Brouillet}, {Herpin}, {Bontemps}, {Aalto}, {Koribalski}, {van der Tak}, {Wiedner}, {R{\"o}llig}, \& {Mookerjea}}]{Gratier2010}
{Gratier}, P., {Braine}, J., {Rodriguez-Fernandez}, N.~J., {et~al.} 2010, \bibinfo{title}{{Molecular and atomic gas in the Local Group galaxy M 33},} \aap, 522, A3, \dodoi{10.1051/0004-6361/201014441}

\bibitem[{B.~E. {Greenawalt}(1998){Greenawalt}}]{Greenawalt1998}
{Greenawalt}, B.~E. 1998, \bibinfo{title}{{Diffuse Ionized Gas in Nearby Spiral Galaxies},} PhD thesis, New Mexico State University

\bibitem[{M. {Grossi} {et~al.}(2007){Grossi}, {Disney}, {Pritzl}, {Knezek}, {Gallagher}, {Minchin}, \& {Freeman}}]{Grossi2007}
{Grossi}, M., {Disney}, M.~J., {Pritzl}, B.~J., {et~al.} 2007, \bibinfo{title}{{Star formation history and evolution of gas-rich dwarf galaxies in the Centaurus A group},} \mnras, 374, 107, \dodoi{10.1111/j.1365-2966.2006.11125.x}

\bibitem[{M. {Grossi} {et~al.}(2015){Grossi}, {Hunt}, {Madden}, {Hughes}, {Auld}, {Baes}, {Bendo}, {Bianchi}, {Bizzocchi}, {Boquien}, {Boselli}, {Clemens}, {Corbelli}, {Cortese}, {Davies}, {De Looze}, {di Serego Alighieri}, {Fritz}, {Pappalardo}, {Pierini}, {R{\'e}my-Ruyer}, {Smith}, {Verstappen}, {Viaene}, \& {Vlahakis}}]{Grossi2015}
{Grossi}, M., {Hunt}, L.~K., {Madden}, S.~C., {et~al.} 2015, \bibinfo{title}{{The Herschel Virgo Cluster Survey. XVIII. Star-forming dwarf galaxies in a cluster environment},} \aap, 574, A126, \dodoi{10.1051/0004-6361/201424866}

\bibitem[{B. {Groves} {et~al.}(2012){Groves}, {Krause}, {Sandstrom}, {Schmiedeke}, {Leroy}, {Linz}, {Kapala}, {Rix}, {Schinnerer}, {Tabatabaei}, {Walter}, \& {da Cunha}}]{Groves2012}
{Groves}, B., {Krause}, O., {Sandstrom}, K., {et~al.} 2012, \bibinfo{title}{{The heating of dust by old stellar populations in the bulge of M31},} \mnras, 426, 892, \dodoi{10.1111/j.1365-2966.2012.21696.x}

\bibitem[{C.~R. {Harris} {et~al.}(2020){Harris}, {Millman}, {van der Walt}, {Gommers}, {Virtanen}, {Cournapeau}, {Wieser}, {Taylor}, {Berg}, {Smith}, {Kern}, {Picus}, {Hoyer}, {van Kerkwijk}, {Brett}, {Haldane}, {del R{\'\i}o}, {Wiebe}, {Peterson}, {G{\'e}rard-Marchant}, {Sheppard}, {Reddy}, {Weckesser}, {Abbasi}, {Gohlke}, \& {Oliphant}}]{NumPy2020}
{Harris}, C.~R., {Millman}, K.~J., {van der Walt}, S.~J., {et~al.} 2020, \bibinfo{title}{{Array programming with NumPy},} \nat, 585, 357, \dodoi{10.1038/s41586-020-2649-2}

\bibitem[{S.~A. {Hashemi} {et~al.}(2019){Hashemi}, {Javadi}, \& {van Loon}}]{Hashemi2019}
{Hashemi}, S.~A., {Javadi}, A., \& {van Loon}, J.~T. 2019, \bibinfo{title}{{From evolved stars to the evolution of IC 1613},} \mnras, 483, 4751, \dodoi{10.1093/mnras/sty3450}

\bibitem[{M. {Heyer} \& T.~M. {Dame}(2015){Heyer} \& {Dame}}]{Heyer2015}
{Heyer}, M., \& {Dame}, T.~M. 2015, \bibinfo{title}{{Molecular Clouds in the Milky Way},} \araa, 53, 583, \dodoi{10.1146/annurev-astro-082214-122324}

\bibitem[{C.~G. {Hoopes} \& R.~A.~M. {Walterbos}(2000){Hoopes} \& {Walterbos}}]{Hoopes2000}
{Hoopes}, C.~G., \& {Walterbos}, R. A.~M. 2000, \bibinfo{title}{{The Contribution of Field OB Stars to the Ionization of the Diffuse Ionized Gas in M33},} \apj, 541, 597, \dodoi{10.1086/309487}

\bibitem[{L.~K. {Hunt} {et~al.}(2015){Hunt}, {Garc{\'\i}a-Burillo}, {Casasola}, {Caselli}, {Combes}, {Henkel}, {Lundgren}, {Maiolino}, {Menten}, {Testi}, \& {Weiss}}]{Hunt2015}
{Hunt}, L.~K., {Garc{\'\i}a-Burillo}, S., {Casasola}, V., {et~al.} 2015, \bibinfo{title}{{Molecular depletion times and the CO-to-H$_{2}$ conversion factor in metal-poor galaxies},} \aap, 583, A114, \dodoi{10.1051/0004-6361/201526553}

\bibitem[{D.~A. {Hunter} \& B.~G. {Elmegreen}(2004){Hunter} \& {Elmegreen}}]{Hunter2004}
{Hunter}, D.~A., \& {Elmegreen}, B.~G. 2004, \bibinfo{title}{{Star Formation Properties of a Large Sample of Irregular Galaxies},} \aj, 128, 2170, \dodoi{10.1086/424615}

\bibitem[{D.~A. {Hunter} {et~al.}(2024){Hunter}, {Elmegreen}, \& {Madden}}]{Hunter2024}
{Hunter}, D.~A., {Elmegreen}, B.~G., \& {Madden}, S.~C. 2024, \bibinfo{title}{{The Interstellar Medium in Dwarf Irregular Galaxies},} \araa, 62, 113, \dodoi{10.1146/annurev-astro-052722-104109}

\bibitem[{D.~A. {Hunter} {et~al.}(2012){Hunter}, {Ficut-Vicas}, {Ashley}, {Brinks}, {Cigan}, {Elmegreen}, {Heesen}, {Herrmann}, {Johnson}, {Oh}, {Rupen}, {Schruba}, {Simpson}, {Walter}, {Westpfahl}, {Young}, \& {Zhang}}]{Hunter2012}
{Hunter}, D.~A., {Ficut-Vicas}, D., {Ashley}, T., {et~al.} 2012, \bibinfo{title}{{Little Things},} \aj, 144, 134, \dodoi{10.1088/0004-6256/144/5/134}

\bibitem[{J.~D. {Hunter}(2007){Hunter}}]{Matplotlib2007}
{Hunter}, J.~D. 2007, \bibinfo{title}{{Matplotlib: A 2D Graphics Environment},} Computing in Science and Engineering, 9, 90, \dodoi{10.1109/MCSE.2007.55}

\bibitem[{S.~Z. {Kam} {et~al.}(2017){Kam}, {Carignan}, {Chemin}, {Foster}, {Elson}, \& {Jarrett}}]{Kam2017}
{Kam}, S.~Z., {Carignan}, C., {Chemin}, L., {et~al.} 2017, \bibinfo{title}{{H I Kinematics and Mass Distribution of Messier 33},} \aj, 154, 41, \dodoi{10.3847/1538-3881/aa79f3}

\bibitem[{R.~C. {Kennicutt} \& N.~J. {Evans}(2012){Kennicutt} \& {Evans}}]{Kennicutt2012}
{Kennicutt}, R.~C., \& {Evans}, N.~J. 2012, \bibinfo{title}{{Star Formation in the Milky Way and Nearby Galaxies},} \araa, 50, 531, \dodoi{10.1146/annurev-astro-081811-125610}

\bibitem[{A.~A. {Kepley} {et~al.}(2018){Kepley}, {Bittle}, {Leroy}, {Jim{\'e}nez-Donaire}, {Schruba}, {Bigiel}, {Gallagher}, {Johnson}, \& {Usero}}]{Kepley2018}
{Kepley}, A.~A., {Bittle}, L., {Leroy}, A.~K., {et~al.} 2018, \bibinfo{title}{{Dense Molecular Gas in the Nearby Low-metallicity Dwarf Starburst Galaxy IC 10},} \apj, 862, 120, \dodoi{10.3847/1538-4357/aacaf4}

\bibitem[{M. {Khademi} {et~al.}(2021){Khademi}, {Yang}, {Hammer}, \& {Nasiri}}]{Khademi2021}
{Khademi}, M., {Yang}, Y., {Hammer}, F., \& {Nasiri}, S. 2021, \bibinfo{title}{{Kinematical asymmetry in the dwarf irregular galaxy WLM and a perturbed halo potential},} \aap, 654, A7, \dodoi{10.1051/0004-6361/202140336}

\bibitem[{E. {Koch} {et~al.}(2025){Koch}, {LGLBS}, {LGLBS}, {LGLBS}, \& {LGLBS}}]{Koch2025}
{Koch}, E., {LGLBS}, C., {LGLBS}, C., {LGLBS}, C., \& {LGLBS}, C. 2025, \bibinfo{title}{{The Karl G. Jansky Very Large Array Local Group L-band Survey (LGLBS)},} \apj, 50, 531

\bibitem[{E.~W. {Koch} {et~al.}(2018){Koch}, {Rosolowsky}, {Lockman}, {Kepley}, {Leroy}, {Schruba}, {Braine}, {Dalcanton}, {Johnson}, \& {Stanimirovi{\'c}}}]{Koch2018}
{Koch}, E.~W., {Rosolowsky}, E.~W., {Lockman}, F.~J., {et~al.} 2018, \bibinfo{title}{{Kinematics of the atomic ISM ifigun M33 on 80 pc scales},} \mnras, 479, 2505, \dodoi{10.1093/mnras/sty1674}

\bibitem[{G. {Krahm} {et~al.}(2026){Krahm}, {Leroy}, {Sun}, {Yim}, {Koch}, {Wong}, {Fisher}, {Rosolowsky}, {Sandstrom}, {Utomo}, {van de Sande}, {Martig}, {Fraser-McKelvie}, \& {Hayden}}]{Krahm2026}
{Krahm}, G., {Leroy}, A.~K., {Sun}, J., {et~al.} 2026, \bibinfo{title}{{The Radial and Vertical Structure of Molecular Gas in the Edge-On Galaxy NGC 4565},} arXiv e-prints, arXiv:2604.14136, \dodoi{10.48550/arXiv.2604.14136}

\bibitem[{P. {Kroupa}(2001){Kroupa}}]{Kroupa2001}
{Kroupa}, P. 2001, \bibinfo{title}{{On the variation of the initial mass function},} \mnras, 322, 231, \dodoi{10.1046/j.1365-8711.2001.04022.x}

\bibitem[{G. {Lake} \& E.~D. {Skillman}(1989){Lake} \& {Skillman}}]{Lake1989}
{Lake}, G., \& {Skillman}, E.~D. 1989, \bibinfo{title}{{The Mass Distribution and the Law of Gravity in the Local Group Dwarf Irregular Galaxy IC 1613},} \aj, 98, 1274, \dodoi{10.1086/115215}

\bibitem[{D. {Lang}(2014){Lang}}]{Lang2014}
{Lang}, D. 2014, \bibinfo{title}{{unWISE: Unblurred Coadds of the WISE Imaging},} \aj, 147, 108, \dodoi{10.1088/0004-6256/147/5/108}

\bibitem[{R. {Leaman} {et~al.}(2012){Leaman}, {Venn}, {Brooks}, {Battaglia}, {Cole}, {Ibata}, {Irwin}, {McConnachie}, {Mendel}, \& {Tolstoy}}]{Leaman2012}
{Leaman}, R., {Venn}, K.~A., {Brooks}, A.~M., {et~al.} 2012, \bibinfo{title}{{The Resolved Structure and Dynamics of an Isolated Dwarf Galaxy: A VLT and Keck Spectroscopic Survey of WLM},} \apj, 750, 33, \dodoi{10.1088/0004-637X/750/1/33}

\bibitem[{A.~J. {Lee} {et~al.}(2021){Lee}, {Freedman}, {Madore}, {Owens}, {Monson}, \& {Hoyt}}]{Lee2021dist}
{Lee}, A.~J., {Freedman}, W.~L., {Madore}, B.~F., {et~al.} 2021, \bibinfo{title}{{The Astrophysical Distance Scale. III. Distance to the Local Group Galaxy WLM Using Multiwavelength Observations of the Tip of the Red Giant Branch, Cepheids, and JAGB Stars},} \apj, 907, 112, \dodoi{10.3847/1538-4357/abd253}

\bibitem[{A.~J. {Lee} {et~al.}(2024){Lee}, {Monson}, {Freedman}, {Madore}, {Owens}, {Beaton}, {Espinoza}, {Ren}, \& {Ren}}]{Lee2024dist}
{Lee}, A.~J., {Monson}, A.~J., {Freedman}, W.~L., {et~al.} 2024, \bibinfo{title}{{Resolved Near-infrared Stellar Photometry from the Magellan Telescope for 13 Nearby Galaxies: J-region Asymptotic Giant Branch Method Distances},} \apj, 967, 22, \dodoi{10.3847/1538-4357/ad32c7}

\bibitem[{J.~C. {Lee} {et~al.}(2016){Lee}, {Veilleux}, {McDonald}, \& {Hilbert}}]{Lee2016}
{Lee}, J.~C., {Veilleux}, S., {McDonald}, M., \& {Hilbert}, B. 2016, \bibinfo{title}{{A Deeper Look at Faint H{\ensuremath{\alpha}} Emission in Nearby Dwarf Galaxies},} \apj, 817, 177, \dodoi{10.3847/0004-637X/817/2/177}

\bibitem[{J.~C. {Lee} {et~al.}(2009){Lee}, {Gil de Paz}, {Tremonti}, {Kennicutt}, {Salim}, {Bothwell}, {Calzetti}, {Dalcanton}, {Dale}, {Engelbracht}, {Funes}, {Johnson}, {Sakai}, {Skillman}, {van Zee}, {Walter}, \& {Weisz}}]{Lee2009}
{Lee}, J.~C., {Gil de Paz}, A., {Tremonti}, C., {et~al.} 2009, \bibinfo{title}{{Comparison of H{\ensuremath{\alpha}} and UV Star Formation Rates in the Local Volume: Systematic Discrepancies for Dwarf Galaxies},} \apj, 706, 599, \dodoi{10.1088/0004-637X/706/1/599}

\bibitem[{N. {Lehner} {et~al.}(2015){Lehner}, {Howk}, \& {Wakker}}]{Lehner2015}
{Lehner}, N., {Howk}, J.~C., \& {Wakker}, B.~P. 2015, \bibinfo{title}{{Evidence for a Massive, Extended Circumgalactic Medium Around the Andromeda Galaxy},} \apj, 804, 79, \dodoi{10.1088/0004-637X/804/2/79}

\bibitem[{C. {Leitherer} {et~al.}(1999){Leitherer}, {Schaerer}, {Goldader}, {Delgado}, {Robert}, {Kune}, {de Mello}, {Devost}, \& {Heckman}}]{Leitherer1999}
{Leitherer}, C., {Schaerer}, D., {Goldader}, J.~D., {et~al.} 1999, \bibinfo{title}{{Starburst99: Synthesis Models for Galaxies with Active Star Formation},} \apjs, 123, 3, \dodoi{10.1086/313233}

\bibitem[{A. {Leroy} {et~al.}(2006){Leroy}, {Bolatto}, {Walter}, \& {Blitz}}]{Leroy2006}
{Leroy}, A., {Bolatto}, A., {Walter}, F., \& {Blitz}, L. 2006, \bibinfo{title}{{Molecular Gas in the Low-Metallicity, Star-forming Dwarf IC 10},} \apj, 643, 825, \dodoi{10.1086/503024}

\bibitem[{A.~K. {Leroy} {et~al.}(2008){Leroy}, {Walter}, {Brinks}, {Bigiel}, {de Blok}, {Madore}, \& {Thornley}}]{Leroy2008}
{Leroy}, A.~K., {Walter}, F., {Brinks}, E., {et~al.} 2008, \bibinfo{title}{{The Star Formation Efficiency in Nearby Galaxies: Measuring Where Gas Forms Stars Effectively},} \aj, 136, 2782, \dodoi{10.1088/0004-6256/136/6/2782}

\bibitem[{A.~K. {Leroy} {et~al.}(2012){Leroy}, {Bigiel}, {de Blok}, {Boissier}, {Bolatto}, {Brinks}, {Madore}, {Munoz-Mateos}, {Murphy}, {Sandstrom}, {Schruba}, \& {Walter}}]{Leroy2012}
{Leroy}, A.~K., {Bigiel}, F., {de Blok}, W.~J.~G., {et~al.} 2012, \bibinfo{title}{{Estimating the Star Formation Rate at 1 kpc Scales in nearby Galaxies},} \aj, 144, 3, \dodoi{10.1088/0004-6256/144/1/3}

\bibitem[{A.~K. {Leroy} {et~al.}(2013){Leroy}, {Walter}, {Sandstrom}, {Schruba}, {Munoz-Mateos}, {Bigiel}, {Bolatto}, {Brinks}, {de Blok}, {Meidt}, {Rix}, {Rosolowsky}, {Schinnerer}, {Schuster}, \& {Usero}}]{Leroy2013}
{Leroy}, A.~K., {Walter}, F., {Sandstrom}, K., {et~al.} 2013, \bibinfo{title}{{Molecular Gas and Star Formation in nearby Disk Galaxies},} \aj, 146, 19, \dodoi{10.1088/0004-6256/146/2/19}

\bibitem[{A.~K. {Leroy} {et~al.}(2016){Leroy}, {Hughes}, {Schruba}, {Rosolowsky}, {Blanc}, {Bolatto}, {Colombo}, {Escala}, {Kramer}, {Kruijssen}, {Meidt}, {Pety}, {Querejeta}, {Sandstrom}, {Schinnerer}, {Sliwa}, \& {Usero}}]{Leroy2016}
{Leroy}, A.~K., {Hughes}, A., {Schruba}, A., {et~al.} 2016, \bibinfo{title}{{A Portrait of Cold Gas in Galaxies at 60 pc Resolution and a Simple Method to Test Hypotheses That Link Small-scale ISM Structure to Galaxy-scale Processes},} \apj, 831, 16, \dodoi{10.3847/0004-637X/831/1/16}

\bibitem[{A.~K. {Leroy} {et~al.}(2019){Leroy}, {Sandstrom}, {Lang}, {Lewis}, {Salim}, {Behrens}, {Chastenet}, {Chiang}, {Gallagher}, {Kessler}, \& {Utomo}}]{Leroy2019}
{Leroy}, A.~K., {Sandstrom}, K.~M., {Lang}, D., {et~al.} 2019, \bibinfo{title}{{A z = 0 Multiwavelength Galaxy Synthesis. I. A WISE and GALEX Atlas of Local Galaxies},} \apjs, 244, 24, \dodoi{10.3847/1538-4365/ab3925}

\bibitem[{A.~K. {Leroy} {et~al.}(2021){Leroy}, {Schinnerer}, {Hughes}, {Rosolowsky}, {Pety}, {Schruba}, {Usero}, {Blanc}, {Chevance}, {Emsellem}, {Faesi}, {Herrera}, {Liu}, {Meidt}, {Querejeta}, {Saito}, {Sandstrom}, {Sun}, {Williams}, {Anand}, {Barnes}, {Behrens}, {Belfiore}, {Benincasa}, {Be{\v{s}}li{\'c}}, {Bigiel}, {Bolatto}, {den Brok}, {Cao}, {Chandar}, {Chastenet}, {Chiang}, {Congiu}, {Dale}, {Deger}, {Eibensteiner}, {Egorov}, {Garc{\'\i}a-Rodr{\'\i}guez}, {Glover}, {Grasha}, {Henshaw}, {Ho}, {Kepley}, {Kim}, {Klessen}, {Kreckel}, {Koch}, {Kruijssen}, {Larson}, {Lee}, {Lopez}, {Machado}, {Mayker}, {McElroy}, {Murphy}, {Ostriker}, {Pan}, {Pessa}, {Puschnig}, {Razza}, {S{\'a}nchez-Bl{\'a}zquez}, {Santoro}, {Sardone}, {Scheuermann}, {Sliwa}, {Sormani}, {Stuber}, {Thilker}, {Turner}, {Utomo}, {Watkins}, \& {Whitmore}}]{Leroy2021_survey}
{Leroy}, A.~K., {Schinnerer}, E., {Hughes}, A., {et~al.} 2021, \bibinfo{title}{{PHANGS-ALMA: Arcsecond CO(2-1) Imaging of Nearby Star-forming Galaxies},} \apjs, 257, 43, \dodoi{10.3847/1538-4365/ac17f3}

\bibitem[{A.~K. {Leroy} {et~al.}(2023){Leroy}, {Sandstrom}, {Rosolowsky}, {Belfiore}, {Bolatto}, {Cao}, {Koch}, {Schinnerer}, {Barnes}, {Be{\v{s}}li{\'c}}, {Bigiel}, {Blanc}, {Chastenet}, {Chen}, {Chevance}, {Chown}, {Congiu}, {Dale}, {Egorov}, {Emsellem}, {Eibensteiner}, {Faesi}, {Glover}, {Grasha}, {Groves}, {Hassani}, {Henshaw}, {Hughes}, {Jim{\'e}nez-Donaire}, {Kim}, {Klessen}, {Kreckel}, {Kruijssen}, {Larson}, {Lee}, {Levy}, {Liu}, {Lopez}, {Meidt}, {Murphy}, {Neumann}, {Pessa}, {Pety}, {Saito}, {Sardone}, {Sun}, {Thilker}, {Usero}, {Watkins}, {Whitcomb}, \& {Williams}}]{Leroy2023}
{Leroy}, A.~K., {Sandstrom}, K., {Rosolowsky}, E., {et~al.} 2023, \bibinfo{title}{{PHANGS-JWST First Results: Mid-infrared Emission Traces Both Gas Column Density and Heating at 100 pc Scales},} \apjl, 944, L9, \dodoi{10.3847/2041-8213/acaf85}

\bibitem[{A.~R. {Lewis} {et~al.}(2015){Lewis}, {Dolphin}, {Dalcanton}, {Weisz}, {Williams}, {Bell}, {Seth}, {Simones}, {Skillman}, {Choi}, {Fouesneau}, {Guhathakurta}, {Johnson}, {Kalirai}, {Leroy}, {Monachesi}, {Rix}, \& {Schruba}}]{Lewis2015}
{Lewis}, A.~R., {Dolphin}, A.~E., {Dalcanton}, J.~J., {et~al.} 2015, \bibinfo{title}{{The Panchromatic Hubble Andromeda Treasury. XI. The Spatially Resolved Recent Star Formation History of M31},} \apj, 805, 183, \dodoi{10.1088/0004-637X/805/2/183}

\bibitem[{A.~R. {Lewis} {et~al.}(2017){Lewis}, {Simones}, {Johnson}, {Dalcanton}, {Skillman}, {Weisz}, {Dolphin}, {Williams}, {Bell}, {Fouesneau}, {Kapala}, {Rosenfield}, \& {Schruba}}]{Lewis2017}
{Lewis}, A.~R., {Simones}, J.~E., {Johnson}, B.~D., {et~al.} 2017, \bibinfo{title}{{The Panchromatic Hubble Andromeda Treasury. XVII. Examining Obscured Star Formation with Synthetic Ultraviolet Flux Maps in M31.},} \apj, 834, 70, \dodoi{10.3847/1538-4357/834/1/70}

\bibitem[{L. {Lin} {et~al.}(2020){Lin}, {Ellison}, {Pan}, {Thorp}, {Su}, {S{\'a}nchez}, {Belfiore}, {Bothwell}, {Bundy}, {Chen}, {Concas}, {Hsieh}, {Hsieh}, {Li}, {Maiolino}, {Masters}, {Newman}, {Rowlands}, {Shi}, {Smethurst}, {Stark}, {Xiao}, \& {Yu}}]{Lin2020}
{Lin}, L., {Ellison}, S.~L., {Pan}, H.-A., {et~al.} 2020, \bibinfo{title}{{ALMaQUEST. IV. The ALMA-MaNGA QUEnching and STar Formation (ALMaQUEST) Survey},} \apj, 903, 145, \dodoi{10.3847/1538-4357/abba3a}

\bibitem[{D. {Makarov} {et~al.}(2014){Makarov}, {Prugniel}, {Terekhova}, {Courtois}, \& {Vauglin}}]{Makarov2014}
{Makarov}, D., {Prugniel}, P., {Terekhova}, N., {Courtois}, H., \& {Vauglin}, I. 2014, \bibinfo{title}{{HyperLEDA. III. The catalogue of extragalactic distances},} \aap, 570, A13, \dodoi{10.1051/0004-6361/201423496}

\bibitem[{D.~C. {Martin} {et~al.}(2005){Martin}, {Fanson}, {Schiminovich}, {Morrissey}, {Friedman}, {Barlow}, {Conrow}, {Grange}, {Jelinsky}, {Milliard}, {Siegmund}, {Bianchi}, {Byun}, {Donas}, {Forster}, {Heckman}, {Lee}, {Madore}, {Malina}, {Neff}, {Rich}, {Small}, {Surber}, {Szalay}, {Welsh}, \& {Wyder}}]{Martin2005}
{Martin}, D.~C., {Fanson}, J., {Schiminovich}, D., {et~al.} 2005, \bibinfo{title}{{The Galaxy Evolution Explorer: A Space Ultraviolet Survey Mission},} \apjl, 619, L1, \dodoi{10.1086/426387}

\bibitem[{P. {Massey} {et~al.}(2006){Massey}, {Olsen}, {Hodge}, {Strong}, {Jacoby}, {Schlingman}, \& {Smith}}]{Massey2006}
{Massey}, P., {Olsen}, K.~A.~G., {Hodge}, P.~W., {et~al.} 2006, \bibinfo{title}{{A Survey of Local Group Galaxies Currently Forming Stars. I. UBVRI Photometry of Stars in M31 and M33},} \aj, 131, 2478, \dodoi{10.1086/503256}

\bibitem[{J. {Meaburn} {et~al.}(1988){Meaburn}, {Clayton}, \& {Whitehead}}]{Meaburn1988}
{Meaburn}, J., {Clayton}, C.~A., \& {Whitehead}, M.~J. 1988, \bibinfo{title}{{The dynamics of a complex of interlocking giant shells in the Local Group dwarf galaxy IC 1613.},} \mnras, 235, 479, \dodoi{10.1093/mnras/235.2.479}

\bibitem[{E.~J. {Murphy} {et~al.}(2011){Murphy}, {Condon}, {Schinnerer}, {Kennicutt}, {Calzetti}, {Armus}, {Helou}, {Turner}, {Aniano}, {Beir{\~a}o}, {Bolatto}, {Brandl}, {Croxall}, {Dale}, {Donovan Meyer}, {Draine}, {Engelbracht}, {Hunt}, {Hao}, {Koda}, {Roussel}, {Skibba}, \& {Smith}}]{Murphy2011}
{Murphy}, E.~J., {Condon}, J.~J., {Schinnerer}, E., {et~al.} 2011, \bibinfo{title}{{Calibrating Extinction-free Star Formation Rate Diagnostics with 33 GHz Free-free Emission in NGC 6946},} \apj, 737, 67, \dodoi{10.1088/0004-637X/737/2/67}

\bibitem[{S.~J. {Mutch} {et~al.}(2011){Mutch}, {Croton}, \& {Poole}}]{Mutch2011}
{Mutch}, S.~J., {Croton}, D.~J., \& {Poole}, G.~B. 2011, \bibinfo{title}{{The Mid-life Crisis of the Milky Way and M31},} \apj, 736, 84, \dodoi{10.1088/0004-637X/736/2/84}

\bibitem[{B. {Namumba} {et~al.}(2017){Namumba}, {Carignan}, {Passmoor}, \& {de Blok}}]{Namumba2017}
{Namumba}, B., {Carignan}, C., {Passmoor}, S., \& {de Blok}, W.~J.~G. 2017, \bibinfo{title}{{H I kinematics, mass distribution and star formation threshold in NGC 6822, using the SKA pathfinder KAT-7},} \mnras, 472, 3761, \dodoi{10.1093/mnras/stx2256}

\bibitem[{C. {Nieten} {et~al.}(2006){Nieten}, {Neininger}, {Gu{\'e}lin}, {Ungerechts}, {Lucas}, {Berkhuijsen}, {Beck}, \& {Wielebinski}}]{Nieten2006}
{Nieten}, C., {Neininger}, N., {Gu{\'e}lin}, M., {et~al.} 2006, \bibinfo{title}{{Molecular gas in the Andromeda galaxy},} \aap, 453, 459, \dodoi{10.1051/0004-6361:20035672}

\bibitem[{S.-H. {Oh} {et~al.}(2015){Oh}, {Hunter}, {Brinks}, {Elmegreen}, {Schruba}, {Walter}, {Rupen}, {Young}, {Simpson}, {Johnson}, {Herrmann}, {Ficut-Vicas}, {Cigan}, {Heesen}, {Ashley}, \& {Zhang}}]{Oh2015}
{Oh}, S.-H., {Hunter}, D.~A., {Brinks}, E., {et~al.} 2015, \bibinfo{title}{{High-resolution Mass Models of Dwarf Galaxies from LITTLE THINGS},} \aj, 149, 180, \dodoi{10.1088/0004-6256/149/6/180}

\bibitem[{J. {Ott} {et~al.}(2012){Ott}, {Stilp}, {Warren}, {Skillman}, {Dalcanton}, {Walter}, {de Blok}, {Koribalski}, \& {West}}]{Ott2012}
{Ott}, J., {Stilp}, A.~M., {Warren}, S.~R., {et~al.} 2012, \bibinfo{title}{{VLA-ANGST: A High-resolution H I Survey of Nearby Dwarf Galaxies},} \aj, 144, 123, \dodoi{10.1088/0004-6256/144/4/123}

\bibitem[{M. {Padave} {et~al.}(2025){Padave}, {Borthakur}, {Jansen}, {Thilker}, {Monkiewicz}, \& {Windhorst}}]{Padave2025}
{Padave}, M., {Borthakur}, S., {Jansen}, R.~A., {et~al.} 2025, \bibinfo{title}{{DIISC-V: Variations in H{\ensuremath{\alpha}}-to-FUV Star Formation Rate Ratios Across Star-forming Regions in Nearby Galaxies},} \apj, 986, 145, \dodoi{10.3847/1538-4357/adce7b}

\bibitem[{J.~E.~G. {Peek} \& D. {Schiminovich}(2013){Peek} \& {Schiminovich}}]{Peek2013}
{Peek}, J.~E.~G., \& {Schiminovich}, D. 2013, \bibinfo{title}{{Ultraviolet Extinction at High Galactic Latitudes},} \apj, 771, 68, \dodoi{10.1088/0004-637X/771/1/68}

\bibitem[{N.~M. {Pingel} {et~al.}(2024){Pingel}, {Chen}, {Stanimirovi{\'c}}, {Koch}, {Leroy}, {Rosolowsky}, {Kim}, {Dalcanton}, {Walter}, {Busch}, {Chown}, {Donovan Meyer}, {Eibensteiner}, {Hunter}, {Sarbadhicary}, {Tarantino}, {Villanueva}, \& {Williams}}]{Pingel2024}
{Pingel}, N.~M., {Chen}, H., {Stanimirovi{\'c}}, S., {et~al.} 2024, \bibinfo{title}{{The Local Group L-band Survey: The First Measurements of Localized Cold Neutral Medium Properties in the Low-metallicity Dwarf Galaxy NGC 6822},} \apj, 974, 93, \dodoi{10.3847/1538-4357/ad6604}

\bibitem[{N.~R. {Pokhrel} {et~al.}(2020){Pokhrel}, {Simpson}, \& {Bagetakos}}]{Pokhrel2020}
{Pokhrel}, N.~R., {Simpson}, C.~E., \& {Bagetakos}, I. 2020, \bibinfo{title}{{A Catalog of Holes and Shells in the Interstellar Medium of the LITTLE THINGS Dwarf Galaxies},} \aj, 160, 66, \dodoi{10.3847/1538-3881/ab9bfa}

\bibitem[{M.~E. {Putman} {et~al.}(2009){Putman}, {Peek}, {Muratov}, {Gnedin}, {Hsu}, {Douglas}, {Heiles}, {Stanimirovic}, {Korpela}, \& {Gibson}}]{Putman2009}
{Putman}, M.~E., {Peek}, J.~E.~G., {Muratov}, A., {et~al.} 2009, \bibinfo{title}{{The Disruption and Fueling of M33},} \apj, 703, 1486, \dodoi{10.1088/0004-637X/703/2/1486}

\bibitem[{P. {Rosenfield} {et~al.}(2012){Rosenfield}, {Johnson}, {Girardi}, {Dalcanton}, {Bressan}, {Lang}, {Williams}, {Guhathakurta}, {Howley}, {Lauer}, {Bell}, {Bianchi}, {Caldwell}, {Dolphin}, {Dorman}, {Gilbert}, {Kalirai}, {Larsen}, {Olsen}, {Rix}, {Seth}, {Skillman}, \& {Weisz}}]{Rosenfield2012}
{Rosenfield}, P., {Johnson}, L.~C., {Girardi}, L., {et~al.} 2012, \bibinfo{title}{{The Panchromatic Hubble Andromeda Treasury. I. Bright UV Stars in the Bulge of M31},} \apj, 755, 131, \dodoi{10.1088/0004-637X/755/2/131}

\bibitem[{E. {Rosolowsky} {et~al.}(2007){Rosolowsky}, {Keto}, {Matsushita}, \& {Willner}}]{Rosolowsky2007}
{Rosolowsky}, E., {Keto}, E., {Matsushita}, S., \& {Willner}, S.~P. 2007, \bibinfo{title}{{High-Resolution Molecular Gas Maps of M33},} \apj, 661, 830, \dodoi{10.1086/516621}

\bibitem[{M. {Rubio} {et~al.}(2015){Rubio}, {Elmegreen}, {Hunter}, {Brinks}, {Cort{\'e}s}, \& {Cigan}}]{Rubio2015}
{Rubio}, M., {Elmegreen}, B.~G., {Hunter}, D.~A., {et~al.} 2015, \bibinfo{title}{{Dense cloud cores revealed by CO in the low metallicity dwarf galaxy WLM},} \nat, 525, 218, \dodoi{10.1038/nature14901}

\bibitem[{A. {Saintonge} \& B. {Catinella}(2022){Saintonge} \& {Catinella}}]{Saintonge2022}
{Saintonge}, A., \& {Catinella}, B. 2022, \bibinfo{title}{{The Cold Interstellar Medium of Galaxies in the Local Universe},} \araa, 60, 319, \dodoi{10.1146/annurev-astro-021022-043545}

\bibitem[{S. {Salim} {et~al.}(2018){Salim}, {Boquien}, \& {Lee}}]{Salim2018}
{Salim}, S., {Boquien}, M., \& {Lee}, J.~C. 2018, \bibinfo{title}{{Dust Attenuation Curves in the Local Universe: Demographics and New Laws for Star-forming Galaxies and High-redshift Analogs},} \apj, 859, 11, \dodoi{10.3847/1538-4357/aabf3c}

\bibitem[{S. {Salim} {et~al.}(2007){Salim}, {Rich}, {Charlot}, {Brinchmann}, {Johnson}, {Schiminovich}, {Seibert}, {Mallery}, {Heckman}, {Forster}, {Friedman}, {Martin}, {Morrissey}, {Neff}, {Small}, {Wyder}, {Bianchi}, {Donas}, {Lee}, {Madore}, {Milliard}, {Szalay}, {Welsh}, \& {Yi}}]{Salim2007}
{Salim}, S., {Rich}, R.~M., {Charlot}, S., {et~al.} 2007, \bibinfo{title}{{UV Star Formation Rates in the Local Universe},} \apjs, 173, 267, \dodoi{10.1086/519218}

\bibitem[{S. {Salim} {et~al.}(2016){Salim}, {Lee}, {Janowiecki}, {da Cunha}, {Dickinson}, {Boquien}, {Burgarella}, {Salzer}, \& {Charlot}}]{Salim2016}
{Salim}, S., {Lee}, J.~C., {Janowiecki}, S., {et~al.} 2016, \bibinfo{title}{{GALEX-SDSS-WISE Legacy Catalog (GSWLC): Star Formation Rates, Stellar Masses, and Dust Attenuations of 700,000 Low-redshift Galaxies},} \apjs, 227, 2, \dodoi{10.3847/0067-0049/227/1/2}

\bibitem[{N. {Sanna} {et~al.}(2008){Sanna}, {Bono}, {Stetson}, {Monelli}, {Pietrinferni}, {Drozdovsky}, {Caputo}, {Cassisi}, {Gennaro}, {Prada Moroni}, {Buonanno}, {Corsi}, {Degl'Innocenti}, {Ferraro}, {Iannicola}, {Nonino}, {Pulone}, {Romaniello}, \& {Walker}}]{Sanna2008}
{Sanna}, N., {Bono}, G., {Stetson}, P.~B., {et~al.} 2008, \bibinfo{title}{{On the Distance and Reddening of the Starburst Galaxy IC 10},} \apjl, 688, L69, \dodoi{10.1086/595551}

\bibitem[{A. {Savino} {et~al.}(2022){Savino}, {Weisz}, {Skillman}, {Dolphin}, {Kallivayalil}, {Wetzel}, {Anderson}, {Besla}, {Boylan-Kolchin}, {Bullock}, {Cole}, {Collins}, {Cooper}, {Deason}, {Dotter}, {Fardal}, {Ferguson}, {Fritz}, {Geha}, {Gilbert}, {Guhathakurta}, {Ibata}, {Irwin}, {Jeon}, {Kirby}, {Lewis}, {Mackey}, {Majewski}, {Martin}, {McConnachie}, {Patel}, {Rich}, {Simon}, {Sohn}, {Tollerud}, \& {van der Marel}}]{Savino2022}
{Savino}, A., {Weisz}, D.~R., {Skillman}, E.~D., {et~al.} 2022, \bibinfo{title}{{The Hubble Space Telescope Survey of M31 Satellite Galaxies. I. RR Lyrae-based Distances and Refined 3D Geometric Structure},} \apj, 938, 101, \dodoi{10.3847/1538-4357/ac91cb}

\bibitem[{E. {Schinnerer} \& A.~K. {Leroy}(2024){Schinnerer} \& {Leroy}}]{Schinnerer2024}
{Schinnerer}, E., \& {Leroy}, A.~K. 2024, \bibinfo{title}{{Molecular Gas and the Star-Formation Process on Cloud Scales in Nearby Galaxies},} \araa, 62, 369, \dodoi{10.1146/annurev-astro-071221-052651}

\bibitem[{D.~J. {Schlegel} {et~al.}(1998){Schlegel}, {Finkbeiner}, \& {Davis}}]{Schlegel1998}
{Schlegel}, D.~J., {Finkbeiner}, D.~P., \& {Davis}, M. 1998, \bibinfo{title}{{Maps of Dust Infrared Emission for Use in Estimation of Reddening and Cosmic Microwave Background Radiation Foregrounds},} \apj, 500, 525, \dodoi{10.1086/305772}

\bibitem[{A. {Schruba} {et~al.}(2011){Schruba}, {Leroy}, {Walter}, {Bigiel}, {Brinks}, {de Blok}, {Dumas}, {Kramer}, {Rosolowsky}, {Sandstrom}, {Schuster}, {Usero}, {Weiss}, \& {Wiesemeyer}}]{Schruba2011}
{Schruba}, A., {Leroy}, A.~K., {Walter}, F., {et~al.} 2011, \bibinfo{title}{{A Molecular Star Formation Law in the Atomic-gas-dominated Regime in Nearby Galaxies},} \aj, 142, 37, \dodoi{10.1088/0004-6256/142/2/37}

\bibitem[{A. {Schruba} {et~al.}(2012){Schruba}, {Leroy}, {Walter}, {Bigiel}, {Brinks}, {de Blok}, {Kramer}, {Rosolowsky}, {Sandstrom}, {Schuster}, {Usero}, {Weiss}, \& {Wiesemeyer}}]{Schruba2012}
{Schruba}, A., {Leroy}, A.~K., {Walter}, F., {et~al.} 2012, \bibinfo{title}{{Low CO Luminosities in Dwarf Galaxies},} \aj, 143, 138, \dodoi{10.1088/0004-6256/143/6/138}

\bibitem[{A. {Schruba} {et~al.}(2017){Schruba}, {Leroy}, {Kruijssen}, {Bigiel}, {Bolatto}, {de Blok}, {Tacconi}, {van Dishoeck}, \& {Walter}}]{Schruba2017}
{Schruba}, A., {Leroy}, A.~K., {Kruijssen}, J.~M.~D., {et~al.} 2017, \bibinfo{title}{{Physical Properties of Molecular Clouds at 2 pc Resolution in the Low-metallicity Dwarf Galaxy NGC 6822 and the Milky Way},} \apj, 835, 278, \dodoi{10.3847/1538-4357/835/2/278}

\bibitem[{G.~S. {Shostak} \& E.~D. {Skillman}(1989){Shostak} \& {Skillman}}]{Shostak1989}
{Shostak}, G.~S., \& {Skillman}, E.~D. 1989, \bibinfo{title}{{Neutral hydrogen observations of the irregular galaxy IC 10.},} \aap, 214, 33

\bibitem[{S. {Silich} {et~al.}(2006){Silich}, {Lozinskaya}, {Moiseev}, {Podorvanuk}, {Rosado}, {Borissova}, \& {Valdez-Gutierrez}}]{Silich2006}
{Silich}, S., {Lozinskaya}, T., {Moiseev}, A., {et~al.} 2006, \bibinfo{title}{{On the neutral gas distribution and kinematics in the dwarf irregular galaxy IC 1613},} \aap, 448, 123, \dodoi{10.1051/0004-6361:20053326}

\bibitem[{G.~V. {Simonian} \& P. {Martini}(2017){Simonian} \& {Martini}}]{Simonian2017}
{Simonian}, G.~V., \& {Martini}, P. 2017, \bibinfo{title}{{Circumstellar dust, PAHs and stellar populations in early-type galaxies: insights from GALEX and WISE},} \mnras, 464, 3920, \dodoi{10.1093/mnras/stw2623}

\bibitem[{A. {Smercina} {et~al.}(2023){Smercina}, {Dalcanton}, {Williams}, {Durbin}, {Lazzarini}, {Bell}, {Choi}, {Dolphin}, {Gilbert}, {Guhathakurta}, {Koch}, {Quirk}, {Rix}, {Rosolowsky}, {Seth}, {Skillman}, \& {Weisz}}]{smercina2023}
{Smercina}, A., {Dalcanton}, J.~J., {Williams}, B.~F., {et~al.} 2023, \bibinfo{title}{{The Panchromatic Hubble Andromeda Treasury: Triangulum Extended Region (PHATTER). V. The Structure of M33 in Resolved Stellar Populations},} \apj, 957, 3, \dodoi{10.3847/1538-4357/acf3e8}

\bibitem[{A. {Sternberg} {et~al.}(2023){Sternberg}, {Bialy}, \& {Gurman}}]{Sternberg2023}
{Sternberg}, A., {Bialy}, S., \& {Gurman}, A. 2023, \bibinfo{title}{{HI in Molecular Clouds: Irradiation by FUV plus Cosmic Rays},} arXiv e-prints, arXiv:2308.13889, \dodoi{10.48550/arXiv.2308.13889}

\bibitem[{A. {Sternberg} {et~al.}(2014){Sternberg}, {Le Petit}, {Roueff}, \& {Le Bourlot}}]{Sternberg2014}
{Sternberg}, A., {Le Petit}, F., {Roueff}, E., \& {Le Bourlot}, J. 2014, \bibinfo{title}{{H I-to-H$_{2}$ Transitions and H I Column Densities in Galaxy Star-forming Regions},} \apj, 790, 10, \dodoi{10.1088/0004-637X/790/1/10}

\bibitem[{J. {Sun} {et~al.}(2022){Sun}, {Leroy}, {Rosolowsky}, {Hughes}, {Schinnerer}, {Schruba}, {Koch}, {Blanc}, {Chiang}, {Groves}, {Liu}, {Meidt}, {Pan}, {Pety}, {Querejeta}, {Saito}, {Sandstrom}, {Sardone}, {Usero}, {Utomo}, {Williams}, {Barnes}, {Benincasa}, {Bigiel}, {Bolatto}, {Boquien}, {Chevance}, {Dale}, {Deger}, {Emsellem}, {Glover}, {Grasha}, {Henshaw}, {Klessen}, {Kreckel}, {Kruijssen}, {Ostriker}, \& {Thilker}}]{Sun2022}
{Sun}, J., {Leroy}, A.~K., {Rosolowsky}, E., {et~al.} 2022, \bibinfo{title}{{Molecular Cloud Populations in the Context of Their Host Galaxy Environments: A Multiwavelength Perspective},} \aj, 164, 43, \dodoi{10.3847/1538-3881/ac74bd}

\bibitem[{J. {Sun} {et~al.}(2023){Sun}, {Leroy}, {Ostriker}, {Meidt}, {Rosolowsky}, {Schinnerer}, {Wilson}, {Utomo}, {Belfiore}, {Blanc}, {Emsellem}, {Faesi}, {Groves}, {Hughes}, {Koch}, {Kreckel}, {Liu}, {Pan}, {Pety}, {Querejeta}, {Razza}, {Saito}, {Sardone}, {Usero}, {Williams}, {Bigiel}, {Bolatto}, {Chevance}, {Dale}, {Gensior}, {Glover}, {Grasha}, {Henshaw}, {Jim{\'e}nez-Donaire}, {Klessen}, {Kruijssen}, {Murphy}, {Neumann}, {Teng}, \& {Thilker}}]{Sun2023}
{Sun}, J., {Leroy}, A.~K., {Ostriker}, E.~C., {et~al.} 2023, \bibinfo{title}{{Star Formation Laws and Efficiencies across 80 Nearby Galaxies},} \apjl, 945, L19, \dodoi{10.3847/2041-8213/acbd9c}

\bibitem[{J. {Sun} {et~al.}(2025){Sun}, {Teng}, {Chiang}, {Leroy}, {Sandstrom}, {den Brok}, {Bolatto}, {Chastenet}, {Chown}, {Hughes}, {Koch}, \& {Williams}}]{Sun2025}
{Sun}, J., {Teng}, Y.-H., {Chiang}, I.-D., {et~al.} 2025, \bibinfo{title}{{Resolved Profiles of Stellar Mass, Star Formation Rate, and Predicted CO-to-H$_{2}$ Conversion Factor Across Thousands of Local Galaxies},} \apj, 994, 263, \dodoi{10.3847/1538-4357/ae10be}

\bibitem[{F.~S. Tabatabaei {et~al.}(2010)Tabatabaei, Schinnerer, Krause, {et~al.}}]{Tabatabaei2010}
Tabatabaei, F.~S., Schinnerer, E., Krause, M., {et~al.} 2010, \bibinfo{title}{Relating dust, gas, and the rate of star formation in M31,} A\&A, 517, A77

\bibitem[{L.~J. {Tacconi} {et~al.}(2020){Tacconi}, {Genzel}, \& {Sternberg}}]{Tacconi2020}
{Tacconi}, L.~J., {Genzel}, R., \& {Sternberg}, A. 2020, \bibinfo{title}{{The Evolution of the Star-Forming Interstellar Medium Across Cosmic Time},} \araa, 58, 157, \dodoi{10.1146/annurev-astro-082812-141034}

\bibitem[{O.~G. {Telford} {et~al.}(2020){Telford}, {Dalcanton}, {Williams}, {Bell}, {Dolphin}, {Durbin}, \& {Choi}}]{Telford2020}
{Telford}, O.~G., {Dalcanton}, J.~J., {Williams}, B.~F., {et~al.} 2020, \bibinfo{title}{{Mass-to-light Ratios of Spatially Resolved Stellar Populations in M31},} \apj, 891, 32, \dodoi{10.3847/1538-4357/ab701c}

\bibitem[{P. Tenjes {et~al.}(2017)Tenjes, Tuvikene, Tihhonova, {et~al.}}]{Tenjes2017}
Tenjes, P., Tuvikene, T., Tihhonova, O., {et~al.} 2017, \bibinfo{title}{Spiral arms and disc stability in the Andromeda galaxy,} A\&A, 600, A34

\bibitem[{ {The pandas development team}(2021){The pandas development team}}]{Pandas_1.3.4}
{The pandas development team}. 2021, Pandas-Dev/Pandas: {{Pandas}}, v1.3.4 Zenodo, \dodoi{10.5281/zenodo.5574486}

\bibitem[{P. {Virtanen} {et~al.}(2020){Virtanen}, {Gommers}, {Oliphant}, {Haberland}, {Reddy}, {Cournapeau}, {Burovski}, {Peterson}, {Weckesser}, {Bright}, {van der Walt}, {Brett}, {Wilson}, {Millman}, {Mayorov}, {Nelson}, {Jones}, {Kern}, {Larson}, {Carey}, {Polat}, {Feng}, {Moore}, {VanderPlas}, {Laxalde}, {Perktold}, {Cimrman}, {Henriksen}, {Quintero}, {Harris}, {Archibald}, {Ribeiro}, {Pedregosa}, {van Mulbregt}, \& {SciPy 1. 0 Contributors}}]{SciPy2020}
{Virtanen}, P., {Gommers}, R., {Oliphant}, T.~E., {et~al.} 2020, \bibinfo{title}{{SciPy 1.0: fundamental algorithms for scientific computing in Python},} Nature Methods, 17, 261, \dodoi{10.1038/s41592-019-0686-2}

\bibitem[{F. {Walter} \& E. {Brinks}(1999){Walter} \& {Brinks}}]{walter1999}
{Walter}, F., \& {Brinks}, E. 1999, \bibinfo{title}{{Holes and Shells in the Interstellar Medium of the Nearby Dwarf Galaxy IC 2574},} \aj, 118, 273, \dodoi{10.1086/300906}

\bibitem[{F. {Walter} {et~al.}(2008){Walter}, {Brinks}, {de Blok}, {Bigiel}, {Kennicutt}, {Thornley}, \& {Leroy}}]{Walter2008}
{Walter}, F., {Brinks}, E., {de Blok}, W.~J.~G., {et~al.} 2008, \bibinfo{title}{{THINGS: The H I Nearby Galaxy Survey},} \aj, 136, 2563, \dodoi{10.1088/0004-6256/136/6/2563}

\bibitem[{F. {Walter} {et~al.}(2020){Walter}, {Carilli}, {Neeleman}, {Decarli}, {Popping}, {Somerville}, {Aravena}, {Bertoldi}, {Boogaard}, {Cox}, {da Cunha}, {Magnelli}, {Obreschkow}, {Riechers}, {Rix}, {Smail}, {Weiss}, {Assef}, {Bauer}, {Bouwens}, {Contini}, {Cortes}, {Daddi}, {Diaz-Santos}, {Gonz{\'a}lez-L{\'o}pez}, {Hennawi}, {Hodge}, {Inami}, {Ivison}, {Oesch}, {Sargent}, {van der Werf}, {Wagg}, \& {Yung}}]{Walter2020}
{Walter}, F., {Carilli}, C., {Neeleman}, M., {et~al.} 2020, \bibinfo{title}{{The Evolution of the Baryons Associated with Galaxies Averaged over Cosmic Time and Space},} \apj, 902, 111, \dodoi{10.3847/1538-4357/abb82e}

\bibitem[{D.~R. {Weisz} {et~al.}(2009){Weisz}, {Skillman}, {Cannon}, {Dolphin}, {Kennicutt}, {Lee}, \& {Walter}}]{Weisz2009}
{Weisz}, D.~R., {Skillman}, E.~D., {Cannon}, J.~M., {et~al.} 2009, \bibinfo{title}{{Does Stellar Feedback Create H I Holes? A Hubble Space Telescope/Very Large Array Study of Holmberg II},} \apj, 704, 1538, \dodoi{10.1088/0004-637X/704/2/1538}

\bibitem[{D.~T.~F. {Weldrake} {et~al.}(2003){Weldrake}, {de Blok}, \& {Walter}}]{Weldrake2003}
{Weldrake}, D.~T.~F., {de Blok}, W.~J.~G., \& {Walter}, F. 2003, \bibinfo{title}{{A high-resolution rotation curve of NGC 6822: a test-case for cold dark matter},} \mnras, 340, 12, \dodoi{10.1046/j.1365-8711.2003.06170.x}

\bibitem[{E.~M. {Wilcots} \& B.~W. {Miller}(1998){Wilcots} \& {Miller}}]{Wilcots1998}
{Wilcots}, E.~M., \& {Miller}, B.~W. 1998, \bibinfo{title}{{The Kinematics and Distribution of H I in IC 10},} \aj, 116, 2363, \dodoi{10.1086/300595}

\bibitem[{C.~N.~A. {Willmer}(2018){Willmer}}]{Willmer2018}
{Willmer}, C. N.~A. 2018, \bibinfo{title}{{The Absolute Magnitude of the Sun in Several Filters},} \apjs, 236, 47, \dodoi{10.3847/1538-4365/aabfdf}

\bibitem[{T. {Wong} \& L. {Blitz}(2002){Wong} \& {Blitz}}]{Wong2002}
{Wong}, T., \& {Blitz}, L. 2002, \bibinfo{title}{{The Relationship between Gas Content and Star Formation in Molecule-rich Spiral Galaxies},} \apj, 569, 157, \dodoi{10.1086/339287}

\bibitem[{E.~L. {Wright} {et~al.}(2010){Wright}, {Eisenhardt}, {Mainzer}, {Ressler}, {Cutri}, {Jarrett}, {Kirkpatrick}, {Padgett}, {McMillan}, {Skrutskie}, {Stanford}, {Cohen}, {Walker}, {Mather}, {Leisawitz}, {Gautier}, {McLean}, {Benford}, {Lonsdale}, {Blain}, {Mendez}, {Irace}, {Duval}, {Liu}, {Royer}, {Heinrichsen}, {Howard}, {Shannon}, {Kendall}, {Walsh}, {Larsen}, {Cardon}, {Schick}, {Schwalm}, {Abid}, {Fabinsky}, {Naes}, \& {Tsai}}]{Wright2010}
{Wright}, E.~L., {Eisenhardt}, P. R.~M., {Mainzer}, A.~K., {et~al.} 2010, \bibinfo{title}{{The Wide-field Infrared Survey Explorer (WISE): Mission Description and Initial On-orbit Performance},} \aj, 140, 1868, \dodoi{10.1088/0004-6256/140/6/1868}

\end{thebibliography}
\bibliographystyle{aasjournalv7}

\appendix

\section{Derivation of physical quantities}
\label{app:physical}

\subsection{Atomic gas surface density}
\label{app:sigatom}

We convert integrated 21-cm line intensity, \Ihi , to \sigatom\ in units of \sigatomunit\ via
\begin{equation}
\label{eq:21cmtosigatom}
\frac{\Sigma_{\rm atom}}{{\rm M_\odot~pc}^{-2}} = 0.020~\frac{I_{\rm 21cm}}{{\rm K~km~s}^{-1}}~\cos i
\end{equation}
\noindent which assumes optically thin 21-cm emission \citep[e.g.,][]{Walter2008} and includes an additional factor of $1.35$ to account for the mass of helium and heavier elements.

\subsection{Mass weighted mean atomic gas surface density} 
\label{app:mass_weighted_sigatom}
We calculate the mass-weighted mean atomic gas surface density within each averaging aperture (e.g., an annulus in the case of a radial profile). This quantity is calculated from the 120 pc-resolution data via: 
\begin{equation}
\label{eq:mass-weighted-sigatom}
\left\langle \Sigma_{\rm atom}^{120{\rm pc}} \right\rangle
=
f_\mathrm{corr}\,
\frac{\sum_{i} \Sigma^2_{{\rm atom},i}}
{\sum_{i} \Sigma_{{\rm atom},i}}
\end{equation}
where the summation is over each of the $i$th pixels above the detection threshold ($S/N > 3$). The prefactor $f_\mathrm{corr}$ corrects for the measurement bias due to finite data sensitivity, following \citet{Sun2022}. It assumes a lognormal distribution of surface densities within each averaging aperture and uses the fractional flux above the detection threshold on 120~pc scale to infer the correction factor.

As detailed in \citet{Leroy2016} and \citet{Sun2022}, this quantity captures the typical surface density from which a 21-cm photon arises within the region. It preserves information from the high resolution map by down-weighting empty or low column density regions.
We also calculate similar quantities for the molecular gas, $\left\langle \Sigma_{\rm mol}^{120{\rm pc}} \right\rangle$, from the CO data for IC 10, M33, and M31.

The ratio between the mass-weighted average and the direct aperture average gas surface densities is also known as the clumping factor \citep[see][]{Leroy2016}:
\begin{equation}
c_\mathrm{HI} = \left\langle \Sigma_{\rm atom}^{120{\rm pc}} \right\rangle / \Sigma_{\rm atom}
\end{equation}
It quantifies the width of the \hi\ gas column density distribution, with larger $c_\mathrm{HI}$ values reflecting wider distributions.

\subsection{Molecular gas surface density}
\label{app:alphaco}

We calculate \sigmol\ using a CO-to-H$_2$ conversion factor and the velocity-integrated \co10 line intensity $I_{\rm CO}$ (in units of \uIco) via
\begin{equation}
\frac{\sigmol}{{\rm M_\odot~pc}^{-2}} = \alphaCOline{1}{0}\frac{I_{\rm CO}}{{\rm K~km~s}^{-1}}\cos{i}~.
\label{eq:Sigma_mol}
\end{equation}
For M33, where we use \co21\ intensities, we apply an additional $R_{21}$ factor of 0.8 \citep[see][]{Druard2014} to convert \co21 to estimated \co10 . We use a \co10-to-H$_2$ conversion factor, $\alphaCOline{1}{0}$, from \citet{Schinnerer2024}:

\begin{equation}
    \alphaCOline{1}{0} = \alphaCOline{1}{0} {_{,\rm MW}}
    \left(\frac{Z}{Z_{\odot}}\right)^{-1.5}
    \left(\rm max\left[\frac{\Sigma_{\star}}{100\;\uSig}, 1 \right] \right)^{-0.25}~.
\end{equation}
Here $\alphaCOline{1}{0} {_{,\rm MW}} = 4.35$~M$_\odot$~pc$^{-2}$ (K~km~s$^{-1}$)$^{-1}$ is the Solar Neighborhood conversion factor \citep{Bolatto2013}. The metallicity ($Z$)-dependent term attempts to correct for the presence of additional CO-dark gas at low metallicity. The $\Sigma_\star$ term, which is only relevant in M31 and M33, attempts to correct for changes in the CO emissivity due to, e.g., variations in temperature and opacity. The \citet{Schinnerer2024} prescriptions are general scaling relations. H. Corbould et al. (in preparation) will present estimates of $\alpha_{\rm CO}$ specifically for LGLBS galaxies using \textit{Herschel} dust emission maps and the same gas maps we show here. We also calculate the molecular mass weighted surface density \mbox{$\langle \Sigma_{\rm mol, pix}^{120{\rm pc}}\rangle$} for IC~10, M33 and M31.

\subsection{Molecular gas expected based on star formation rate surface density}\label{app:expectedSigmol}
We estimate the molecular gas surface density expected based on the star formation rate surface density in each ring. This approach has significant associated uncertainty, but is useful to infer \sigmol\ in cases where $\alphaCOline{1}{0}$ is uncertain or where CO data are missing or have insufficient sensitivity \citep[e.g., see][]{Leroy2008,Schruba2012,Genzel2012}, a case that applies to the LGLBS dwarf galaxies.
For each annulus, we convert the star-formation rate surface density, $\Sigma_{\rm SFR}$ into an expected molecular surface density via
\begin{equation}
\frac{\Sigma_{\rm mol} (\tau_{\rm dep}^{\rm mol}) }{{\rm M_\odot~pc}^{-2}} 
\;=\; 10^{3} \frac{\sigsfr }{\uSigSFR }~\frac{\tau_{\rm dep}^{\rm mol}}{\rm Gyr}
\end{equation}
where $\tau^{\rm mol}_{\rm depl}$ is the assumed or estimate molecular gas depletion time. We choose a single $\tau^{\rm mol}_{\rm depl}$ per galaxy based on the $\tau_{\rm dep}^{\rm mol}${-}$M_\star$ relation in \citet[][their Eq. 12 and 13]{Sun2025}. This involves extrapolating their relation, which they derive over the range $M_\star = 10^{9.5-11}~{\rm M}_\odot$. IC~1613  and NGC~6822 have similar stellar masses ${\sim}10^8$ M$_{\odot}$, which imply $\tau_{\rm dep}^{\rm mol} \sim 0.7$ Gyr. For WLM, with $M_\star {\sim}10^{7.5}$ M$_{\odot}$, we estimate $\tau_{\rm dep}^{\rm mol} = 0.65$ Gyr. 

\subsection{Star formation rate surface density}\label{app:sigsfr}

\paragraph{Map construction} 

We trace stellar mass and star formation rate surface density using a combination GALEX \citep{Martin2005} UV and WISE/unWISE \citep{Wright2010,Lang2014} near- and mid-IR images constructed following $z0$MGS \citep{Leroy2019}. M31 and M33 were in the original $z0$MGS atlas but were not given special treatment and we improve on the processing here. Our four dwarf galaxy targets do not meet the $z0$MGS selection criteria because of their low absolute magnitude.

We built intensity maps from the GALEX tiles and the unWISE processing of neoWISE year 9 data (for W1 and W2) and the unWISE processing of the original WISE mission (for W3 and W4). Foreground stars were identified and masked using a combination of the 2MASS point source catalog for bright stars and Gaia DR3 for fainter stars. For Gaia, we required a $>3.5\sigma$ detection in either parallax or proper motion to classify sources as foreground. We predicted the flux of the stars in the 2MASS and WISE bands based on their Gaia G magnitudes and masked an appropriate area. Background galaxies were identified and masked based on HyperLEDA \citep{Makarov2014}. We also visually inspected each image to identify and mask instrumental artifacts. We estimated the background level by constructing and fitting a radial profile. The background level was identified as the value at which the profile reaches a plateau away from the galaxy. One target, IC 1613, first required a plane fit to the W3 and W4 emission. Finally, the GALEX UV images were corrected for Milky Way foreground extinction using the \citet{Schlegel1998} dust maps and the \citet{Peek2013} extinction curve. We used the following values for foreground $E(B{-}V)$: IC~10 0.70~mag, IC~1613 0.03~mag, NGC~6822 0.24~mag, WLM 0.038~mag, M33 0.04~mag, M31 0.07~mag. See \citet{Leroy2019} for more details on each step.

The resulting images are background-subtracted, extinction-corrected, and star-masked, and have units of MJy~sr$^{-1}$. We also created a set of images with a FWHM of 15, 20, and 30\arcsec for all bands and FWHM 7.5\arcsec for WISE1 through WISE3 and GALEX NUV and FUV. We use these in combination with the H$\alpha$ data (\autoref{tab:literature_obs}) to make several estimates of the star formation rate surface density (\sigsfr) in units of \sigsfrunit . 

\paragraph{FUV}~In star-forming environments, photospheric emission from young stars often dominates the FUV and NUV light. We convert FUV intensities, \Ifuv, corrected for Galactic but not internal extinction into estimates of \sigsfr\ via
\begin{equation}\label{eq:SFR_FUV}
    \frac{\sigsfr_{, \rm FUV}}{\uSigSFR } = 1.04 \times10^{-44.24}\times\frac{I_{\rm FUV}}{\rm MJy~sr^{-1} }~.
\end{equation}
The conversion factor comes from \citet{Leroy2019} based on the SED modeling of SDSS galaxies by \citet{Salim2016,Salim2018} and is within $\sim 0.1$~dex of the conversion suggested by \citet{Salim2007} and \citet{Kennicutt2012}.

\paragraph{WISE4} Mid-IR emission is often associated with recent star formation \citep[e.g., see][]{Kennicutt2012}. We use WISE4 data to estimate
\begin{equation}\label{eq:SigSFR_W4ONLY}
    \frac{{\sigsfr _{, \rm W4}}}{{\uSigSFR}} = 3.8\times10^{-3}\,\left(\frac{I\sbsc{22\,\mu m}}{\uI}\right) \cos{i}~.
\end{equation}
following \cite{Leroy2019} based on SED modeling by \citet{Salim2016,Salim2018}. In environments with little dust the mid-IR alone is expected to underestimate the SFR.

\paragraph{FUV+WISE4}~Following, e.g., \citet{Leroy2008} FUV and mid-IR emission can be combined to estimate the total obscured and unobscured star formation activity over large regions in galaxies. We estimate \sigsfr $_{, \rm FUV+W4}$ following the calibration of \citet[][]{Belfiore2023}:

\begin{equation}
\label{eq:sigsfr_fuvw4}
\frac{\sigsfr _{, \rm FUV+W4}}{\uSigSFR }
 = \Sigma_{\rm SFR,FUV} + 3.39 \times 10^{-3}\,
     \left( \frac{C^{\rm FUV}_{\rm W4}}{10^{-42.68}} \right)
     \frac{I\sbsc{22\,\mu m}}{\uI} \cos{i}~ 
\end{equation}

Here $C_{\rm W4}^{\rm FUV}$ is an empirically calibrated coefficient designed to yield a total \sigsfr\ that matches some set of known \sigsfr\ values. \citet{Belfiore2023} calibrates $C_{\rm W4}^{\rm FUV}$ to match H$\alpha$ emission corrected for extinction using the Balmer decrement, i.e., combining $H\beta$ and H$\alpha$. In the \citet{Belfiore2023} calibration, $C_{\rm W4}^{\rm FUV}$ varies as a function of the local specific star formation rate in a way that accounts for contamination of the W4 band by infrared cirrus. The fiducial $C_{\rm W4}^{\rm FUV}$ quoted above is the maximum value in the \citet{Belfiore2023} calibration.

\paragraph{H$\alpha$}~
We estimate the SFR surface density from the observed H$\alpha$ intensity without any correction for internal extinction via
\begin{equation}
    \frac{\sigsfr_{, \rm H\alpha}}{\uSigSFR } =
    2.7\times10^{13}\,\frac{I\sbsc{H\alpha}}{\uIha}\,\cos i~,
    \label{eq:SigSFR_HaONLY}
\end{equation}
where the prefactor comes from \citet{Murphy2011}, who used STARBURST99 \citep{Leitherer1999} and assumed continuous star formation and a \citet{Kroupa2001} IMF with maximum stellar mass $100$~M$_\odot$.

\paragraph{H$_{\alpha}$+WISE4}~To account for internal extinction, we combine H$\alpha$ with mid-IR data following \citet{Calzetti2007}. Specifically, we use the WISE4 22~\micron\ data with the prescription of \citet{Belfiore2023} to estimate the amount of H$\alpha$ emission obscured by dust,

\begin{equation}
    \frac{\sigsfr _{, \rm H\alpha+W4}}{\uSigSFR } = 
     \Sigma_{\rm SFR, H\alpha} + \\
     2.14 \times10^{-3}\,
     \left( \frac{C^{\rm H\alpha}_{\rm W4}}{10^{-42.88}} \right)
     \frac{I\sbsc{22\,\mu m}}{\uI} \cos{i}~,
\label{eq:SigSFR_HaW4}
\end{equation}

\noindent Here $C^{\rm H\alpha}_{\rm W4}$ is an empirically calibrated factor. In \citet{Belfiore2023}, $C^{\rm H\alpha}_{\rm W4}$ varies as a function of the H$\alpha$-to-W1 ratio to account for a varying IR cirrus contribution. The fiducial $10^{-42.88}$ is their maximum value, approached in regions dominated by star formation activity, and corresponds well to the \citet{Calzetti2007} or \citet{Belfiore2023b} calibrations for individual star-forming regions.

\subsection{Stellar mass surface density} \label{app:sigstar}

We estimate the stellar mass surface density, \sigstar, from near-infrared emission at 3.4~$\mu$m (W1). We adopt a single mass-to-light ratio for each galaxy, $\MtoLwiseone$, which we estimate based on the photometry in \autoref{tab:fluxes} and report in the same table. The stellar mass surface density is then
\begin{equation}
    \frac{\sigstar}{\uSig} =
    330 \times
    \left(\frac{\MtoLwiseone}{0.5}\right)
    \frac{\Iwiseone}{\uI}
    \cos{i}~,
    \label{eq:Sigstar_3p4um}
\end{equation}
where $\MtoLwiseone$ is the mass-to-light ratio at 3.4~$\mu$m in units of Solar mass per Solar 3.4~$\mu$m luminosity.
We follow the calibration in \citet{Leroy2019}, which is based on \citet{Salim2016,Salim2018}. 
For comparison, we also compute alternative \sigstar\ profiles using the direct point-by-point SFR-to-W1 prescription, the W4-to-W1 prescription, and the global GSWLC-based prescription. These offer the prospect to account for mass-to-light ratio gradients, e.g., as seen in M31 by \citet{Telford2020}, but because we lack SFR/M$_\star$ estimates over the outer disks of most of our targets, we defer applying these to future work.

\section{M31: Correction for old stellar contamination in IR and UV Tracers}
\label{app:oldstar_correction}

The inner region of M31 has high stellar surface density and little gas, dust, or recent star formation. As a result, the observed IR and UV emission are affected by the older stellar populations. Interstellar dust heated by older stars, the ``IR cirrus,'' can contribute a large fraction of the mid-IR intensity in this regime \citep{Groves2012,Leroy2012,Davis2014}, together with circumstellar dust  \citep{Simonian2017}. Meanwhile, the UV bands can include contributions from evolved populations that can become dominant in the absence of recent star formation \citep[e.g.,][]{Rosenfield2012}. If uncorrected, these effects can cause the apparent SFR surface density profile to closely track the stellar mass surface density in the inner part of M31, which is incorrect based on, e.g., the distribution of H$\alpha$ emission \citep{Massey2006} or modeling of the recent star formation history \citep{Lewis2015}.

To account for this, we followed the approach of \citet{Davis2014} and used the WISE 3.4\,$\mu$m band (W1), which traces the old stellar mass distribution, as a reference for the contamination. We fit a linear relation with a fixed $y$-intercept of $0$ between the $22\mu$m W4 intensity and W1 intensity restricted to the inner region where emission from star formation, gas, and dust are small compared to the old stellar surface population. We then used the best-fit slope $\alpha$ to scale W1 and estimate the contamination affecting W4 over the full map:
\begin{equation}
  I^{\rm corr}_{\rm W4} = I^{\rm obs}_{\rm W4} - \alpha \, I_{\rm W1} .
\end{equation}
We applied an analogous procedure to relate FUV to W1 and subtract a W1-based estimate of FUV contamination. $\alpha$ is 0.135 and 0.001 for W4 and FUV, respectively.  Formally, we expect that using W1 as a template is more appropriate for the FUV contamination, which arises from the stellar population itself. The W4 contamination will be a mixture of circumstellar dust, which will be well-traced by W1, and interstellar dust heated by an intense but soft radiation field related to the old stars \citep[e.g.,][]{Groves2012}. In that case, the contamination will have a distribution that reflects both the location of the ISM material and the radiation field \citep[e.g., see][]{Leroy2012,Leroy2023}. We proceed using the W1 as a template for W4 for simplicity and under the assumption that the radiation field variations dominate the contamination.

\begin{figure}
    \centering
    \includegraphics[width=0.9\linewidth]{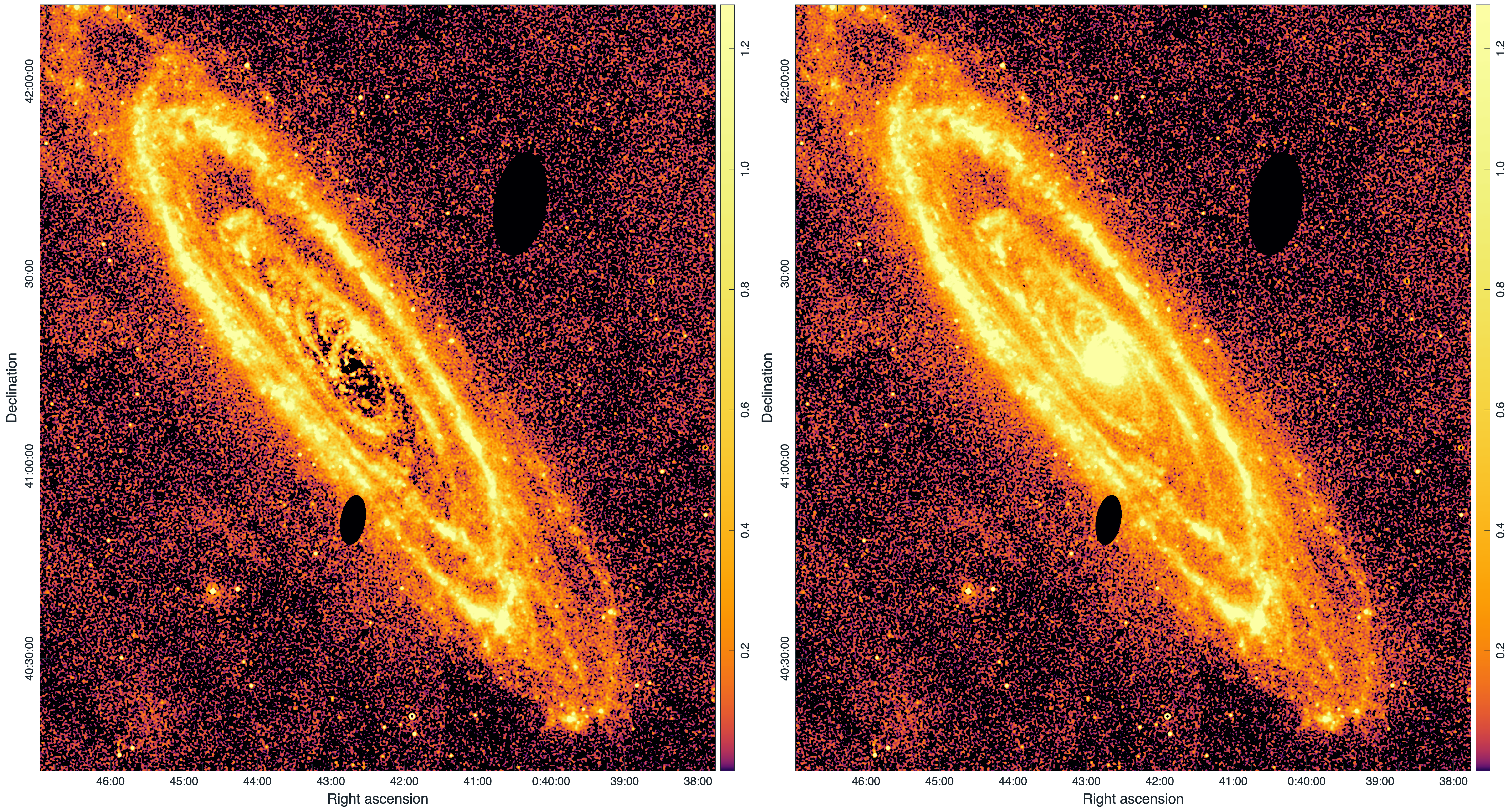}
    \caption{Correction for IR cirrus and stellar contamination in M31. The left panel shows the W4 map corrected by scaling and subtracting the W1 map. The right panel shows the uncorrected W4 map. The correction removes a large amount of emission not associated with recent star formation from the central region of M31 but does not significantly affect the disk (see \autoref{app:m31_sfr_plot}). We apply an analogous procedure to the FUV emission.}
    \label{app:m31_w1corr}
\end{figure}

\begin{figure}
    \centering
    \includegraphics[width=0.9\linewidth]{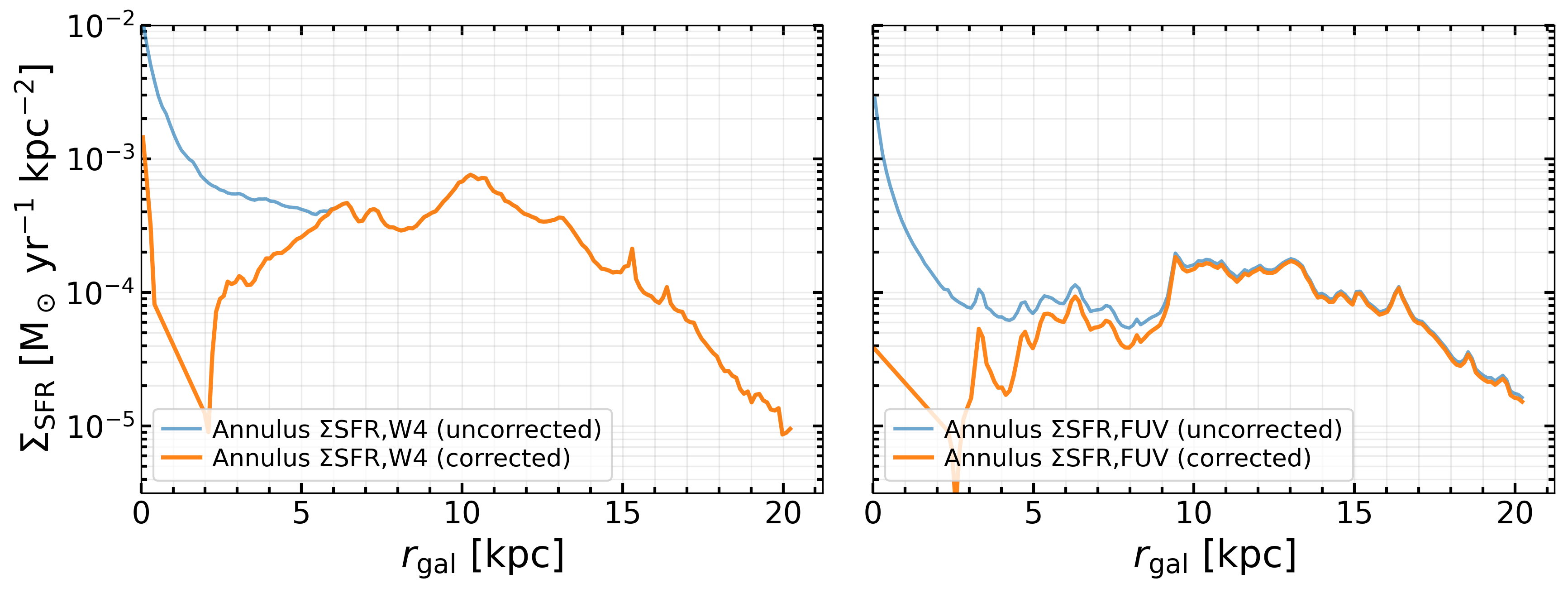}
    \caption{Left: Curves show the W4 ($22\,\mu$m) annular profiles for M31: uncorrected (blue) and after subtracting the W1-based contaminant estimate (orange). The W1-based subtraction suppresses central emission unassociated with star formation while preserving the star-forming ring/outer disk. Right: The same for FUV profiles. See Appendix \ref{app:oldstar_correction}}
    \label{app:m31_sfr_plot}
\end{figure}

We show the corrected and uncorrected W4 maps in \autoref{app:m31_w1corr} and the corrected and uncorrected radial profiles in \autoref{app:m31_sfr_plot}. The $\Sigma_{\rm SFR}$ presented in the main text use these contamination-corrected estimates. This W1-based correction has a significant impact on the central regions of M31. It has only a small effect at larger radii, where star-forming regions are the main source of IR and UV emission. In the outer parts of M31 and the other targets in our sample, the relevant W4 cirrus term will be more related to the distribution of the dust column density, not variations in the interstellar radiation field. We leave treating this additional cirrus term for future work.

\startlongtable
\begin{deluxetable*}{lllcl}
\tabletypesize{\footnotesize}
\tablecaption{\label{tab:radii} Selected integration radii for fluxes, luminosities, and physical quantities measurements used in Table \ref{tab:fluxes} and shown in vertical dashed lines in radial profiles of \sigstar, \sigsfr\ and depletion times.}
\tablewidth{0pt}
\tablehead{
\colhead{Galaxy} & \colhead{Tracer} & \colhead{Radius} & \colhead{Value [kpc]} & \colhead{Selection basis}
}
\startdata
IC10 & W1 & $r_{\rm conv}$ & $2.52$ & cumulative-profile convergence threshold\\
IC10 & W2 & $r_{\rm conv}$ & $2.52$ & cumulative-profile convergence threshold\\
IC10 & W3 & $r_{\rm conv}$ & $1.32$ & selected by-eye from profile\\
IC10 & W4 & $r_{\rm conv}$ & $1.20$ & cumulative-profile convergence threshold\\
IC10 & H$\alpha$ & $r_{\rm conv}$ & $0.96$ & cumulative-profile convergence threshold\\
IC10 & HI & $r_{\rm conv}$ & $3.84$ & profile still rising; adopt last reliable radius\\
IC10 & CO(1-0) & $r_{\rm conv}$ & $1.32$ & cumulative-profile convergence threshold\\
IC1613 & W1 & $r_{\rm bkg}$ & $3.60$ & background-corrected profile\\
IC1613 & W2 & $r_{\rm bkg}$ & $3.36$ & background-corrected profile\\
IC1613 & W3 & $r_{\rm conv}$ & $2.28$ & selected by-eye from profile\\
IC1613 & W4 & $r_{\rm conv}$ & $1.08$ & selected by-eye from profile\\
IC1613 & FUV & $r_{\rm conv}$ & $2.52$ & cumulative-profile convergence threshold\\
IC1613 & NUV & $r_{\rm conv}$ & $3.00$ & cumulative-profile convergence threshold\\
IC1613 & H$\alpha$ & $r_{\rm conv}$ & $1.32$ & cumulative-profile convergence threshold\\
IC1613 & HI & $r_{\rm conv}$ & $4.08$ & profile still rising; adopt last reliable radius\\
WLM & W1 & $r_{\rm bkg}$ & $3.60$ & background-corrected profile\\
WLM & W2 & $r_{\rm bkg}$ & $3.84$ & background-corrected profile\\
WLM & W3 & $r_{\rm conv}$ & $3.24$ & selected by-eye from profile\\
WLM & W4 & $r_{\rm conv}$ & $1.44$ & selected by-eye from profile\\
WLM & FUV & $r_{\rm conv}$ & $1.80$ & cumulative-profile convergence threshold\\
WLM & NUV & $r_{\rm conv}$ & $3.00$ & cumulative-profile convergence threshold\\
WLM & H$\alpha$ & $r_{\rm conv}$ & $2.04$ & cumulative-profile convergence threshold\\
WLM & HI & $r_{\rm conv}$ & $4.56$ & profile still rising; adopt last reliable radius\\
WLM & CO(2-1) & $r_{\rm conv}$ & $0.96$ & cumulative-profile convergence threshold\\
NGC6822 & W1 & $r_{\rm bkg}$ & $2.64$ & background-corrected profile\\
NGC6822 & W2 & $r_{\rm bkg}$ & $2.52$ & background-corrected profile\\
NGC6822 & W3 & $r_{\rm conv}$ & $2.04$ & selected by-eye from profile\\
NGC6822 & W4 & $r_{\rm conv}$ & $2.16$ & selected by-eye from profile\\
NGC6822 & FUV & $r_{\rm conv}$ & $2.52$ & cumulative-profile convergence threshold\\
NGC6822 & NUV & $r_{\rm conv}$ & $2.64$ & cumulative-profile convergence threshold\\
NGC6822 & H$\alpha$ & $r_{\rm conv}$ & $2.28$ & cumulative-profile convergence threshold\\
NGC6822 & HI & $r_{\rm conv}$ & $3.00$ & profile still rising; adopt last reliable radius\\
M33 & W1 & $r_{\rm conv}$ & $7.20$ & background-corrected profile\\
M33 & W2 & $r_{\rm conv}$ & $7.08$ & background-corrected profile\\
M33 & W3 & $r_{\rm conv}$ & $6.48$ & cumulative-profile convergence threshold\\
M33 & W4 & $r_{\rm conv}$ & $6.48$ & cumulative-profile convergence threshold\\
M33 & FUV & $r_{\rm conv}$ & $7.56$ & cumulative-profile convergence threshold\\
M33 & NUV & $r_{\rm conv}$ & $7.32$ & cumulative-profile convergence threshold\\
M33 & H$\alpha$ & $r_{\rm conv}$ & $6.84$ & cumulative-profile convergence threshold\\
M33 & HI & $r_{\rm conv}$ & $12.60$ & cumulative-profile convergence threshold\\
M33 & CO(2-1) & $r_{\rm conv}$ & $7.68$ & cumulative-profile convergence threshold\\
M31 & W1 & $r_{\rm conv}$ & $18.48$ & cumulative-profile convergence threshold\\
M31 & W2 & $r_{\rm conv}$ & $19.20$ & cumulative-profile convergence threshold\\
M31 & W3 & $r_{\rm conv}$ & $18.60$ & cumulative-profile convergence threshold\\
M31 & W4 & $r_{\rm conv}$ & $18.84$ & cumulative-profile convergence threshold\\
M31 & FUV & $r_{\rm conv}$ & $19.44$ & cumulative-profile convergence threshold\\
M31 & NUV & $r_{\rm conv}$ & $19.56$ & cumulative-profile convergence threshold\\
M31 & HI & $r_{\rm conv}$ & $20.28$ & cumulative-profile convergence threshold\\
M31 & CO(1-0) & $r_{\rm conv}$ & $16.56$ & cumulative-profile convergence threshold\\
\enddata
\tablecomments{We use $r_{\rm conv}$ for radii selected from cumulative-profile convergence and $r_{\rm bkg}$ for background-corrected W1/W2 profiles where the adopted integration radius is set after subtracting an outer-profile pedestal.}
\end{deluxetable*}






\end{document}